 \documentclass[final,5p,times,twocolumn]{elsarticle}

\usepackage[colorlinks]{hyperref}
\hypersetup{linkcolor=blue,citecolor=blue,urlcolor=blue}
\usepackage{dcolumn}
\usepackage{braket}
\usepackage{amssymb}
\usepackage{amsmath}
\usepackage{lineno}

\journal{Physics Letters B}

\biboptions{square,comma,sort&compress}

\begin{document}
\begin{frontmatter}



\title{Purifying one-neutron removal as a probe of single-particle strength}


\author[1]{Erxi Xiao}
\author[2]{Guangshuai Li}
\author[1]{Yu Yang}
\author[1]{Long Zhu}
\author[2]{Jianwei Zhao}
\author[1]{Jun Su \corref{cor1}}
\ead{sujun3@mail.sysu.edu.cn}
\author[2]{Baohua Sun \corref{cor1}}
\ead{bhsun@buaa.edu.cn}

\address[1]{Sino-French Institute of Nuclear Engineering and Technology, Sun Yat-sen University, Zhuhai 519082, China}
\address[2]{School of Physics, Beihang University, Beijing 100191, China}

\cortext[cor1]{Corresponding author.}

\begin{abstract}

One-neutron removal reactions exhibit a strong proton--neutron asymmetry dependence in the inclusive reduction factor $R_s$, a long-standing issue that has been discussed in terms of both possible intrinsic isospin dependence of single-particle strength and reaction-mechanism effects.
We address this issue by reframing inclusive removal as a coupled fast-dynamics and deexcitation process, and by validating this transport--deexcitation chain against a global, mutually constraining data set.
Confronting 73 one-neutron removal cross sections and 28 residue parallel-momentum distributions with isospin-dependent quantum molecular dynamics followed by GEMINI evaporation shows that the apparent $R_s$--$\Delta S$ trend is correlated with evaporation feeding and evaporation loss.
By subtracting the feeding contribution and correcting for the loss component in the measured cross sections, we construct a purified reduction factor $R_{\rm dir}$, that more closely reflects single-particle strength than the inclusive $R_s$.
The resulting $R_{\rm dir}$ exhibits a much weaker $\Delta S$ dependence within current uncertainties, consistent with the weak isospin-asymmetry dependence observed in nucleon-transfer and quasifree-knockout systematics.

\end{abstract}




\end{frontmatter}



\section{Introduction}
\label{introduction}

Correlations are a defining feature of nuclei that quench single-particle strength and delimit the independent-particle picture, with direct implications for nuclear spectroscopy. Quasifree $(e,e^{\prime}p)$ measurements on stable nuclei established a 30--40\% suppression of valence strength relative to independent-particle estimates \cite{lapikas1993quasi}, commonly interpreted as arising from correlations beyond the mean field \cite{piasetzky2006evidence, subedi2008probing, barbieri2009role, paschalis2020nucleon}. Hadronic direct reactions with radioactive-ion beams extend these spectroscopic constraints to rare isotopes with extreme proton--neutron asymmetry, but have produced different quenching systematics across probes \cite{gade2008reduction, lee2010neutron, aumann2021quenching}. These differences raise a broader question of how reliably reaction observables can be converted into spectroscopic information across large proton--neutron asymmetries.

A particularly influential quantity is the inclusive reduction factor $R_s=\sigma_{\rm exp}/\sigma_{\rm eik}$ in intermediate-energy nucleon-removal reactions, where $\sigma_{\rm eik}$ is computed in an eikonal picture with shell-model spectroscopic-factor input \cite{tostevin2001single,hansen2003direct}. Systematic analyses of one-neutron and one-proton removal show a pronounced negative trend of $R_s$ with the proton--neutron Fermi-surface asymmetry $\Delta S$, defined from the one-neutron and one-proton separation energies $S_n$ and $S_p$ (neutron removal: $S_n-S_p$; proton removal: $S_p-S_n$) \cite{tostevin2014systematics,tostevin2021updated}. In contrast, nucleon transfer and quasifree $(p,pN)$ knockout yield reduction factors that are moderately quenched yet nearly $\Delta S$ independent \cite{flavigny2013limited,kay2013quenching,atar2018quasifree,gomez2018binding,aumann2013quasifree}, fueling debate over whether the apparent $R_s$--$\Delta S$ trend is intrinsic or generated by mechanism-dependent biases.

One source of such mechanism-dependent bias is the breakdown of the inert-core spectator picture in intermediate-energy nucleon removal, as indicated for deeply bound removal and reactions populating resonant continua by distorted residue parallel-momentum distributions and unusually strong quenching \cite{grinyer2011knockout,flavigny2012nonsudden,charity2020single}. Such dissipative dynamics can redirect flux both out of the observed $(A-1)$ channel through reduced residue survival (including ``core destruction'') \cite{bertulani2023core,gomez2023isospin} and into the channel via inelastic-scattering pathways followed by evaporation \cite{sun2016experimental,pohl2023multiple,li2024single}. Because these loss and feeding contributions evolve systematically with separation energies, comparing inclusive $\sigma_{\rm exp}$ directly with standard eikonal $\sigma_{\rm eik}$ can bias $R_s$ and mimic an apparent $R_s$--$\Delta S$ trend unrelated to intrinsic correlations \cite{li2024single}. The key issue is therefore whether these contributions can be quantified and removed to isolate a direct spectroscopic component.

In this Letter we use an isospin-dependent quantum molecular dynamics calculation coupled to GEMINI evaporation model (IQMD+GEMINI) to analyze all published one-neutron removal cross sections (73 systems) and residue parallel-momentum distributions (28 systems). The transport--deexcitation chain is used here as a reaction-mechanism filter for the inclusive yield, rather than as a replacement for quantum direct-reaction theory. IQMD+GEMINI perform a global mechanism decomposition of evaporation feeding and evaporation loss, to purify the inclusive cross sections, and to define a purified reduction factor $R_{\rm dir}$ relative to an external direct-reaction baseline with shell-model spectroscopic factors. We show that removing the estimated feeding and loss contributions strongly suppresses the inclusive $R_s$--$\Delta S$ trend and brings the resulting systematics closer to transfer and quasifree benchmarks.

\section{\label{sec:theory}Theoretical framework and implementation}

The starting point of the present framework is not to replace the quantum knockout baseline, but to distinguish the measured inclusive residue yield from the primary removal yield to which that baseline is normally compared.
In standard eikonal analyses, the calculated cross section, folded with shell-model spectroscopic factors, provides the direct-reaction reference.
However, dissipative dynamics and subsequent deexcitation can feed flux into, or remove flux from, the observed $A-1$ residue channel, so that the inclusive yield need not be identical to the primary removal component conventionally benchmarked against single-particle strength.
Schematically, the observed $A-1$ residue can therefore be produced through two classes of pathways,
\begin{equation}
{}^{A}_{Z}X \rightarrow {}^{A-1}_{\ \ Z}X^\ast 
\rightarrow {}^{A-1}_{\ \ Z}X ,
\label{eq:path-surv}
\end{equation}
and
\begin{equation}
{}^{A}_{Z}X \rightarrow {}^{A}_{Z}X^\ast 
\rightarrow {}^{A-1}_{\ \ Z}X+n ,
\label{eq:path-feed}
\end{equation}
where the first pathway represents fast-stage $A-1$ prefragment production followed by survival to the observed residue, while the second represents evaporation feeding from an excited intact projectile-like prefragment.
The excited $A-1$ prefragment in Eq.~\eqref{eq:path-surv} may also undergo further particle emission and be lost from the observed $A-1$ residue channel.

To implement this schematic separation quantitatively, we group projectile-like events at the end of the fast dynamical stage into two operational branches: (i) intact excited prefragments ${}^{A}_{Z}X^\ast$, and (ii) excited $A-1$ prefragments ${}^{A-1}_{\ \ Z}X^\ast$ produced during the dynamical stage.
We denote the corresponding transport-stage cross sections as
\begin{equation}
\sigma_{\rm dyn}^{(Z,A)}\ \big[{}^{A}_{Z}X\!\rightarrow\!{}^{A}_{Z}X^\ast\big],\qquad
\sigma_{\rm dyn}^{(Z,A-1)}\ \big[{}^{A}_{Z}X\!\rightarrow\!{}^{A-1}_{\ \ Z}X^\ast\big].
\label{eq:transport}
\end{equation}
Each prefragment ensemble is then deexcited event by event.
The intact ${}^{A}_{Z}X^\ast$ branch can feed the observed ${}^{A-1}_{\ \ Z}X$ residue through neutron evaporation with probability $P_{\rm feed}$.
The dynamical ${}^{A-1}_{\ \ Z}X^\ast$ branch can either survive as the observed residue with probability $1-P_{\rm loss}$ or be lost from the observed residue channel through additional particle emission with probability $P_{\rm loss}$.
This defines three contributions,
\begin{equation}
\begin{aligned}
\sigma_{\rm feed} &= P_{\rm feed}\,\sigma_{\rm dyn}^{(Z,A)},\\
\sigma_{\rm surv} &= (1-P_{\rm loss})\,\sigma_{\rm dyn}^{(Z,A-1)},\\
\sigma_{\rm loss} &= P_{\rm loss}\,\sigma_{\rm dyn}^{(Z,A-1)} .
\end{aligned}
\label{eq:channels}
\end{equation}
Thus, the observed residue channel contains evaporation feeding from the intact branch and only the surviving part of the dynamical $A-1$ branch.
The corresponding inclusive one-neutron removal cross section is
\begin{equation}
\sigma_{\rm incl}=\sigma_{\rm feed}+\sigma_{\rm surv}.
\label{eq:inclusive}
\end{equation}

In the above decomposition, the dynamical $A-1$ prefragment-production cross section is
$\sigma_{\rm dyn}^{(Z,A-1)}=\sigma_{\rm surv}+\sigma_{\rm loss}$,
whereas the measured inclusive residue yield contains the surviving part $\sigma_{\rm surv}$ together with the evaporation-feeding contribution $\sigma_{\rm feed}$.
Accordingly, the measured inclusive cross section is purified by subtracting the feeding contribution and unfolding the survival probability,
\begin{equation}
  \sigma_{\rm pur}^{\rm exp} \equiv \frac{\sigma_{\rm incl}^{\rm exp}\,(1-f_{\rm feed})}{1-P_{\rm loss}} \,,
\label{eq:sig-final}
\end{equation}
where the evaporation-feeding fraction is
\begin{equation}
f_{\rm feed}=\frac{\sigma_{\rm feed}}{\sigma_{\rm incl}}\,.
\label{eq:fractions}
\end{equation}
This purified cross section is the experimental estimate of the non-feeding removal yield before secondary evaporation loss.
It is therefore closer than the measured inclusive yield to the removal component conventionally compared with the eikonal direct-reaction baseline.
The purified cross section is then compared with the same eikonal direct-reaction baseline to define
\begin{equation}
  R_{\rm dir} \equiv \frac{\sigma_{\rm pur}^{\rm exp}}{\sigma_{\rm eik}}
  = R_s\,\frac{1-f_{\rm feed}}{1-P_{\rm loss}}\,,
\label{eq:Rdir-final}
\end{equation}
with $R_s\equiv\sigma_{\rm incl}^{\rm exp}/\sigma_{\rm eik}$.
Thus, $R_{\rm dir}$ does not introduce a new microscopic structure calculation.
Rather, it is a purified reduction factor intended to be closer than the inclusive $R_s$ to the spectroscopic-strength reduction factor conventionally extracted from one-nucleon removal.

The remaining task is to estimate the feeding and loss corrections, $f_{\rm feed}$ and $P_{\rm loss}$.
Before a fully experimental event-by-event separation becomes available, we use IQMD+GEMINI as a transport--deexcitation framework for this purpose.
The eikonal calculation is kept as the direct-reaction reference $\sigma_{\rm eik}$, while IQMD+GEMINI is used only to estimate the feeding and loss corrections applied to the measured inclusive yield.
Specifically, we use IQMD for the fast dynamical stage, followed by GEMINI for statistical deexcitation.
In IQMD, the time evolution is propagated self-consistently in a Skyrme-like energy-density functional with stochastic in-medium $NN$ scatterings subject to Pauli blocking \cite{aichelin1991quantum,wolter2022transport}.
In the present implementation, the phase-space density constraint (PSDC) is adopted to improve the fermionic properties, which has been shown in previous applications to give a good description of peripheral collisions \cite{su2018dynamical,su2019uniform,su2022fluctuations}.
GEMINI then simulates the statistical deexcitation of the excited prefragments by Monte Carlo sampling of sequential decay channels from partial widths until further decay is energetically forbidden (see Refs.~\cite{charity1988systematics, li2024single}).
A single standard IQMD+GEMINI parameter set is used for all 73 systems, without tuning to individual reactions.

\section{Results and discussions}

\begin{figure}[t]
  \centering
  \includegraphics[width=\columnwidth]{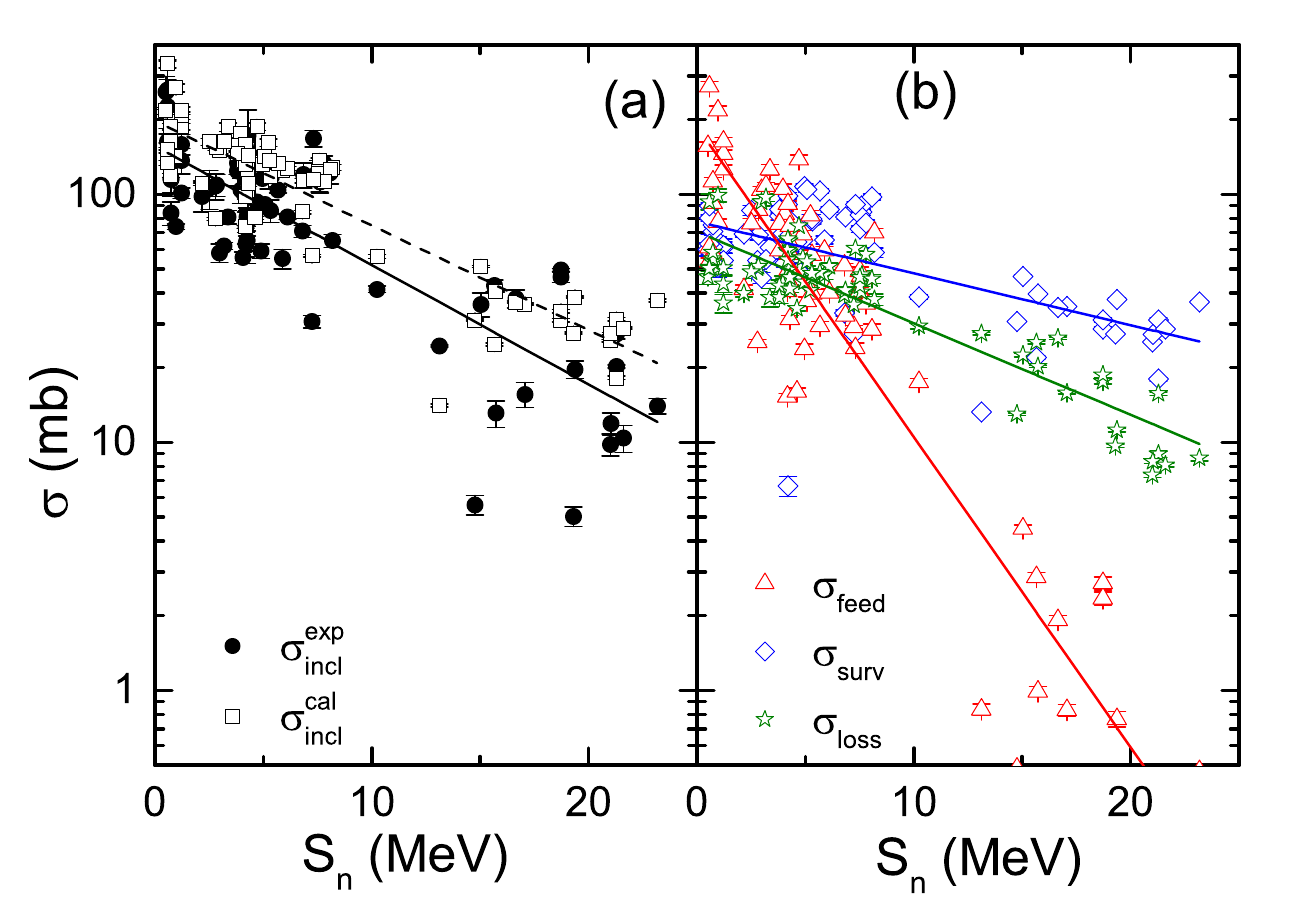}
\caption{\label{fig:sigmaSn}
One-neutron removal cross sections as a function of the neutron separation energy $S_n$.
(a) Comparison between the measured inclusive cross sections $\sigma_{\mathrm{incl}}^{\mathrm{exp}}$ and the calculated inclusive cross sections $\sigma_{\mathrm{incl}}^{\mathrm{cal}}$ from the IQMD+GEMINI model.
(b) Decomposition of $\sigma_{\mathrm{incl}}^{\mathrm{cal}}$ into the evaporation-feeding contribution $\sigma_{\mathrm{feed}}$ and the surviving component $\sigma_{\mathrm{surv}}$ of the dynamical $A-1$ branch.
The evaporation-loss component $\sigma_{\mathrm{loss}}$ of the same branch is also shown.
The curves illustrate the exponential-like $S_n$ systematics and are used only to guide the eye.
Error bars on calculated points denote Monte-Carlo counting statistics.
The full data set is listed in Table~I of the Supplemental Material.}
\end{figure}

Figure~\ref{fig:sigmaSn}(a) summarizes all published inclusive cross sections as a function of $S_n$.
The IQMD+GEMINI inclusive cross section $\sigma_{\mathrm{incl}}^{\mathrm{cal}}(S_n)$ reproduces the measured envelope $\sigma_{\mathrm{incl}}^{\mathrm{exp}}(S_n)$, both decreasing approximately exponentially with $S_n$.
As shown in Fig.~\ref{fig:sigmaSn}(b), the decomposition of $\sigma_{\mathrm{incl}}^{\mathrm{cal}}$ highlights the mechanism content of the global trend:
(i) the evaporation-feeding term $\sigma_{\mathrm{feed}}$ shows the strongest $S_n$ dependence, as expected from Weisskopf--Ewing-type evaporation widths that scale as $\exp(-S_n/T)$~\cite{weisskopf1940yield}; $\sigma_{\mathrm{feed}}$ therefore dominates at the most neutron-rich (small-$S_n$) end.
(ii) The surviving component of the dynamical $A-1$ yield, $\sigma_{\mathrm{surv}}$, becomes increasingly important toward the neutron-deficient side.
(iii) A substantial fraction of the dynamical $A-1$ yield is removed from the inclusive channel by evaporation loss, quantified by $\sigma_{\mathrm{loss}}$.
The complementary evolution of $\sigma_{\mathrm{feed}}$ and $\sigma_{\mathrm{loss}}$ across the data set indicates that the observed $S_n$ systematics are largely reaction-mechanism driven.

\begin{figure}[t]
  \centering
  \includegraphics[width=\columnwidth]{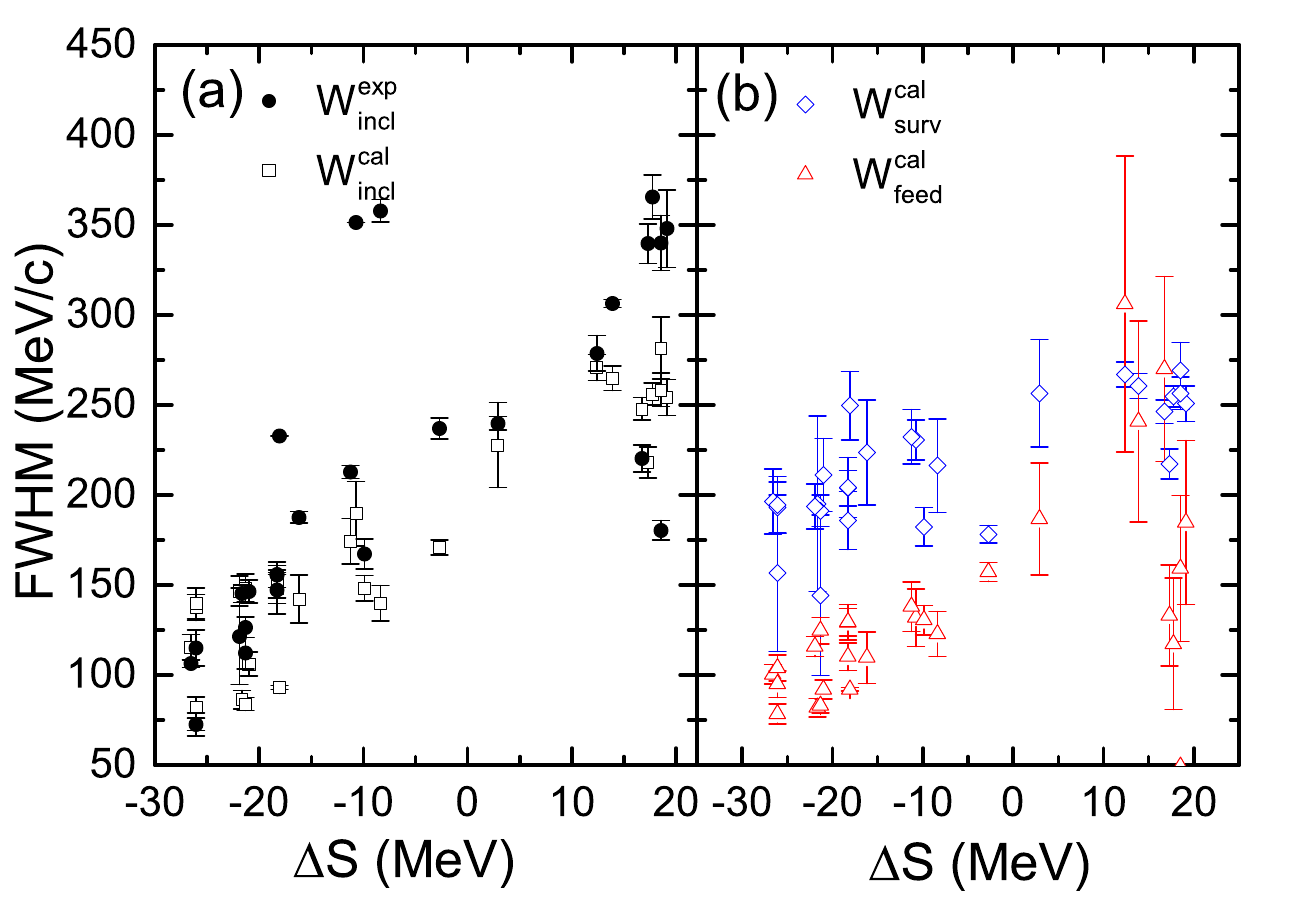}
\caption{\label{fig:FWHM}
The full widths at half maximum (FWHM) of the residue parallel-momentum distributions for the 28 systems with measured spectra.
(a) Experimental inclusive widths $W_{\rm incl}^{\rm exp}$ compared with the corresponding IQMD+GEMINI results $W_{\rm incl}^{\rm cal}$.
(b) Decomposed widths of the surviving component of the dynamical $A-1$ yield, $W_{\rm surv}^{\rm cal}$, and the evaporation-feeding component, $W_{\rm feed}^{\rm cal}$, highlighting their distinct spectral scales.
The corresponding momentum distributions are shown in Appendix~B of the Supplemental Material.}
\end{figure}

A complementary constraint on the mechanism decomposition is provided by the residue parallel-momentum distributions, which probe the reaction dynamics through their spectral shapes.
As shown in Fig.~\ref{fig:FWHM}(a), the measured inclusive width $W_{\rm incl}^{\rm exp}$ shows an overall increase with $\Delta S$.
The IQMD+GEMINI calculation reproduces this global behavior and yields a robust hierarchy $W_{\rm surv}^{\rm cal}>W_{\rm feed}^{\rm cal}$ [Fig.~\ref{fig:FWHM}(b)].
The extracted component widths depend only weakly on $\Delta S$, with the remaining scatter reflecting variations of the reaction system and beam energy.
In the model, events classified as survival from the dynamical $A-1$ branch involve more violent dynamical interactions (including in-medium $NN$ scatterings and a non-sudden core response), leading to larger recoil fluctuations and broader spectra.
In contrast, evaporation feeding proceeds from projectile-like prefragments with smaller net recoil, with the evaporation recoil providing only a secondary broadening, and thus results in a narrower residue spectrum.
Consequently, neutron-rich systems (small $S_n$, negative $\Delta S$) tend to exhibit narrower $W_{\rm incl}^{\rm exp}$ due to feeding dominance, whereas toward neutron-deficient nuclei the increasing survival fraction of the dynamical $A-1$ branch broadens $W_{\rm incl}^{\rm exp}$.

\begin{figure}[t]
  \centering
  \includegraphics[width=\columnwidth]{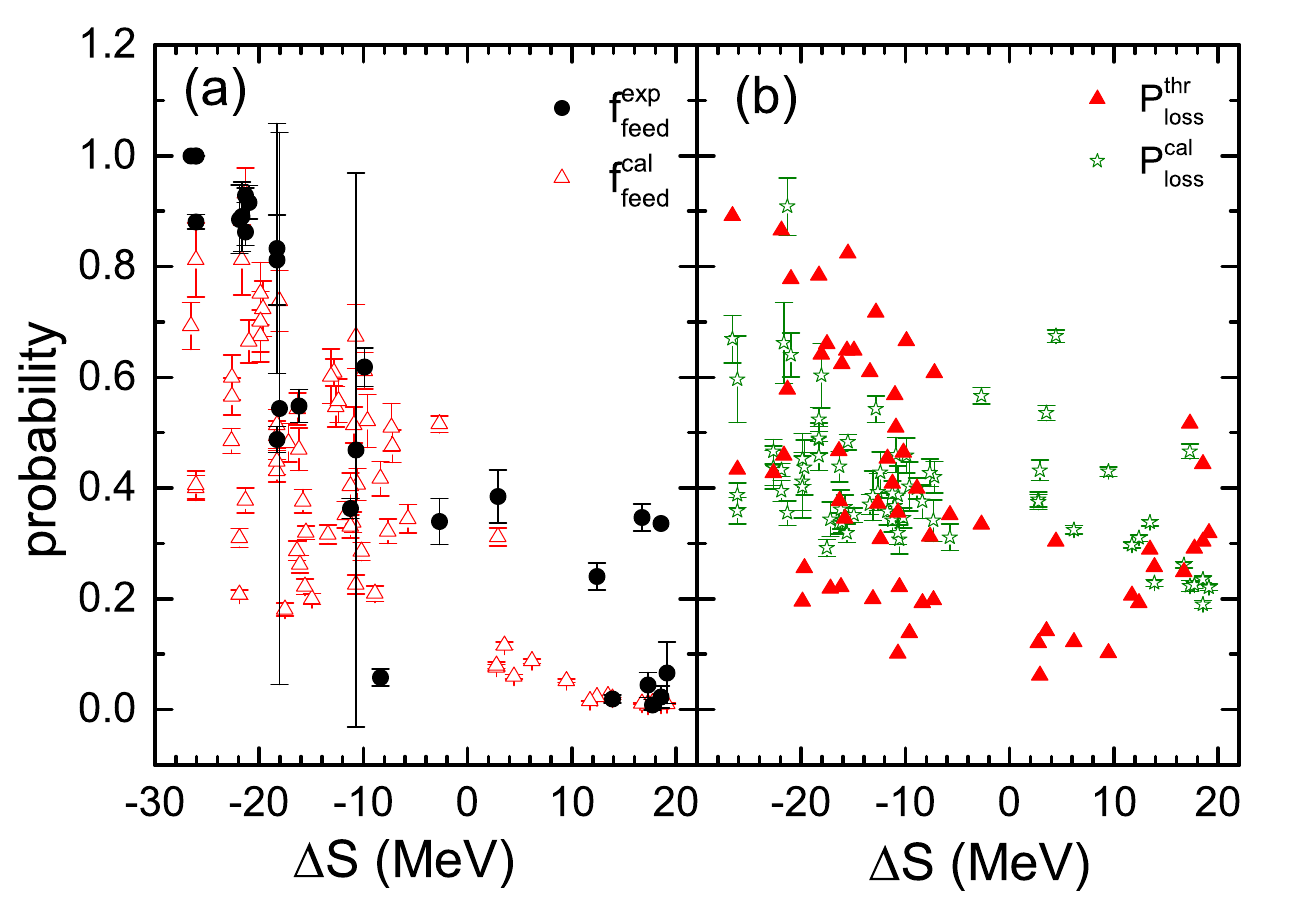}
  \caption{\label{fig:fis}
(a) Evaporation-feeding fraction $f^{\rm exp}_{\rm feed}$ extracted from residue parallel-momentum spectra and calculated $f^{\rm cal}_{\rm feed}$ by the IQMD+GEMINI model as a function of the asymmetry $\Delta S$.
Details of the $f^{\rm exp}_{\rm feed}$ extraction and representative spectral decompositions are provided in Appendix~C of the Supplemental Material.
(b) Evaporation-loss probability $P^{\rm thr}_{\rm loss}$ by the minimal threshold-tail model (details in Appendix~D of the Supplemental Material) and calculated $P^{\rm cal}_{\rm loss}$ by the IQMD+GEMINI model as a function of the separation-energy asymmetry $\Delta S$.}
\end{figure}

The mechanism decomposition in Figs.~\ref{fig:sigmaSn} and \ref{fig:FWHM} suggests that evaporation feeding and dynamical survival differ not only in their $S_n$ systematics but also in their residue momentum scales.
This motivates a data-driven attempt to constrain the evaporation-feeding contribution directly from the measured parallel-momentum spectra.
We therefore extract a semi-experimental feeding fraction by fitting each spectrum with a non-negative mixture of two fixed template shapes, $T_{\rm feed}$ and $T_{\rm surv}$, constructed from the anchor spectra at the extreme $\Delta S$ values in the dataset ($^{20}$C and $^{32}$Ar).
These anchors are not assumed to represent pure mechanisms; rather, $T_{\rm feed}$ and $T_{\rm surv}$ provide limiting spectral shapes for an empirical separation:
\begin{equation}
\frac{{\rm d}\sigma}{{\rm d}p}\propto
\frac{(1-f^{\rm exp}_{\rm feed})}{\lambda_{\rm surv}}\,T_{\rm surv}\!\left(\dfrac{p-p_1}{\lambda_{\rm surv}}\right)
+
\frac{f^{\rm exp}_{\rm feed}}{\lambda_{\rm feed}}\,T_{\rm feed}\!\left(\dfrac{p-p_2}{\lambda_{\rm feed}}\right),
\label{twocomp}
\end{equation}
normalized to the spectrum integral (details in the Supplemental Material).

The uncertainty of $f^{\rm exp}_{\rm feed}$ is obtained by Monte-Carlo resampling of the measured momentum spectra.
In each pseudo-spectrum, the measured bins are fluctuated within their experimental uncertainties, the spectrum is renormalized to the inclusive cross section, and the same two-template fit is repeated.
The quoted uncertainty is then taken from the central spread of the resulting $f^{\rm exp}_{\rm feed}$ distribution.

As shown in Fig.~\ref{fig:fis}(a), the extracted $f^{\rm exp}_{\rm feed}$ decreases smoothly from $\simeq 1$ at the neutron-rich end to $\simeq 0$ at the neutron-deficient end.
The calculated feeding fraction $f^{\rm cal}_{\rm feed}$ exhibits the same global trend, decreasing from $\sim0.9$ to $0$, providing a consistency check of the mechanism picture against the measured spectral shapes.
We have also repeated the extraction with alternative template pairs in the Supplemental Material; the resulting absolute values of $f^{\rm exp}_{\rm feed}$ vary only weakly, and all tested choices preserve the dominant $\Delta S$ trend.

Although this two-template construction is not a unique mechanism decomposition, it provides a practical empirical diagnostic of feeding from existing inclusive momentum spectra, for which no event-by-event feeding tag is available.
Therefore, the two-template analysis is used not to determine a unique feeding fraction for each system, but to test whether the measured spectral shapes support the feeding trend suggested by the transport--deexcitation calculation.

To provide a complementary consistency check and a parametric envelope for the leakage correction, we also evaluate $P_{\rm loss}$ with a minimal threshold--tail model (see Appendix~D of the Supplemental Material).
We approximate the excitation-energy distribution of the dynamical prefragment $^{A-1}_{~~~Z}X^\ast$ by an exponential tail, $w(E^\ast)=(1/E_{0})e^{-E^\ast/E_{0}}$, and adopt a step-like emission criterion with an effective threshold $E_{\rm th}=\min\!\big(S_n,\;S_p+V_C^{(p)}\big)$: $E^\ast<E_{\rm th}$ is taken as no particle evaporation, whereas $E^\ast\ge E_{\rm th}$ implies at least one nucleon emission.
This gives
\begin{equation}
P^{\rm thr}_{\rm loss}\approx
\exp\!\left[-\frac{1}{E_{0}}\min\!\Big(S_n,\;S_p+V_C^{(p)}\Big)\right].
\label{eq:Ploss_compact_text_short}
\end{equation}

As shown in Fig.~\ref{fig:fis}(b), for a representative choice $E_{0}=5$~MeV, $P^{\rm thr}_{\rm loss}$ reproduces the qualitative $\Delta S$ dependence obtained from the transport--deexcitation chain.
We therefore use the threshold--tail model as a conservative cross-check of the scale and $\Delta S$ dependence of $P_{\rm loss}$, rather than as an independent microscopic determination of the loss probability.
A quantitative constraint on $E_{0}$ and thus on $P_{\rm loss}$ can ultimately be obtained from measurements of evaporated-particle energy spectra.

The momentum-template analysis and the threshold--tail estimate should be viewed as complementary consistency and sensitivity tests, not as fully independent microscopic calculations of $f_{\rm feed}$ and $P_{\rm loss}$.
The former tests the feeding trend through measured spectral-shape information, whereas the latter tests the separation-energy dependence of the evaporation-loss correction.
A full inter-model quantification of the transport uncertainty would require applying independent transport codes to the same feeding/loss classification and is beyond the scope of the present work.

\begin{figure}[t]
  \centering
  \includegraphics[width=\columnwidth]{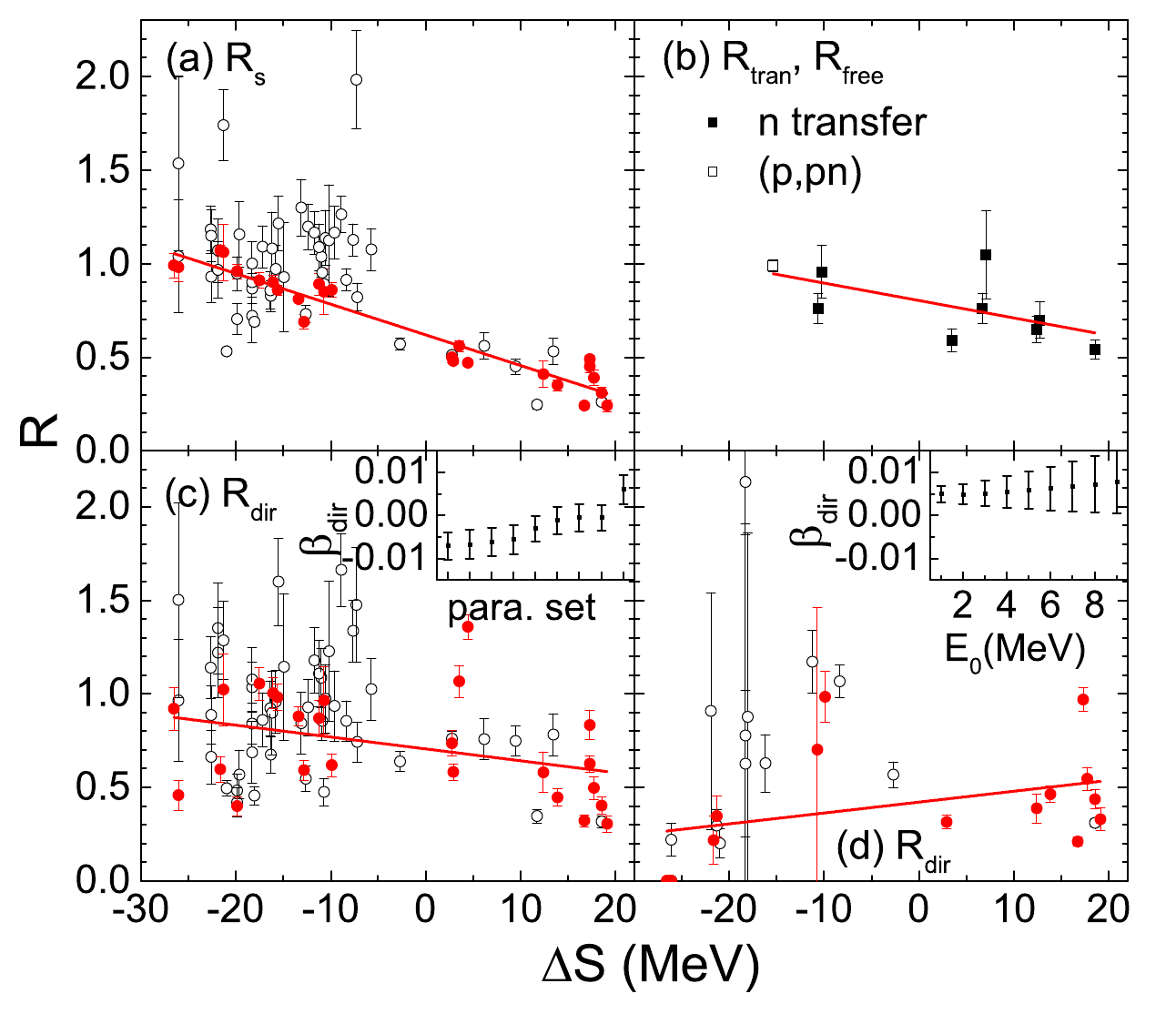}
\caption{\label{fig:Rs}
Reduction factors as a function of the separation-energy asymmetry $\Delta S$
(filled circles: the 25 systems compiled in Ref.~\cite{tostevin2021updated};
open circles: additional systems).
(a) Inclusive $R_s$; a linear fit to the 25-system subset gives
$\beta_s \equiv dR_s/d\Delta S = -0.016\pm0.001$.
(b) Reduction factors from transfer and quasifree reactions~\cite{aumann2021quenching}, $R_{\rm tran}$ and $R_{\rm free}$;
a linear fit gives $\beta_{\rm tqf}\equiv dR/d\Delta S = -0.009\pm0.004$.
(c) Purified $R_{\rm dir}$ using IQMD+GEMINI $f_{\rm feed}^{\rm cal}$ and $P_{\rm loss}^{\rm cal}$;
the fit gives $\beta_{\rm dir}=-0.006\pm0.003$.
Inset: variations of the mean field and in-medium $NN$ cross sections yield
$\beta_{\rm dir}$ in the range $-0.007$ to $0.006$.
(d) $R_{\rm dir}$ using semi-experimental $f_{\rm feed}^{\rm exp}$ and threshold--tail $P_{\rm loss}^{\rm thr}$;
the fit gives $\beta_{\rm dir}=0.006\pm0.004$.
Inset: varying $E_0$ from 1 to 9~MeV yields $\beta_{\rm dir}$ in the range $0.005$ to $0.008$.
}
\end{figure}

Figure~\ref{fig:Rs} summarizes the $\Delta S$ systematics of the inclusive and purified reduction factors.
To limit baseline heterogeneity, we quote slopes from linear fits to the 25 systems compiled in Ref.~\cite{tostevin2021updated} (filled circles); additional systems are shown as open circles.

For this subset, the inclusive $R_s$ shows a statistically significant negative slope,
$\beta_s\equiv dR_s/d\Delta S=-0.016\pm0.001$ [Fig.~\ref{fig:Rs}(a)].
In contrast, reduction factors from transfer and quasifree probes exhibit a weaker dependence~\cite{aumann2021quenching}, with $\beta_{\rm tqf}=-0.009\pm0.004$ [Fig.~\ref{fig:Rs}(b)].
After correcting the inclusive cross sections for evaporation feeding and evaporation loss using IQMD+GEMINI, the residual slope is reduced to $\beta_{\rm dir}=-0.006\pm0.003$ [Fig.~\ref{fig:Rs}(c)].
To assess the systematic sensitivity within the IQMD+GEMINI framework, we vary the mean field, characterized by the incompressibility $K=200$--270~MeV and the symmetry-energy slope $L=67.9$--139.3~MeV, together with five in-medium $NN$ prescriptions~\cite{su2016effects}.
As shown in the inset of Fig.~\ref{fig:Rs}(c), these variations give $\beta_{\rm dir}$ in the range $-0.007$ to $0.006$, indicating that the reduced slope after purification is weakly sensitive to the parameter-set changes.
An independent cross-check based on $f_{\rm feed}^{\rm exp}$ and $P_{\rm loss}^{\rm thr}$ gives a comparably small slope, $\beta_{\rm dir}=0.006\pm0.004$, with only modest sensitivity to $E_0$ (inset) [Fig.~\ref{fig:Rs}(d)].
Varying $E_0$ from 1 to 9~MeV yields $\beta_{\rm dir}$ in the range $0.005$ to $0.008$ [inset of Fig.~\ref{fig:Rs}(d)], showing that this cross-check is also weakly sensitive to the excitation-tail scale.

The model-based and semi-experimental corrections give slightly negative and slightly positive slopes, respectively, suggesting that the present uncertainties do not allow a robust determination of whether $R_{\rm dir}$ is strictly asymmetry-independent.
However, these comparisons indicate that the apparent $R_s$--$\Delta S$ trend is largely driven by $\Delta S$-dependent admixtures of evaporation feeding on the neutron-rich side and evaporation loss toward neutron-deficient systems; after purification, $R_{\rm dir}$ is compatible, within uncertainties, with transfer and quasifree benchmarks.

It should be noted that $R_s$ is commonly interpreted as an empirical measure of single-particle-strength quenching relative to shell-model spectroscopic factors folded with an eikonal direct-reaction calculation, providing the conventional link to microscopic calculations of spectroscopic strength.
The present analysis clarifies that the measured inclusive cross section used in $R_s$ can contain evaporation feeding and evaporation loss, so the extracted reduction factor may not reflect spectroscopic quenching alone.
Keeping the eikonal direct-reaction baseline unchanged, we purify the measured inclusive cross section for the estimated feeding and loss contributions.
In this sense, $R_{\rm dir}$ is intended to restore, as far as possible, the conventional spectroscopic-strength interpretation of the reduction factor, rather than to introduce a new connection to microscopic structure calculations.

\section{Summary}
\label{summary}

In summary, the transport--deexcitation analysis of inclusive one-neutron removal supports a mechanism-based interpretation in which the apparent $R_s$--$\Delta S$ trend is correlated with evaporation feeding and evaporation loss.
By subtracting the feeding contribution and correcting for the loss component in the measured cross sections, we define a purified reduction factor $R_{\rm dir}$, that provides a cleaner empirical measure of single-particle strength than the inclusive $R_s$.
After this purification, $R_{\rm dir}$ exhibits a much weaker $\Delta S$ dependence, consistent with the weak isospin-asymmetry dependence observed in nucleon-transfer and quasifree-knockout systematics.
Looking forward, coincidence measurements that tag residue mass/charge changes and/or detect evaporated light particles can directly constrain the loss probability $P_{\rm loss}$ and further reduce the uncertainty, enabling one-neutron removal to provide controlled constraints on quenched single-particle strength across asymmetry.

\section*{Acknowledgements}
This work is supported by the National Natural Science Foundation of China underGrant No. 12475136.



\bibliographystyle{elsarticle-num} 
\bibliography{onr_ref}





\end{document}


\begin{frontmatter}
\title{Supplemental Material to: Purifying one-neutron removal as a probe of single-particle strength}

\FloatBarrier

\end{frontmatter}

\section*{Supplement A. Dataset and construction of $R_{\rm dir}$}
\addcontentsline{toc}{section}{Appendix A. Dataset and construction of $R_{\rm dir}$}

Table~\ref{table} provides the full numerical dataset used in this work, enabling reproduction of all derived quantities reported in the main text.
For each system we list the measured inclusive yield, the IQMD+GEMINI decomposition, and the associated reduction factors $R_s$ and $R_{\rm dir}$.

\begin{sidewaystable}[p]
\centering
\caption{\label{table}
Compilation of the experimental and calculated quantities entering the dataset and the construction of $R_{\rm dir}$.
Columns list, in order: the incident energy $E_{\rm in}$, reaction system, reference, neutron separation energy $S_n$,
asymmetry $\Delta S$, the measured inclusive one-neutron-removal cross section $\sigma_{\rm incl}^{\rm exp}$, the inclusive IQMD+GEMINI yield
$\sigma_{\rm incl}^{\rm cal}$, and its mechanism-resolved components $\sigma_{\rm surv}$ (dynamical $A-1$ residue surviving without further particle emission),
$\sigma_{\rm feed}$ (evaporation feeding from intact prefragments), and $\sigma_{\rm loss}$ (leakage due to additional particle emission following dynamical removal).
The last two columns give the literature inclusive reduction factor $R_s$ and the purified reduction factor $R_{\rm dir}$ obtained with the IQMD+GEMINI correction.
Quoted cross sections and reduction factors are reported as value $\pm$ uncertainty.}
\begin{tabular}{cccccccccccc}
\hline\hline
Beam energy &
Reactions &
Ref. &
$S_n$ &
$\Delta S$ &
$\sigma_{\rm incl}^{\rm exp}$ &
$\sigma_{\rm incl}^{\rm cal}$ &
$\sigma_{\rm surv}$ &
$\sigma_{\rm feed}$ &
$\sigma_{\rm loss}$ &
$R_{s}$ &
$R_{\rm dir}$\\
(MeV/n.) &  &  & (MeV) & (MeV) & (mb) & (mb) & (mb) & (mb) & (mb)  &  &  \\
\hline
60 & $^{9}$Be($^{11}$Be, $^{10}$Be) & \cite{aumann2000one}  & 0.50  &  -19.67  & 259$\pm$39  & $216.5 \pm 6.9$ & $59.9 \pm 3.6$ & $156.7 \pm 5.9$ & $46.5 \pm 3.2$ & $1.16 \pm 0.17$ & $0.57 \pm 0.13$\\
243 & $^{12}$C($^{19}$C, $^{18}$C) & \cite{kobayashi2012one}  & 0.58  &  -26.09  & 163$\pm$12  & $335.8 \pm 13.1$ & $63.2 \pm 5.7$ & $272.6 \pm 11.8$ & $93.3 \pm 6.9$ & $0.98 \pm 0.08$ & $0.46 \pm 0.08$\\
57 & $^{9}$Be($^{19}$C, $^{18}$C) & \cite{maddalena2001single}  & 0.58  &  -26.09  & 264$\pm$80  & $133.7 \pm 2.9$ & $80.2 \pm 2.3$ & $53.5 \pm 1.8$ & $50.7 \pm 1.8$ & $1.53 \pm 0.47$ & $1.50 \pm 0.52$\\
64 & $^{9}$Be($^{19}$C, $^{18}$C) & \cite{simpson2009one}  & 0.58  &  -26.09  & 226$\pm$65 & $150.6 \pm 3.6$ & $89.6 \pm 2.8$ & $61.0 \pm 2.3$ & $50.1 \pm 2.1$ & $1.04 \pm 0.30$ & $0.96 \pm 0.33$\\
49 & $^{12}$C($^{17}$C, $^{16}$C)  & \cite{sauvan2000one} & 0.73 &  -22.64 & 84$\pm$9 & $119.2 \pm 2.3$ & $61.4 \pm 1.6$ & $57.8 \pm 1.6$ & $53.5 \pm 1.5$ & $1.18 \pm 0.13$ & $1.14 \pm 0.17$\\
79 & $^{12}$C($^{17}$C, $^{16}$C)  & \cite{wu2004neutron} & 0.73 &  -22.64 & 116$\pm$18 & $187.4 \pm 5.6$ & $75.0 \pm 3.5$ & $112.4 \pm 4.3$ & $58.4 \pm 3.1$ & $0.93 \pm 0.14$ & $0.66 \pm 0.14$\\
62 & $^{9}$Be($^{17}$C, $^{16}$C)  & \cite{maddalena2001single} & 0.73 &  -22.64 & $115.0 \pm 14.0$& $163.2 \pm 4.2$ & $70.9 \pm 2.7$ & $92.3 \pm 3.1$ & $54.8 \pm 2.4$ & $1.15 \pm 0.14$ & $0.89 \pm 0.16$\\
50 & $^{12}$C($^{14}$B, $^{13}$B)  & \cite{sauvan2000one} & 0.97 &  -16.31 & 153$\pm$15 & $142.2 \pm 3.2$ & $65.0 \pm 2.2$ & $77.2 \pm 2.4$ & $50.7 \pm 1.9$ & $0.83 \pm 0.08$ & $0.67 \pm 0.10$\\
240 & $^{12}$C($^{29}$Ne,$^{28}$Ne)  & \cite{kobayashi2016one} & 0.97  &  -21.66  & 74$\pm$2 & $268.8 \pm 9.9$ & $50.6 \pm 4.3$ & $218.2 \pm 8.9$ & $99.0 \pm 6.0$ & $1.07 \pm 0.03$ & $0.60 \pm 0.07$ \\
103 & $^{9}$Be($^{15}$C, $^{14}$C)  & \cite{terry2004absolute} & 1.22  &  -19.86  & 101$\pm$5 & $217.2 \pm 7.6$ & $54.0 \pm 3.8$ & $163.2 \pm 6.6$ & $36.4 \pm 3.1$ & $0.96 \pm 0.04$ & $0.40 \pm 0.05$ \\
62 & $^{12}$C($^{15}$C, $^{14}$C)  & \cite{sauvan2000one} & 1.22  &  -19.86  & 159$\pm$15 & $207.2 \pm 7.3$ & $62.1 \pm 4.0$ & $145.1 \pm 6.1$ & $43.5 \pm 3.4$ & $0.95 \pm 0.09$ & $0.48 \pm 0.09$\\
54 & $^{9}$Be($^{15}$C, $^{14}$C)  & \cite{tostevin2002single} & 1.22  &  -19.86  & 137$\pm$16 & $187.2 \pm 5.9$ & $61.0 \pm 3.4$ & $126.2 \pm 4.8$ & $50.4 \pm 3.1$ & $0.70 \pm 0.08$ & $0.42 \pm 0.08$\\
48 & $^{12}$C($^{20}$N, $^{19}$N)  & \cite{sauvan2000one} & 2.16 &  -15.78  & 98$\pm$13 & $110.7 \pm 2.4$ & $69.1 \pm 1.9$ & $41.6 \pm 1.5$ & $39.9 \pm 1.4$ & $0.97 \pm 0.13$ & $0.96 \pm 0.17$\\
73 & $^{9}$Be($^{35}$Si, $^{34}$Si) & \cite{enders2002single} & 2.51  & -16.17  & 106$\pm$19 & $162.4 \pm 5.4$ & $86.1 \pm 3.9$ & $76.3 \pm 3.7$ & $48.7 \pm 3.0$ & $1.08 \pm 0.19$ & $0.90 \pm 0.22$\\
43 & $^{12}$C($^{15}$B, $^{14}$B) & \cite{sauvan2000one} & 2.78 &  -15.51  & 108$\pm$13 & $79.5 \pm 1.1$ & $54.1 \pm 0.9$ & $25.4 \pm 0.6$ & $50.6 \pm 0.9$ & $1.21 \pm 0.15$ & $1.60 \pm 0.23$ \\
59 & $^{12}$C($^{18}$N, $^{17}$N) & \cite{sauvan2000one} & 2.83 &  -12.38  & 109$\pm$11 & $154.7 \pm 4.7$ & $68.4 \pm 3.1$ & $86.3 \pm 3.5$ & $51.2 \pm 2.7$ & $1.20 \pm 0.12$ & $0.93 \pm 0.15$\\
241 & $^{12}$C($^{20}$C, $^{19}$C) & \cite{kobayashi2012one}  & 2.98  &  -26.57  & 58$\pm$5  & $149.9 \pm 4.2$ & $46.1 \pm 2.3$ & $103.8 \pm 3.5$ & $93.1 \pm 3.3$ & $0.99 \pm 0.07$ & $0.92 \pm 0.11$\\
228 & $^{12}$C($^{30}$Ne, $^{29}$Ne)  & \cite{liu2017intruder} & 3.19  &  -20.97  & 62$\pm$2  & $162.6 \pm 4.3$ & $54.5 \pm 2.5$ & $108.1 \pm 3.5$ & $97.0 \pm 3.3$ & $0.53 \pm 0.02$ & $0.49 \pm 0.04$ \\
67 & $^{12}$C($^{12}$B, $^{11}$B) & \cite{sauvan2000one} & 3.37 &  -10.73  & 81$\pm$5 & $186.7 \pm 7.2$ & $60.9 \pm 4.1$ & $125.8 \pm 5.9$ & $38.4 \pm 3.3$ & $0.89 \pm 0.05$ & $0.47 \pm 0.07$ \\
56 & $^{12}$C($^{21}$O, $^{20}$O)  & \cite{sauvan2000one} & 3.81 & -17.19  & 134$\pm$14 & $158.0 \pm 4.6$ & $81.6 \pm 3.3$ & $75.9 \pm 3.2$ & $42.7 \pm 2.4$ & $1.09 \pm 0.11$ & $0.86 \pm 0.14$\\
54 & $^{12}$C($^{24}$F, $^{23}$F) & \cite{sauvan2000one} & 3.81 &  -10.55  & 124$\pm$16 & $146.1 \pm 4.1$ & $86.7 \pm 3.2$ & $59.4 \pm 2.6$ & $38.4 \pm 2.1$ & $1.14 \pm 0.15$ & $0.97 \pm 0.18$ \\
68 & $^{12}$C($^{19}$O, $^{18}$O) & \cite{sauvan2000one} & 3.96 & -13.11  & 104$\pm$12 & $174.7 \pm 6.3$ & $69.6 \pm 4.0$ & $105.1 \pm 4.9$ & $43.7 \pm 3.1$ & $1.30 \pm 0.15$ & $0.84 \pm 0.17$ \\
80 & $^{9}$Be($^{9}$Li, $^{8}$Li) & \cite{grinyer2012systematic} & 4.06  & -9.88  & 55.6$\pm$2.9 & $136.0 \pm 3.2$ & $52.8 \pm 2.0$ & $83.2 \pm 2.5$ & $44.8 \pm 1.8$ & $0.86 \pm 0.04$ & $0.62 \pm 0.06$\\
43 & $^{12}$C($^{18}$C, $^{17}$C)  & \cite{sauvan2000one} & 4.18 & -21.9  & 115$\pm$18 & $73.8 \pm 0.9$ & $58.5 \pm 0.8$ & $15.3 \pm 0.4$ & $44.6 \pm 0.7$ & $0.97 \pm 0.15$ & $1.35 \pm 0.24$\\
80 & $^{12}$C($^{18}$C, $^{17}$C)  & \cite{ozawa2008measurement} & 4.18 & -21.9  & 155$\pm$24 & $131.8 \pm 2.6$ & $91.0 \pm 2.1$ & $40.8 \pm 1.4$ & $59.3 \pm 1.7$ & $1.07 \pm 0.17$ & $1.22 \pm 0.24$\\
920 & $^{9}$Be($^{24}$O, $^{23}$O) & \cite{kanungo2009one} & 4.19 & -21.32 & 63$\pm$7 & $98.2 \pm 2.4$ & $6.7 \pm 0.6$ & $91.5 \pm 2.3$ & $65.9 \pm 1.9$ & $1.74 \pm 0.19$ & $1.29 \pm 0.21$ \\
92.3 & $^{9}$Be($^{24}$O, $^{23}$O) & \cite{divaratne2018one} & 4.19  &  -21.32  & $74.0 \pm 11.0$ & $158.5 \pm 3.7$ & $98.7 \pm 2.9$ & $59.8 \pm 2.2$ & $54.3 \pm 2.1$ & $1.06 \pm 0.15$ & $1.02 \pm 0.19$\\
75 & $^{9}$Be($^{16}$C, $^{15}$C) & \cite{flavigny2012nonsudden} & 4.25  &  -18.3  & 81$\pm$7 & $116.0 \pm 2.6$ & $58.6 \pm 1.8$ & $57.3 \pm 1.8$ & $49.5 \pm 1.7$ & $0.90 \pm 0.08$ & $0.84 \pm 0.11$ \\
55 & $^{12}$C($^{16}$C, $^{15}$C) & \cite{sauvan2000one} & 4.25  & -18.3  & 65$\pm$6 & $103.5 \pm 1.9$ & $58.9 \pm 1.5$ & $44.6 \pm 1.3$ & $64.7 \pm 1.5$ & $0.87 \pm 0.08$ & $1.04 \pm 0.13$ \\
83 & $^{12}$C($^{16}$C, $^{15}$C) & \cite{yamaguchi2004neutron} & 4.25  & -18.30  & 65$\pm$12.5 & $125.3 \pm 2.9$ & $60.9 \pm 2.0$ & $64.4 \pm 2.1$ & $58.5 \pm 2.0$ & $0.72 \pm 0.14$ & $0.69 \pm 0.17$ \\
$continue$ &   &   &   &    &  &  &  & & & &  \\
\hline\hline
\end{tabular}
\end{sidewaystable}

\begin{sidewaystable}[p]
\centering
\begin{tabular}{cccccccccccc}
\hline\hline
Beam energy &
Reactions &
Ref. &
$S_n$ &
$\Delta S$ &
$\sigma_{\rm incl}^{\rm exp}$ &
$\sigma_{\rm incl}^{\rm cal}$ &
$\sigma_{\rm surv}$ &
$\sigma_{\rm feed}$ &
$\sigma_{\rm loss}$ &
$R_{s}$ &
$R_{\rm dir}$\\
(MeV/n.) &  &  & (MeV) & (MeV) & (mb) & (mb) & (mb) & (mb) & (mb)  &  &  \\
\hline
62 & $^{9}$Be($^{16}$C, $^{15}$C) & \cite{maddalena2001single} & 4.25  &  -18.3  & 77$\pm$9 & $111.2 \pm 2.3$ & $61.4 \pm 1.7$ & $49.8 \pm 1.6$ & $58.4 \pm 1.7$ & $1.00 \pm 0.12$ & $1.08 \pm 0.17$ \\
50 & $^{12}$C($^{25}$F, $^{24}$F) & \cite{sauvan2000one} & 4.28  & -10.17  & 173$\pm$46 & $109.8 \pm 2.2$ & $78.4 \pm 1.9$ & $31.4 \pm 1.2$ & $41.5 \pm 1.3$ & $1.12 \pm 0.30$ & $1.23 \pm 0.37$ \\
69 & $^{9}$Be($^{37}$S, $^{36}$S) & \cite{enders2002single} & 4.3  & -9.63  & 99$\pm$12 & $143.1 \pm 5.5$ & $68.5 \pm 3.8$ & $74.6 \pm 3.9$ & $46.2 \pm 3.1$ & $1.16 \pm 0.14$ & $0.93 \pm 0.19$ \\
43 & $^{12}$C($^{21}$N, $^{20}$N) & \cite{sauvan2000one}  & 4.61  & -14.95  & 140$\pm$44 & $80.2 \pm 1.2$ & $64.2 \pm 1.1$ & $16.0 \pm 0.5$ & $34.7 \pm 0.8$ & $0.93 \pm 0.29$ & $1.14 \pm 0.39$ \\
242.5 & $^{9}$Be($^{34}$Mg,$^{33}$Mg) & \cite{bazin2021spectroscopy} & 4.71  & -18.04  & 93.0$\pm$2.0 & $187.4 \pm 6.5$ & $49.2 \pm 3.3$ & $138.3 \pm 5.6$ & $74.6 \pm 4.1$ & $0.69 \pm 0.02$ & $0.46 \pm 0.05$ \\
57 & $^{12}$C($^{13}$B, $^{12}$B) & \cite{sauvan2000one}  & 4.88  & -10.93  & 59$\pm$4 & $134.1 \pm 3.5$ & $65.2 \pm 2.4$ & $68.9 \pm 2.5$ & $55.4 \pm 2.3$ & $0.95 \pm 0.06$ & $0.86 \pm 0.10$ \\
79 & $^{9}$Be($^{40}$Si, $^{39}$Si) & \cite{stroberg2014single} & 4.96  & -17.55  & 116$\pm$5 & $131.5 \pm 2.6$ & $107.8 \pm 2.4$ & $23.8 \pm 1.1$ & $44.4 \pm 1.5$ & $0.91 \pm 0.04$ & $1.05 \pm 0.09$ \\
93.7 & $^{9}$Be($^{44}$S, $^{43}$S) & \cite{momiyama2020shell} & 5.08  & -16.09  & 91$\pm$4 & $141.6 \pm 2.9$ & $104.5 \pm 2.5$ & $37.1 \pm 1.5$ & $53.3 \pm 1.8$ & $0.90 \pm 0.04$ & $1.00 \pm 0.08$ \\
64 & $^{12}$C($^{22}$F, $^{21}$F) & \cite{sauvan2000one} & 5.23  & -7.33  & 121$\pm$16 & $161.2 \pm 5.6$ & $79.0 \pm 3.9$ & $82.2 \pm 4.0$ & $40.9 \pm 2.8$ & $1.98 \pm 0.26$ & $1.48 \pm 0.31$ \\
77 & $^{9}$Be($^{57}$Cr, $^{56}$Cr) & \cite{gade2006one} & 5.31  & -8.35  & 122$\pm$8 & $136.7 \pm 4.0$ & $79.7 \pm 3.1$ & $57.0 \pm 2.6$ & $47.9 \pm 2.4$ & $0.91 \pm 0.06$ & $0.85 \pm 0.11$\\
53 & $^{12}$C($^{19}$N, $^{18}$N) & \cite{sauvan2000one} & 5.33  & -11.02  & 86$\pm$9 & $117.8 \pm 2.4$ & $77.9 \pm 2.0$ & $39.8 \pm 1.4$ & $45.4 \pm 1.5$ & $1.04 \pm 0.11$ & $1.08 \pm 0.16$ \\
86 & $^{9}$Be($^{38}$Si, $^{37}$Si) & \cite{stroberg2014single} & 5.67  & -15.6  & 104$\pm$3 & $132.8 \pm 2.6$ & $103.3 \pm 2.3$ & $29.5 \pm 1.2$ & $48.3 \pm 1.6$ & $0.86 \pm 0.03$ & $0.98 \pm 0.07$ \\
65 & $^{12}$C($^{17}$N, $^{16}$N) & \cite{sauvan2000one} & 5.89  & -7.23  & 55$\pm$5 & $124.9 \pm 3.2$ & $65.5 \pm 2.3$ & $59.4 \pm 2.2$ & $47.3 \pm 1.9$ & $0.82 \pm 0.07$ & $0.74 \pm 0.11$ \\
98 & $^{9}$Be($^{36}$Si, $^{35}$Si) & \cite{stroberg2014single} & 6.12  & -13.4  & 81$\pm$2 & $127.2 \pm 2.5$ & $87.0 \pm 2.0$ & $40.2 \pm 1.4$ & $51.2 \pm 1.6$ & $0.81 \pm 0.01$ & $0.88 \pm 0.05$ \\
120 & $^{9}$Be($^{10}$Be, $^{9}$Be) & \cite{grinyer2012systematic}  & 6.8  & -12.82  & 71$\pm$4 & $84.7 \pm 1.5$ & $33.2 \pm 0.9$ & $51.6 \pm 1.1$ & $39.4 \pm 1.0$ & $0.69 \pm 0.04$ & $0.59 \pm 0.05$ \\
51 & $^{12}$C($^{22}$O, $^{21}$O) & \cite{sauvan2000one} & 6.85  & -16.39  & 120$\pm$14 & $113.5 \pm 2.4$ & $81.0 \pm 2.0$ & $32.5 \pm 1.3$ & $41.0 \pm 1.5$ & $0.86 \pm 0.10$ & $0.92 \pm 0.15$ \\
120 & $^{9}$Be($^{7}$Li, $^{6}$Li) & \cite{grinyer2012systematic}  & 7.25  & -2.72  & 30.7$\pm$1.8 & $56.5 \pm 0.7$ & $27.4 \pm 0.5$ & $29.1 \pm 0.5$ & $35.8 \pm 0.5$ & $0.57 \pm 0.03$ & $0.64 \pm 0.05$ \\
74 & $^{9}$Be($^{70}$Ni, $^{69}$Ni) & \cite{recchia2016neutron} & 7.31  & -8.91  & 168$\pm$13 & $114.9 \pm 2.4$ & $90.9 \pm 2.2$ & $24.1 \pm 1.1$ & $60.4 \pm 1.8$ & $1.26 \pm 0.10$ & $1.66 \pm 0.19$\\
98.5 & $^{9}$Be($^{34}$Si, $^{33}$Si) & \cite{jongile2020structure} & 7.51  & -11.23  & 116$\pm$6 & $121.5 \pm 2.7$ & $72.4 \pm 2.1$ & $49.1 \pm 1.7$ & $46.3 \pm 1.7$ & $0.89 \pm 0.06$ & $0.87 \pm 0.09$ \\
73 & $^{9}$Be($^{34}$Si, $^{33}$Si) & \cite{enders2002single}  & 7.51  & -11.23  & 123$\pm$14 & $123.1 \pm 2.9$ & $82.4 \pm 2.3$ & $40.7 \pm 1.7$ & $42.8 \pm 1.7$ & $1.09 \pm 0.12$ & $1.11 \pm 0.18$ \\
59 & $^{12}$C($^{23}$F, $^{22}$F) & \cite{sauvan2000one} & 7.58 & -5.71 & 114$\pm$12 & $137.0 \pm 3.7$ & $89.5 \pm 3.0$ & $47.1 \pm 2.2$ & $40.4 \pm 2.0$ & $1.08 \pm 0.11$ & $1.02 \pm 0.16$ \\
62 & $^{12}$C($^{20}$O, $^{19}$O) & \cite{sauvan2000one} & 7.61 & -11.74 & 112$\pm$11 & $130.6 \pm 3.3$ & $84.7 \pm 2.7$ & $45.9 \pm 2.0$ & $47.2 \pm 2.0$ & $1.17 \pm 0.11$ & $1.18 \pm 0.18$ \\
85 & $^{9}$Be($^{68}$Ni, $^{67}$Ni) & \cite{recchia2016neutron} & 7.79  & -7.64  & 133$\pm$10 & $112.8 \pm 2.8$ & $76.5 \pm 2.3$ & $36.3 \pm 1.6$ & $57.1 \pm 2.0$ & $1.13 \pm 0.08$ & $1.34 \pm 0.16$\\
70 & $^{9}$Be($^{46}$Ar, $^{45}$Ar) & \cite{gade2005knockout} & 8.07  & -10.73  & 122$\pm$12 & $126.2 \pm 3.1$ & $97.6 \pm 2.7$ & $28.6 \pm 1.5$ & $45.6 \pm 1.9$ & $0.85 \pm 0.12$ & $0.96 \pm 0.18$ \\
71 & $^{12}$C($^{14}$C, $^{13}$C) & \cite{sauvan2000one}  & 8.18 & -12.65 & 65$\pm$4 & $128.3 \pm 3.6$ & $58.2 \pm 2.5$ & $70.1 \pm 2.7$ & $37.6 \pm 2.0$ & $0.73 \pm 0.04$ & $0.55 \pm 0.07$ \\
73 & $^{9}$Be($^{57}$Ni, $^{56}$Ni) & \cite{yurkewicz2006one} & 10.25  & 2.92  & 41.4$\pm$1.2 & $56.1 \pm 1.0$ & $38.6 \pm 0.9$ & $17.5 \pm 0.6$ & $29.4 \pm 0.7$ & $0.48 \pm 0.02$ & $0.58 \pm 0.04$ \\
1670 & $^{12}$C($^{11}$C, $^{10}$C) & \cite{tostevin2021updated} & 13.12  & 4.43  & 24.4$\pm$0.2 & $14.1 \pm 0.2$ & $13.2 \pm 0.2$ & $0.8 \pm 0.0$ & $27.4 \pm 0.3$ & $0.47 \pm 0.01$ & $1.36 \pm 0.07$ \\
$continue$ &   &   &   &    &  &  &  & & & &  \\
\hline\hline
\end{tabular}
\end{sidewaystable}

\begin{sidewaystable}[htbp]
\centering
\begin{tabular}{cccccccccccc}
\hline\hline
Beam energy &
Reactions &
Ref. &
$S_n$ &
$\Delta S$ &
$\sigma_{\rm incl}^{\rm exp}$ &
$\sigma_{\rm incl}^{\rm cal}$ &
$\sigma_{\rm surv}$ &
$\sigma_{\rm feed}$ &
$\sigma_{\rm loss}$ &
$R_{s}$ &
$R_{\rm dir}$\\
(MeV/n.) &  &  & (MeV) & (MeV) & (mb) & (mb) & (mb) & (mb) & (mb)  &  &  \\
\hline
60 & $^{9}$Be($^{37}$Ca, $^{36}$Ca) & \cite{burger2012cross} & 14.76  & 11.75  & 5.6$\pm$0.5 & $31.1 \pm 0.2$ & $30.7 \pm 0.2$ & $0.5 \pm 0.0$ & $13.0 \pm 0.2$ & $0.25 \pm 0.02$ & $0.34 \pm 0.04$ \\
62.8 & $^{9}$Be($^{32}$S, $^{31}$S) & \cite{gade2004one} & 15.04  &  6.18  & 36$\pm$4 & $51.2 \pm 0.5$ & $46.7 \pm 0.5$ & $4.5 \pm 0.2$ & $22.5 \pm 0.4$ & $0.56 \pm 0.07$ & $0.76 \pm 0.11$ \\
2100 & $^{12}$C($^{16}$O, $^{15}$O) & \cite{brown2002absolute}  & 15.66  & 3.54  & 42.9$\pm$2.3 & $24.8 \pm 0.4$ & $21.9 \pm 0.3$ & $2.9 \pm 0.1$ & $25.3 \pm 0.4$ & $0.56 \pm 0.03$ & $1.07 \pm 0.08$ \\
66.4 & $^{9}$Be($^{33}$Cl, $^{32}$Cl) & \cite{gade2004one} & 15.74  &  13.46  & 13.1$\pm$1.6 & $40.7 \pm 0.3$ & $39.7 \pm 0.3$ & $1.0 \pm 0.0$ & $20.2 \pm 0.2$ & $0.53 \pm 0.07$ & $0.78 \pm 0.11$ \\
85.9 & $^{9}$Be($^{56}$Ni, $^{55}$Ni) & \cite{spieker2019one} & 16.64  & 9.48  & 38.0$\pm$3.2 & $36.9 \pm 0.4$ & $34.9 \pm 0.4$ & $1.9 \pm 0.1$ & $26.3 \pm 0.3$ & $0.45 \pm 0.04$ & $0.75 \pm 0.08$ \\
70 & $^{9}$Be($^{34}$Ar,$^{33}$Ar) & \cite{gade2004one} & 17.07  & 12.4  & 15.6$\pm$1.8 & $36.1 \pm 0.3$ & $35.3 \pm 0.3$ & $0.8 \pm 0.0$ & $15.8 \pm 0.2$ & $0.41 \pm 0.07$ & $0.58 \pm 0.11$ \\
2100 & $^{12}$C($^{12}$C, $^{11}$C) & \cite{brown2002absolute} & 18.72 &  2.76  & 46.5$\pm$2.3 & $33.8 \pm 0.6$ & $31.1 \pm 0.5$ & $2.7 \pm 0.2$ & $18.7 \pm 0.4$ & $0.50 \pm 0.03$ & $0.74 \pm 0.07$ \\
1670 & $^{12}$C($^{12}$C, $^{11}$C) & \cite{tostevin2021updated} & 18.72  & 2.76  & 49.4$\pm$0.9 & $31.0 \pm 0.5$ & $28.7 \pm 0.5$ & $2.3 \pm 0.1$ & $17.6 \pm 0.4$ & $0.51 \pm 0.01$ & $0.76 \pm 0.04$ \\
70 & $^{9}$Be($^{36}$Ca, $^{35}$Ca) & \cite{shane2012proton} & 19.31  & 16.74  & 5.0$\pm$0.5 & $27.6 \pm 0.2$ & $27.3 \pm 0.2$ & $0.3 \pm 0.0$ & $9.7 \pm 0.1$ & $0.24 \pm 0.02$ & $0.32 \pm 0.03$ \\
74 & $^{9}$Be($^{22}$Mg,$^{21}$Mg) & \cite{diget2008structure} & 19.37  & 13.87  & 19.7$\pm$1.6 & $38.5 \pm 0.4$ & $37.7 \pm 0.4$ & $0.8 \pm 0.1$ & $11.2 \pm 0.2$ & $0.35 \pm 0.03$ & $0.44 \pm 0.05$ \\
85 & $^{9}$Be($^{24}$Si, $^{23}$Si) & \cite{gade2008reduction} & 21.02  & 17.73  & 9.8$\pm$1 & $25.7 \pm 0.3$ & $25.4 \pm 0.3$ & $0.3 \pm 0.0$ & $7.4 \pm 0.1$ & $0.39 \pm 0.04$ & $0.50 \pm 0.06$ \\
80 & $^{9}$Be($^{28}$S, $^{27}$S) & \cite{gade2008reduction} & 21.03  & 18.53 & 11.9$\pm$1.2 & $27.5 \pm 0.3$ & $27.3 \pm 0.3$ & $0.3 \pm 0.0$ & $8.4 \pm 0.1$ & $0.31 \pm 0.03$ & $0.40 \pm 0.05$ \\
1670 &  $^{12}$C($^{10}$C,$^{9}$C) & \cite{tostevin2021updated} & 21.28  & 17.26  & 20.2$\pm$0.3 & $18.2 \pm 0.3$ & $18.0 \pm 0.3$ & $0.2 \pm 0.0$ & $15.7 \pm 0.3$ & $0.45 \pm 0.03$ & $0.83 \pm 0.08$ \\
120 &  $^{12}$C($^{10}$C,$^{9}$C) & \cite{grinyer2012systematic} & 21.28  & 17.26  & 27.4$\pm$1.3 & $31.4 \pm 0.5$ & $31.2 \pm 0.5$ & $0.2 \pm 0.0$ & $9.0 \pm 0.3$ & $0.49 \pm 0.02$ & $0.63 \pm 0.04$ \\
65.1 & $^{9}$Be($^{32}$Ar, $^{31}$Ar) & \cite{gade2004reduced}  & 21.6  &  19.14  & 10.4$\pm$1.3 & $28.9 \pm 0.3$ & $28.6 \pm 0.3$ & $0.3 \pm 0.0$ & $8.1 \pm 0.1$ & $0.24 \pm 0.03$ & $0.30 \pm 0.04$ \\
53 & $^{9}$Be($^{14}$O, $^{13}$O) & \cite{flavigny2012nonsudden}  & 23.18  & 18.55  & 14$\pm$1 & $37.3 \pm 0.5$ & $36.8 \pm 0.5$ & $0.5 \pm 0.1$ & $8.6 \pm 0.2$ & $0.26 \pm 0.02$ & $0.32 \pm 0.03$ \\
\hline\hline
\end{tabular}
\end{sidewaystable}

\clearpage   

\section*{Supplement B. Longitudinal-momentum widths}
\addcontentsline{toc}{section}{Appendix B. Longitudinal-momentum widths}

\setcounter{table}{0}
\setcounter{figure}{0}
\renewcommand{\thetable}{B\arabic{table}}
\renewcommand{\thefigure}{B\arabic{figure}}

\begin{figure}[H]   
\centering
    \includegraphics[width=0.45\textwidth,angle=0]{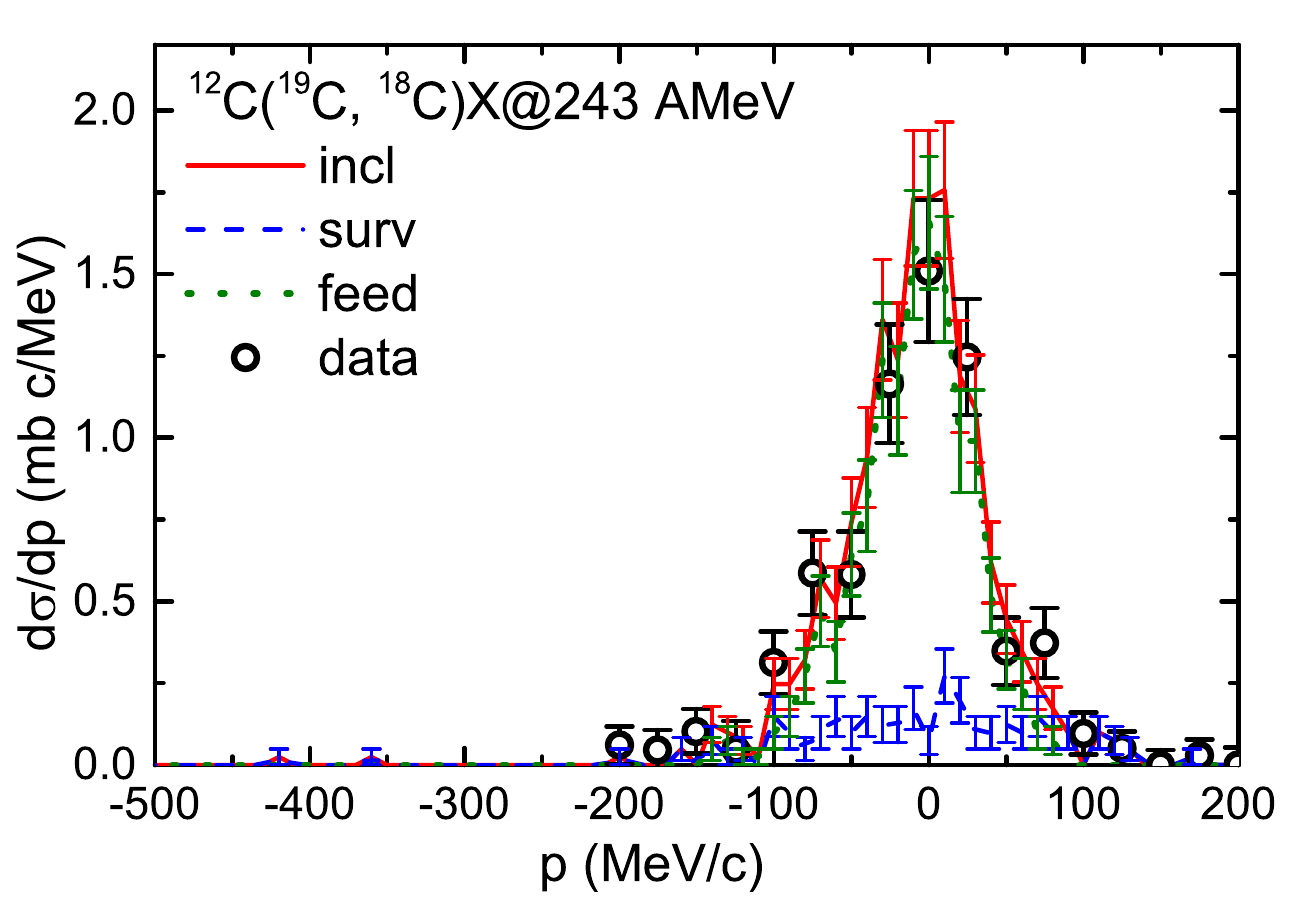}
\caption{Parallel-momentum distribution of the one-neutron-removal residue from $^{19}$C + $^{12}$C collisions at 243~AMeV \cite{kobayashi2012one}.
The experimental inclusive spectrum and the IQMD+GEMINI inclusive result are shown by circles and a solid line, respectively.
The decomposed components are the surviving part of the dynamical $A-1$ yield (dashed line) and the evaporation-feeding contribution (dotted line).}
    \label{19C}
\end{figure}

\begin{figure}[H]   
\centering
    \includegraphics[width=0.45\textwidth,angle=0]{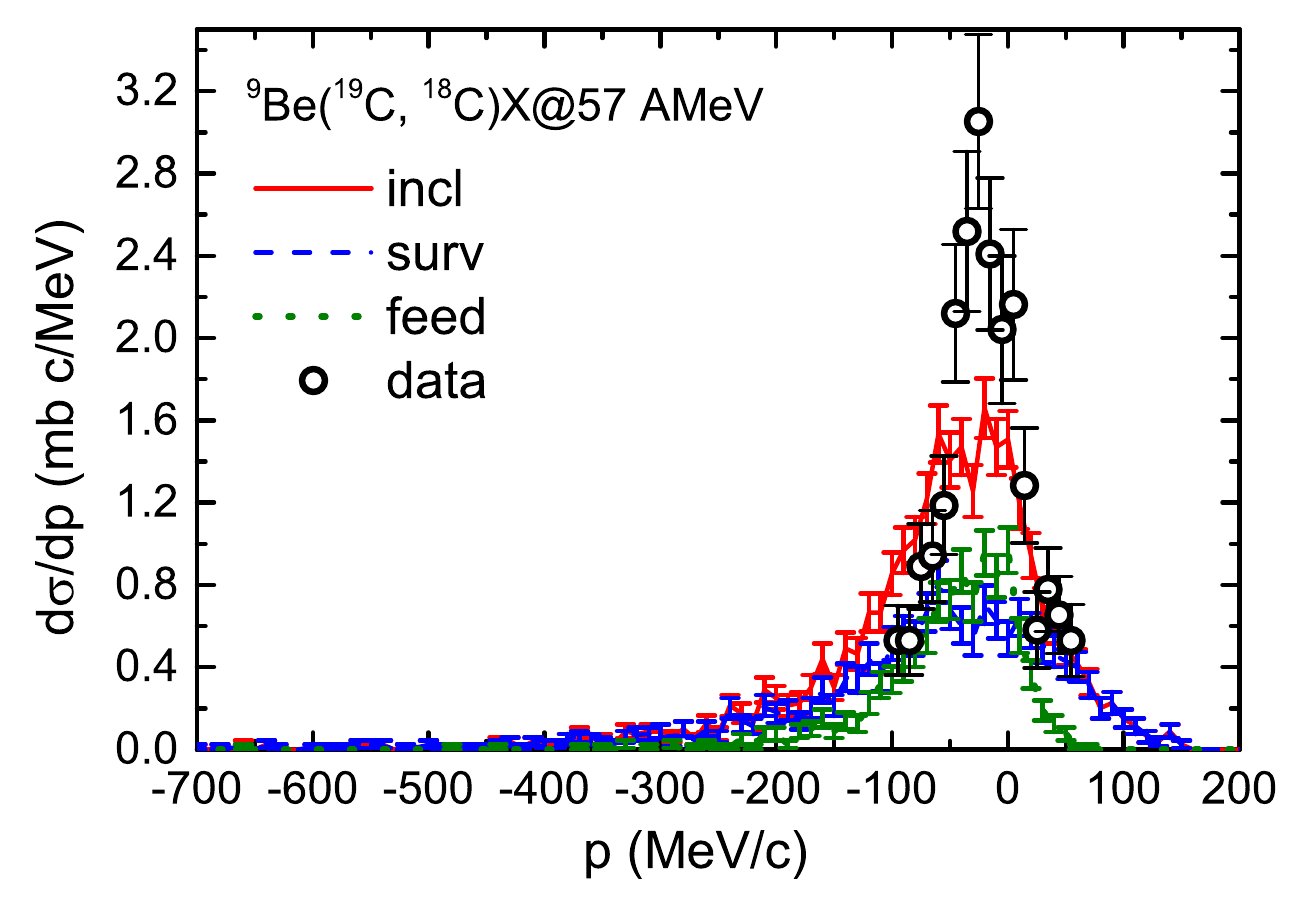}
\caption{Same as Fig. \ref{19C} but for $^{19}$C + $^{9}$Be collisions at 57 AMeV \cite{maddalena2001single}. }
    \label{19C57}
\end{figure}

\begin{figure}[H]   
\centering
    \includegraphics[width=0.45\textwidth,angle=0]{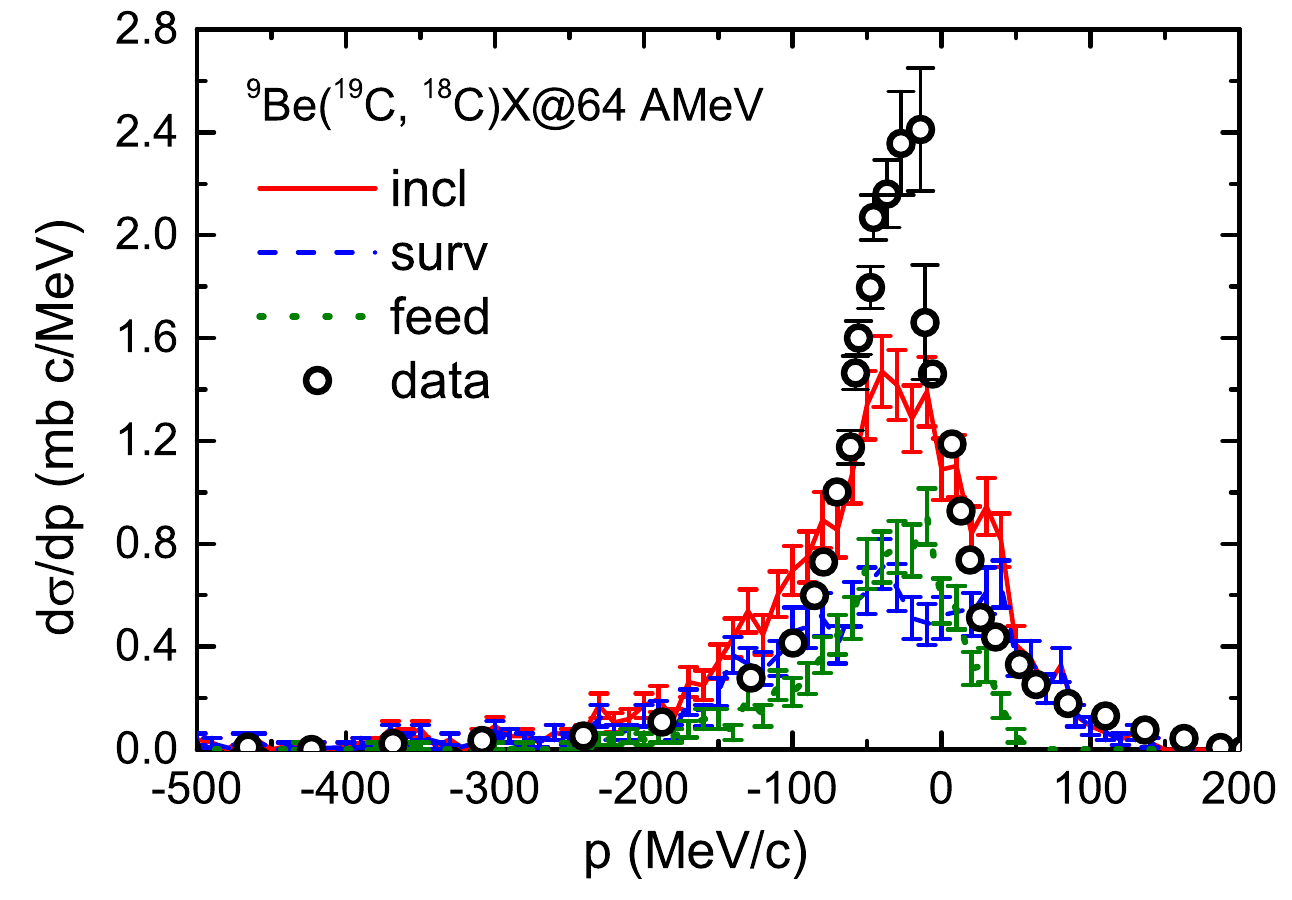}
\caption{Same as Fig. \ref{19C} but for $^{19}$C + $^{9}$Be collisions at 64 AMeV \cite{simpson2009one}. }
    \label{19C64}
\end{figure}

\begin{figure}[H]   
\centering
    \includegraphics[width=0.45\textwidth,angle=0]{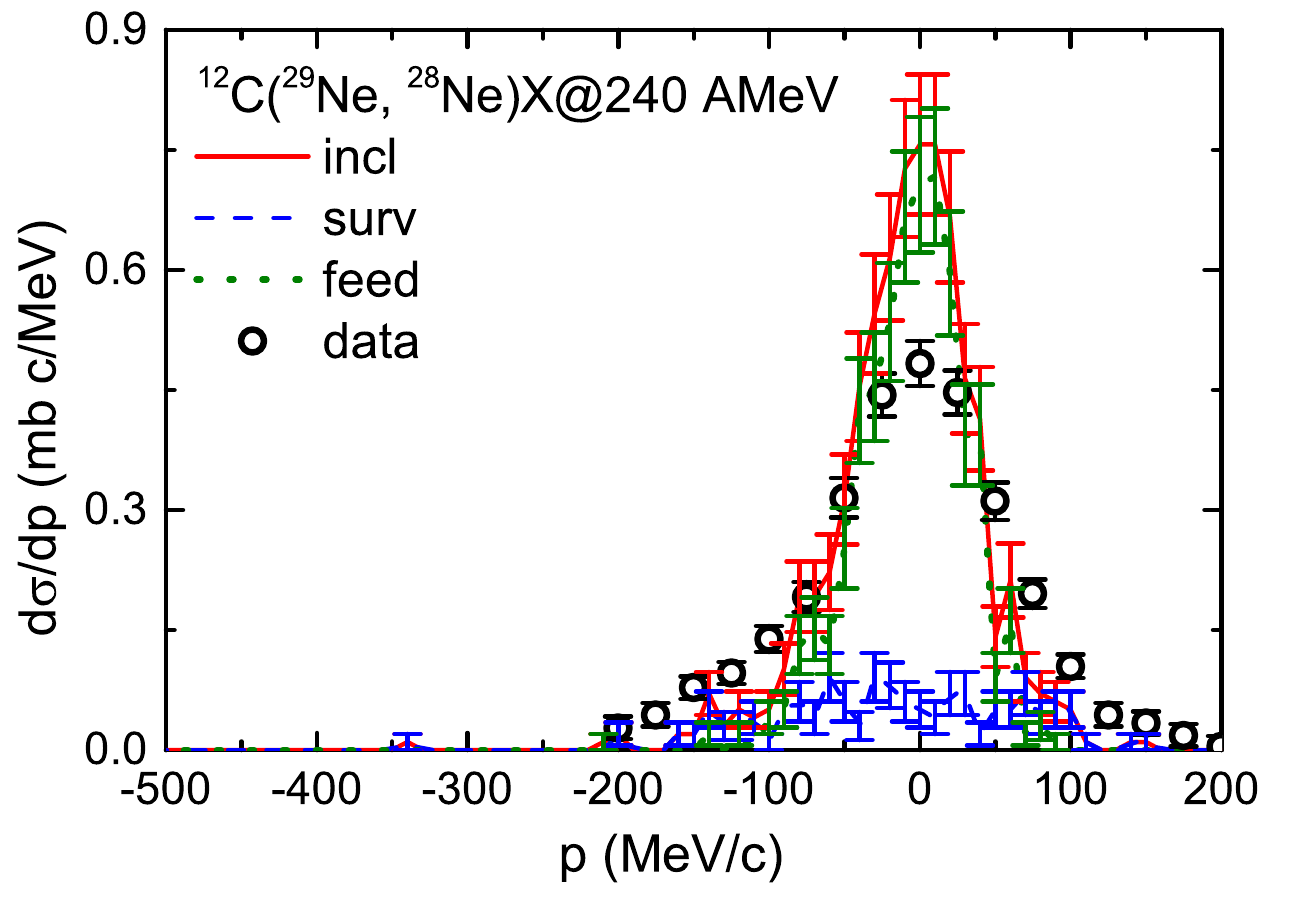}
\caption{Same as Fig. \ref{19C} but for $^{29}$Ne + $^{12}$C collisions at 240 AMeV \cite{kobayashi2016one}.}
    \label{29Ne}
\end{figure}

\begin{figure}[H]   
\centering
    \includegraphics[width=0.45\textwidth,angle=0]{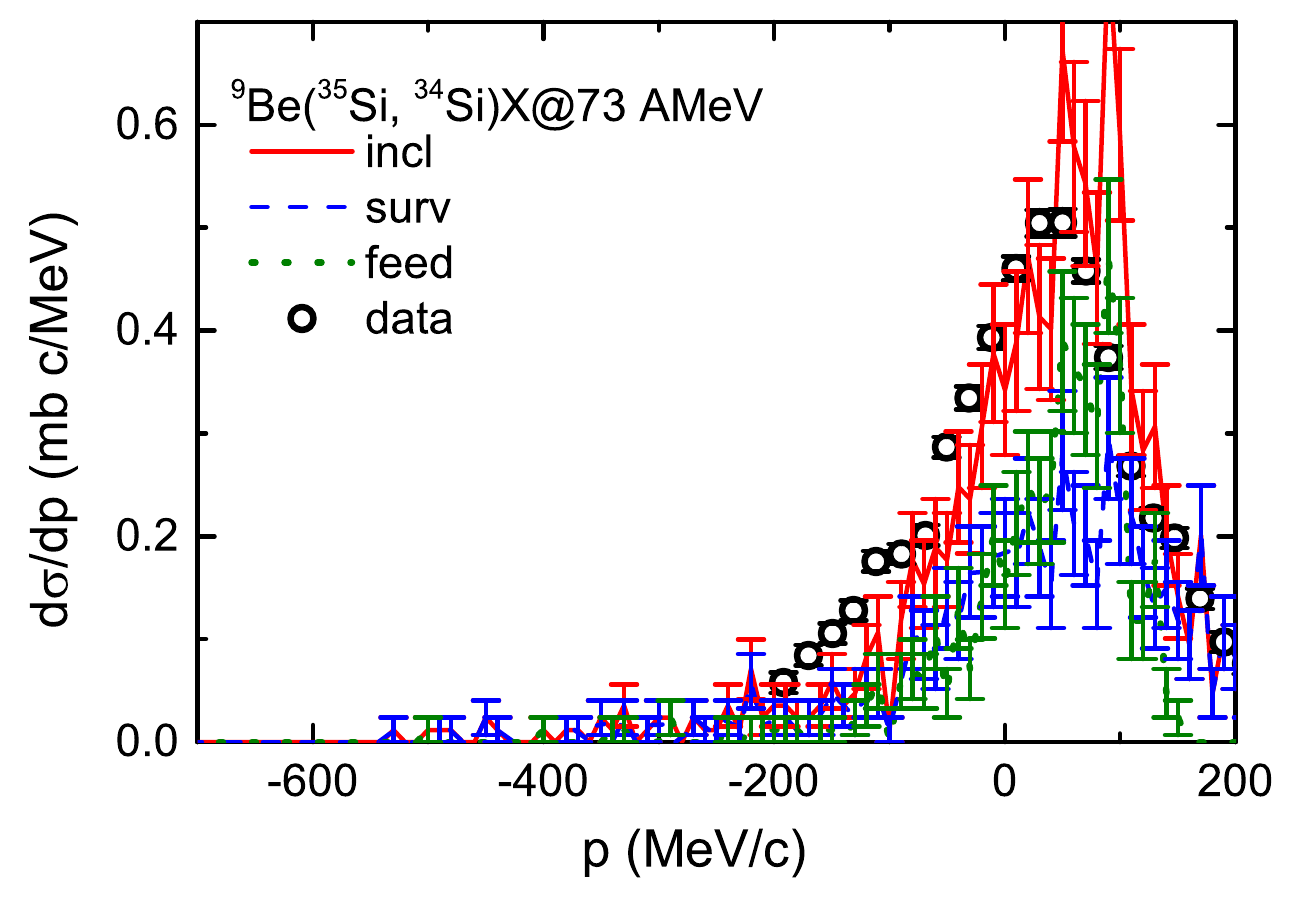}
\caption{Same as Fig. \ref{19C} but for $^{35}$Si + $^{9}$Be collisions at 73 AMeV \cite{enders2002single}.}
    \label{35Si}
\end{figure}

\begin{figure}[H]   
\centering
    \includegraphics[width=0.45\textwidth,angle=0]{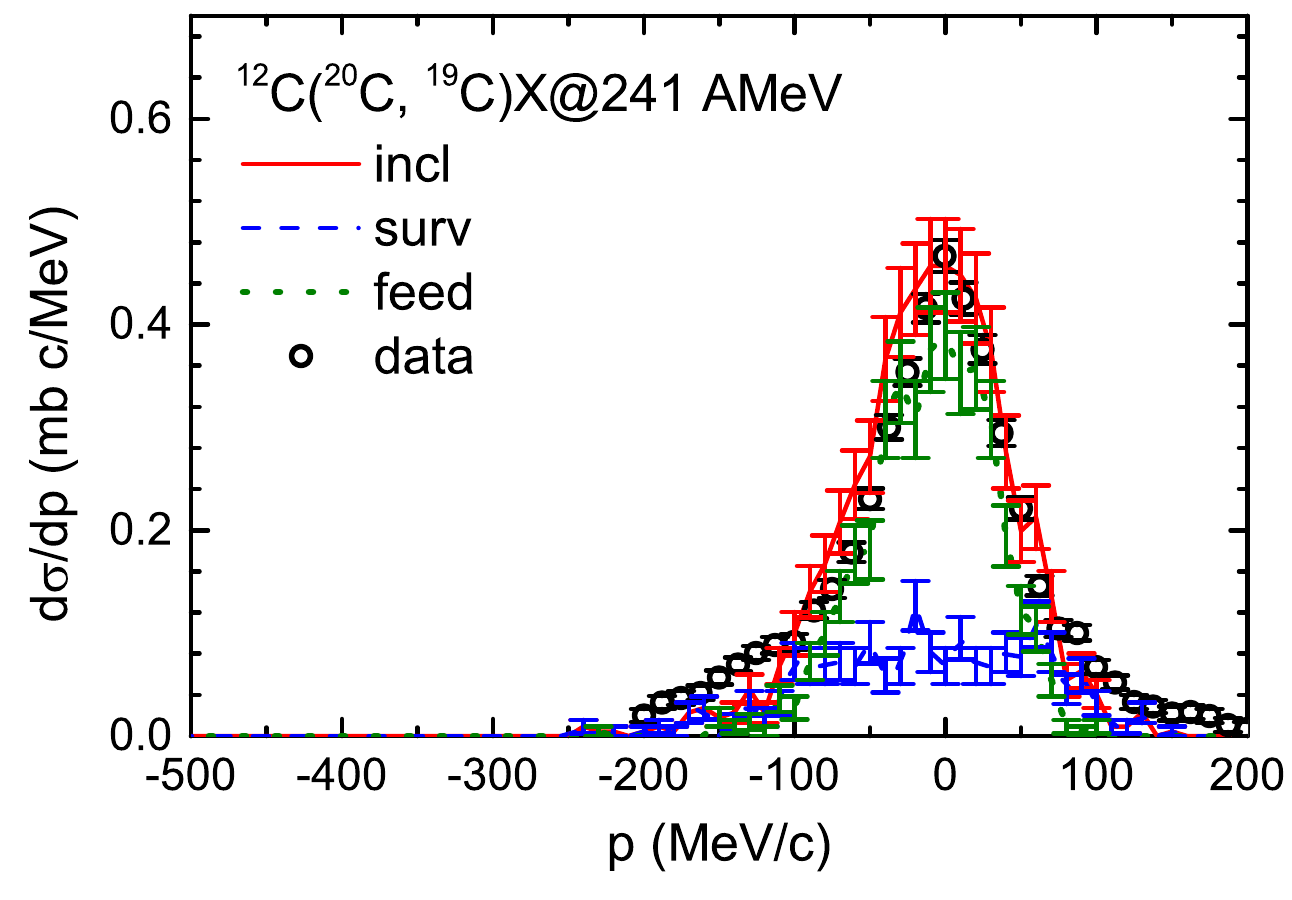}
\caption{Same as Fig. \ref{19C} but for $^{20}$C + $^{12}$C collisions at 241 AMeV \cite{kobayashi2012one}.}
    \label{20C}
\end{figure}

\begin{figure}[H]   
\centering
    \includegraphics[width=0.45\textwidth,angle=0]{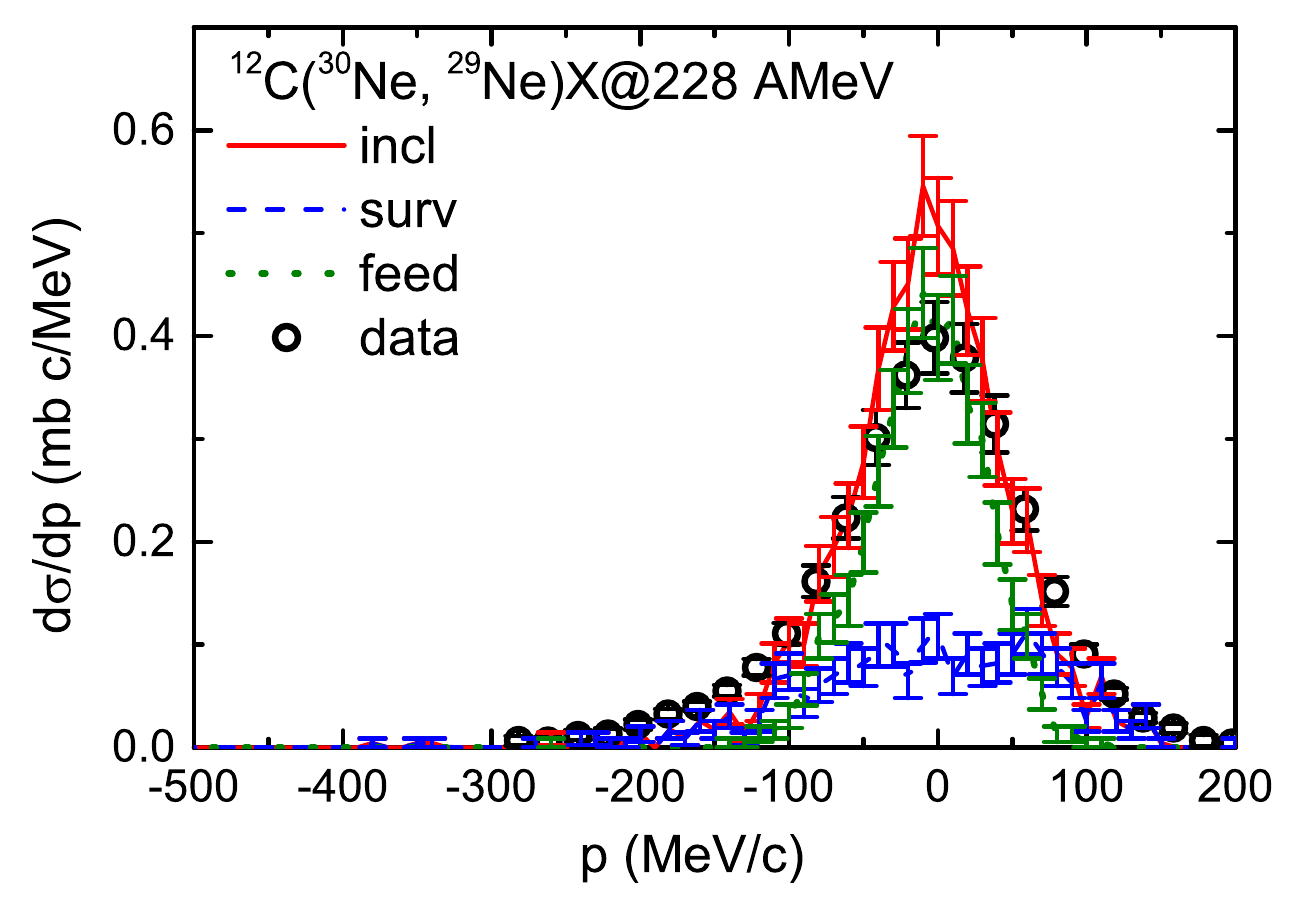}
\caption{Same as Fig. \ref{19C} but for $^{30}$Ne + $^{12}$C collisions at 228 AMeV \cite{liu2017intruder}.}
    \label{30Ne}
\end{figure}

\begin{figure}[H]   
\centering
    \includegraphics[width=0.45\textwidth,angle=0]{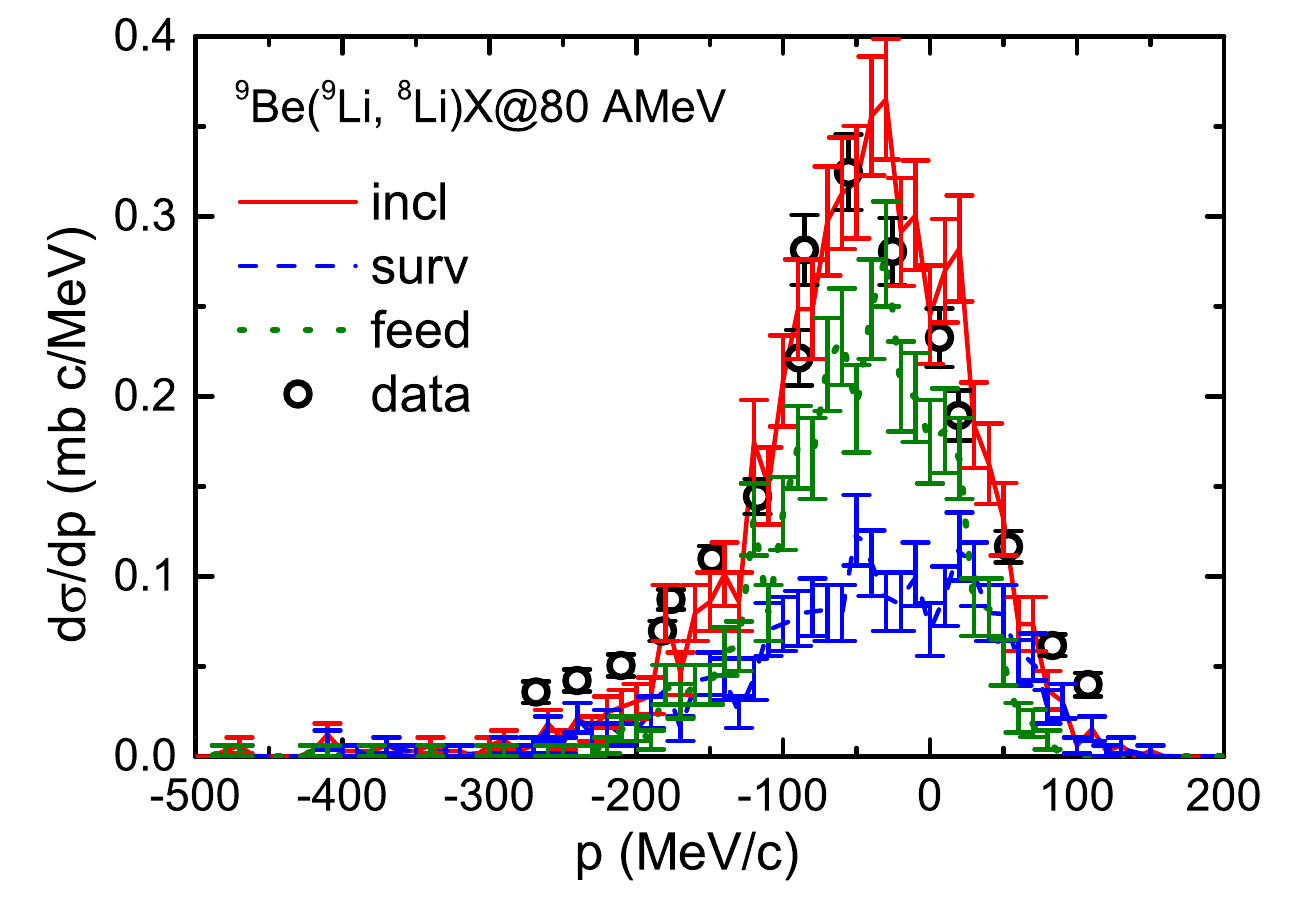}
\caption{Same as Fig. \ref{19C} but for $^{9}$Li + $^{9}$Be collisions at 80 AMeV \cite{grinyer2012systematic}.}
    \label{9Li}
\end{figure}

\begin{figure}[H]   
\centering
    \includegraphics[width=0.45\textwidth,angle=0]{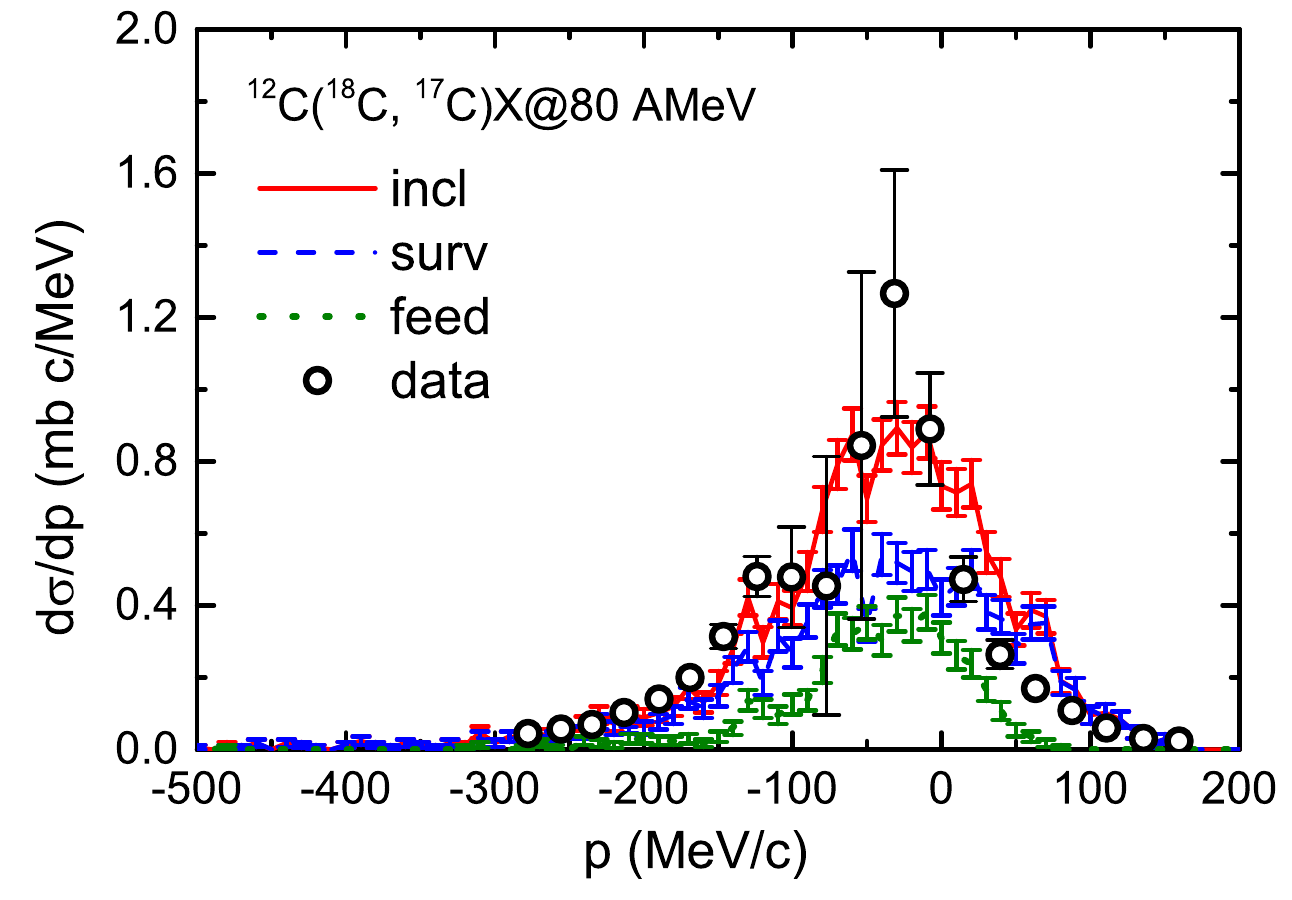}
\caption{Same as Fig. \ref{19C} but for $^{18}$C + $^{12}$C collisions at 80 AMeV \cite{ozawa2008measurement}.}
    \label{18C80}
\end{figure}

\begin{figure}[H]   
\centering
    \includegraphics[width=0.45\textwidth,angle=0]{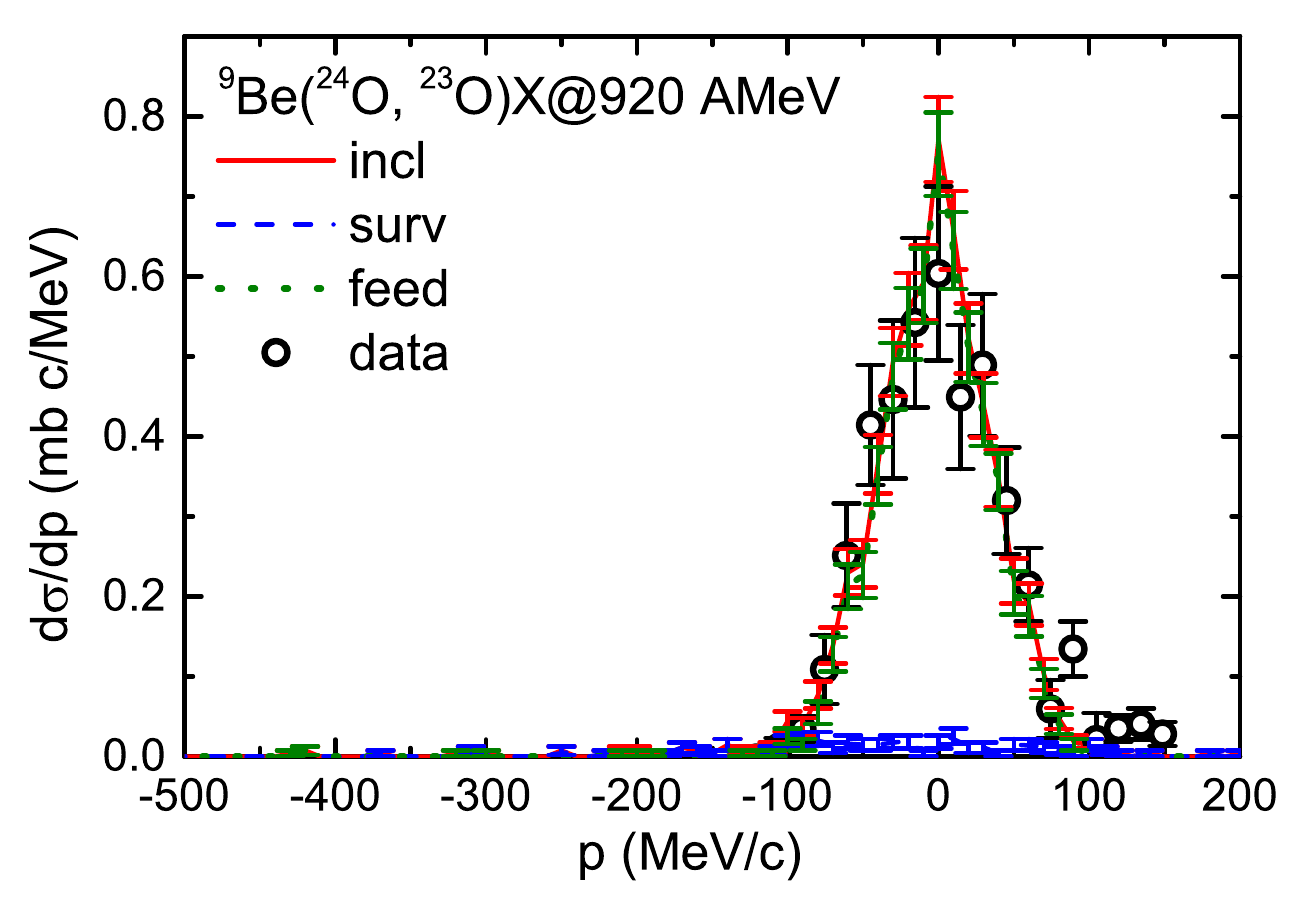}
\caption{Same as Fig. \ref{19C} but for $^{24}$O + $^{12}$C collisions at 920 AMeV \cite{kanungo2009one}.}
    \label{24O}
\end{figure}

\begin{figure}[H]   
\centering
    \includegraphics[width=0.45\textwidth,angle=0]{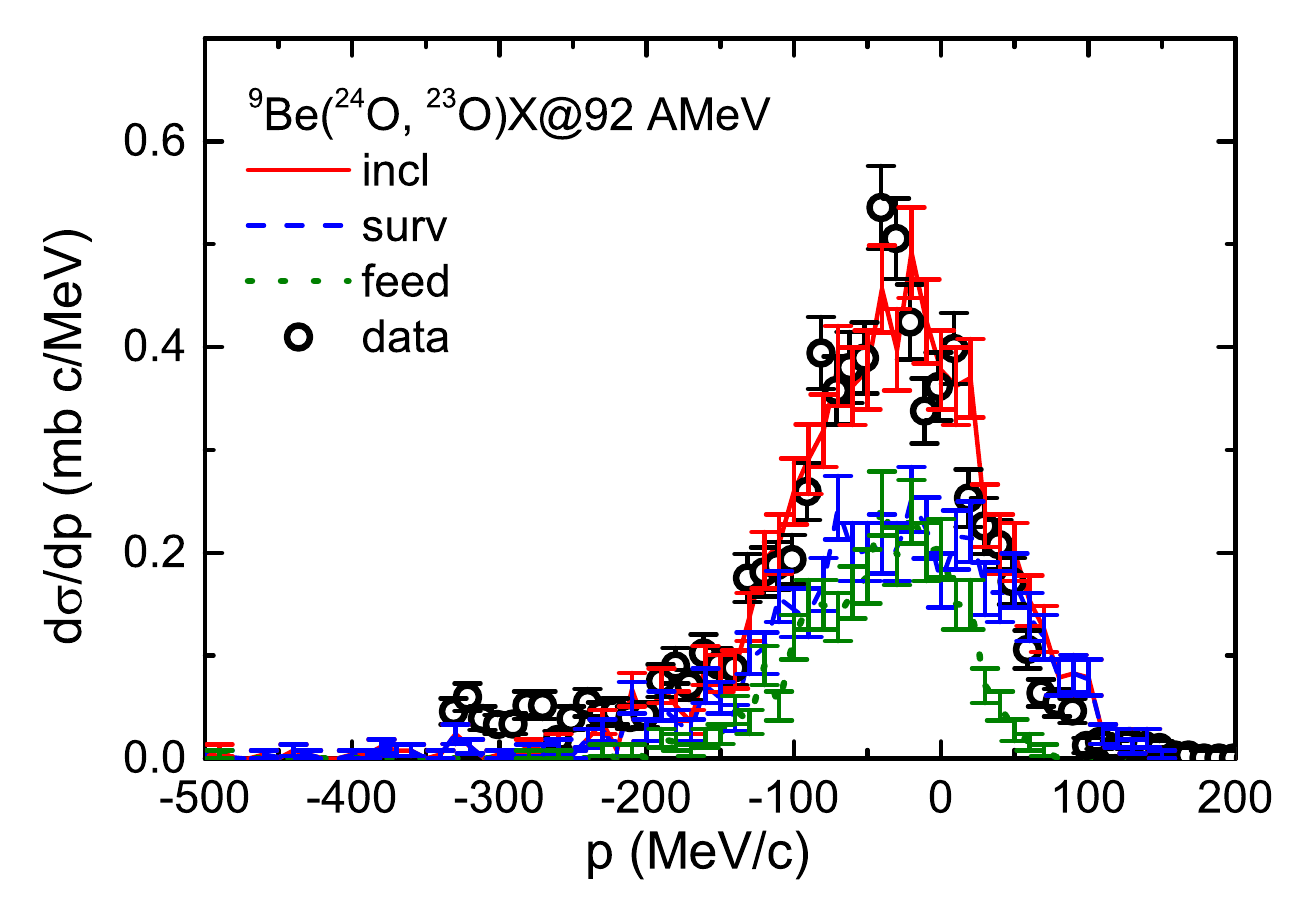}
\caption{Same as Fig. \ref{19C} but for $^{24}$O + $^{9}$Be collisions at 92 AMeV \cite{divaratne2018one}.}
    \label{24O92}
\end{figure}

\begin{figure}[H]   
\centering
    \includegraphics[width=0.45\textwidth,angle=0]{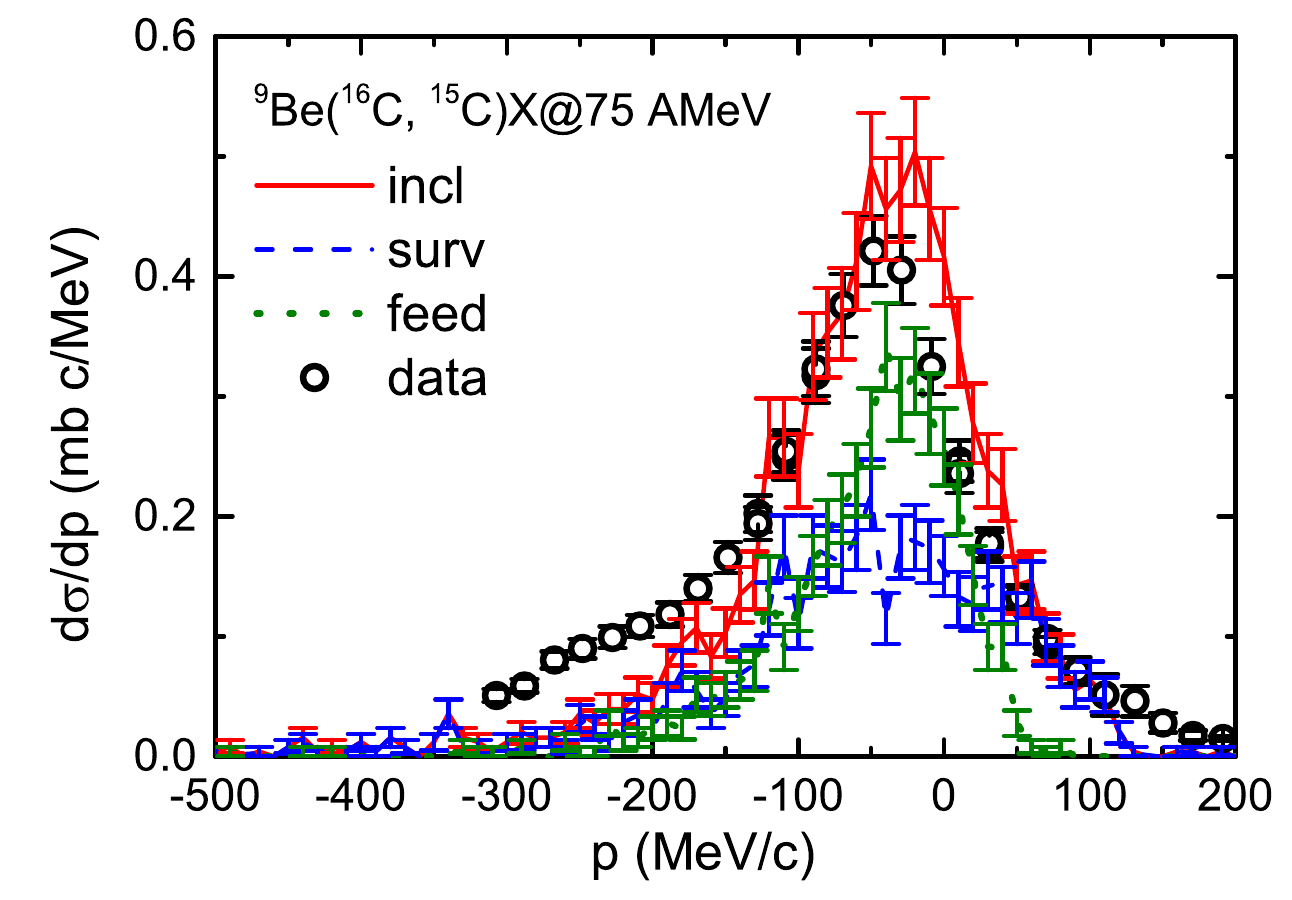}
\caption{Same as Fig. \ref{19C} but for  $^{16}$C + $^{9}$Be collisions at 75 AMeV \cite{flavigny2012nonsudden}.}
    \label{16C}
\end{figure}

\begin{figure}[H]   
\centering
    \includegraphics[width=0.45\textwidth,angle=0]{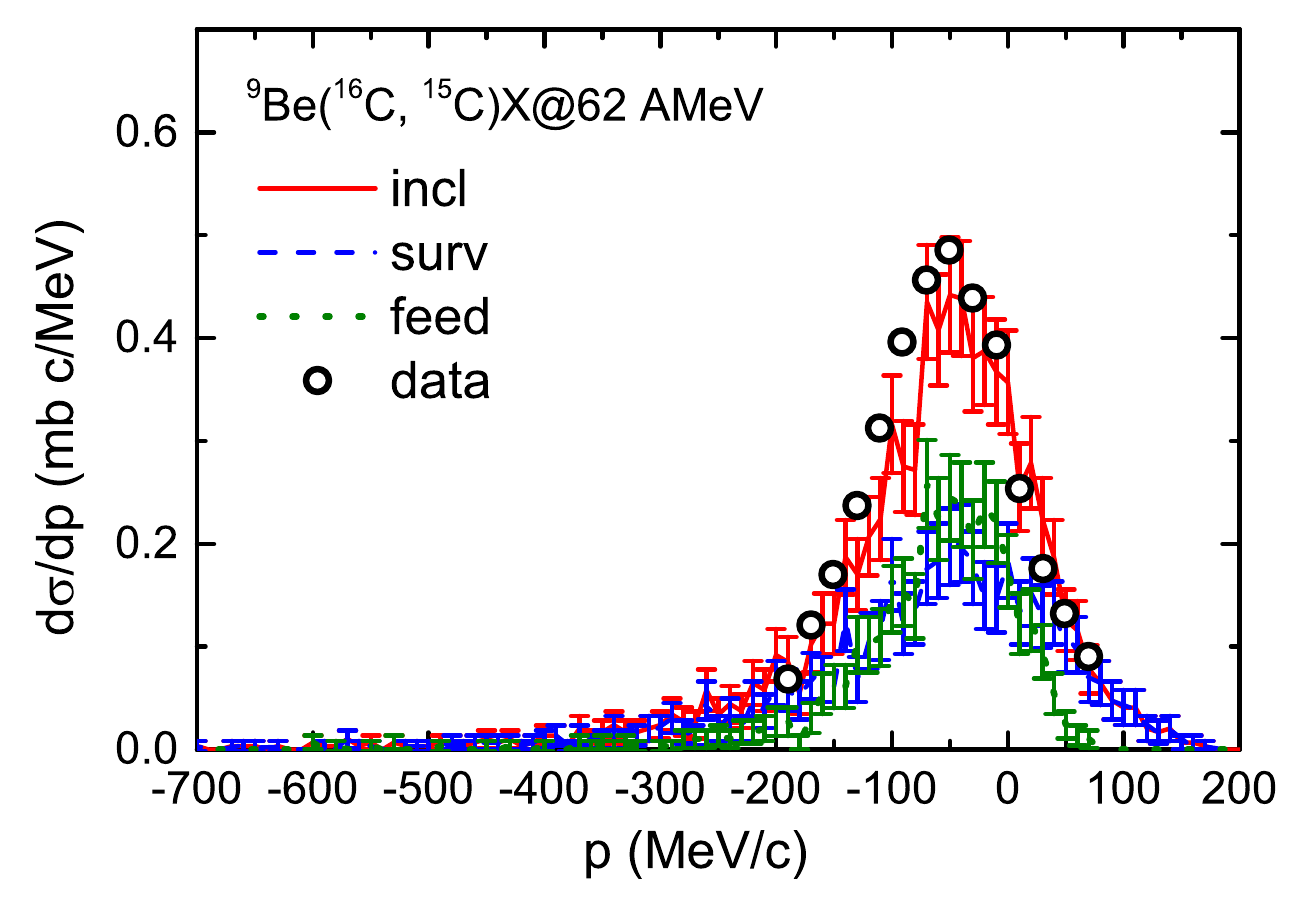}
\caption{Same as Fig. \ref{19C} but for  $^{16}$C + $^{9}$Be collisions at 62 AMeV \cite{maddalena2001single}.}
    \label{16C62}
\end{figure}

\begin{figure}[H]   
\centering
    \includegraphics[width=0.45\textwidth,angle=0]{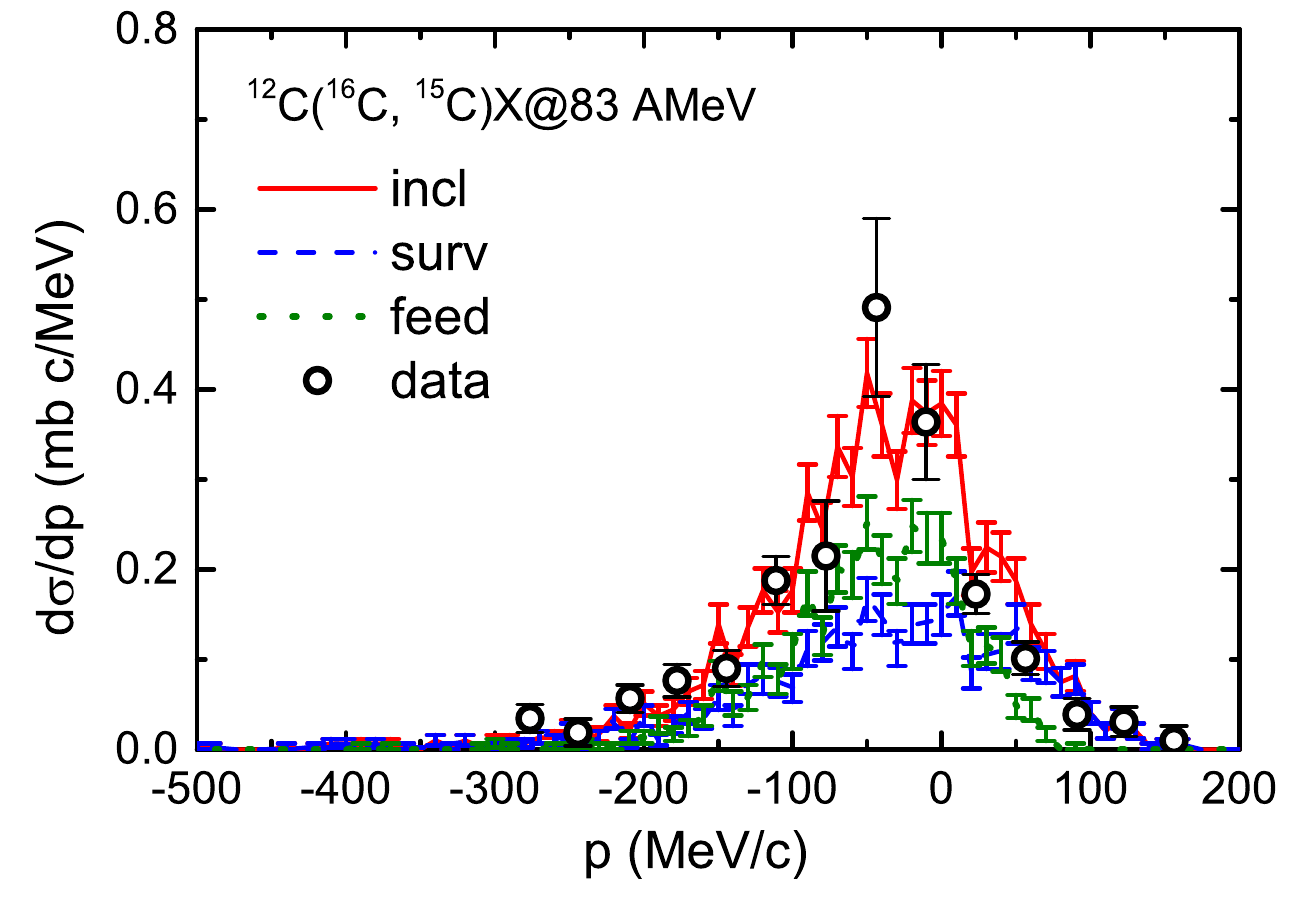}
\caption{Same as Fig. \ref{19C} but for  $^{16}$C + $^{12}$C collisions at 83 AMeV \cite{yamaguchi2004neutron}.}
    \label{16C83}
\end{figure}

\begin{figure}[H]   
\centering
    \includegraphics[width=0.45\textwidth,angle=0]{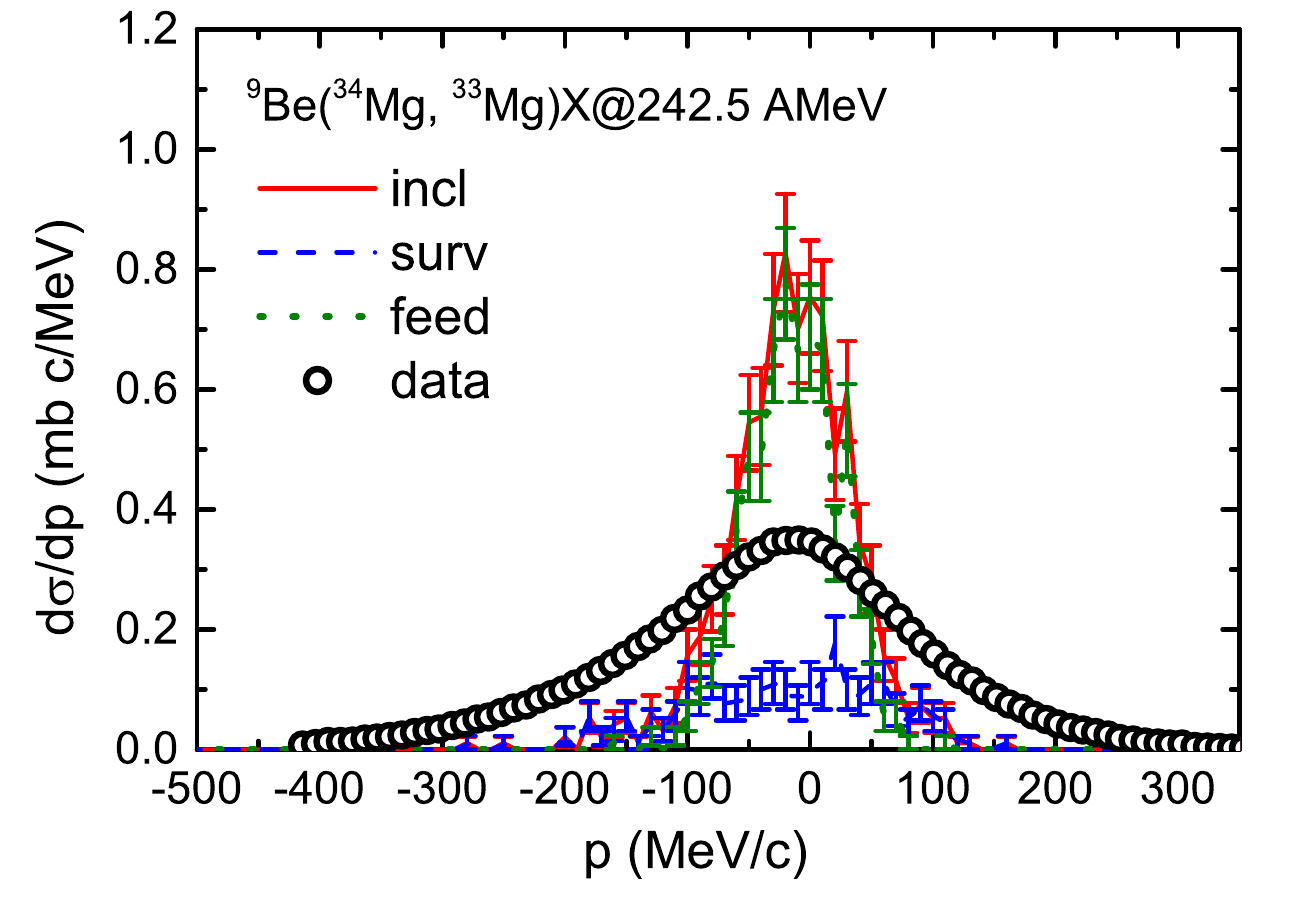}
\caption{Same as Fig. \ref{19C} but for  $^{34}$Mg + $^{9}$Be collisions at 242.5 AMeV \cite{bazin2021spectroscopy}.}
    \label{34Mg242.5}
\end{figure}

\begin{figure}[H]   
\centering
    \includegraphics[width=0.45\textwidth,angle=0]{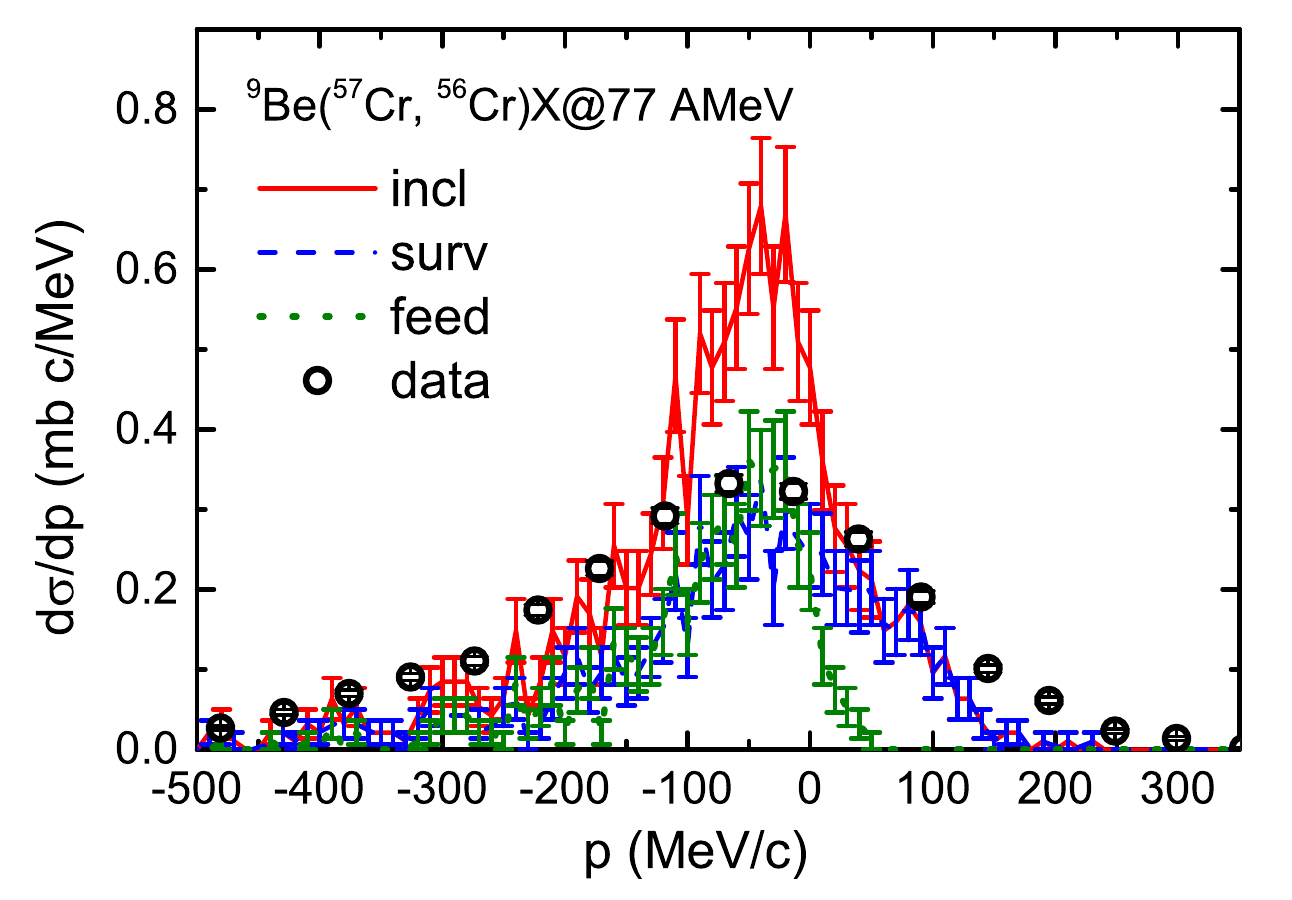}
\caption{Same as Fig. \ref{19C} but for  $^{57}$Cr + $^{9}$Be collisions at 77 AMeV \cite{gade2006one}.}
    \label{57Cr77}
\end{figure}

\begin{figure}[H]   
\centering
    \includegraphics[width=0.45\textwidth,angle=0]{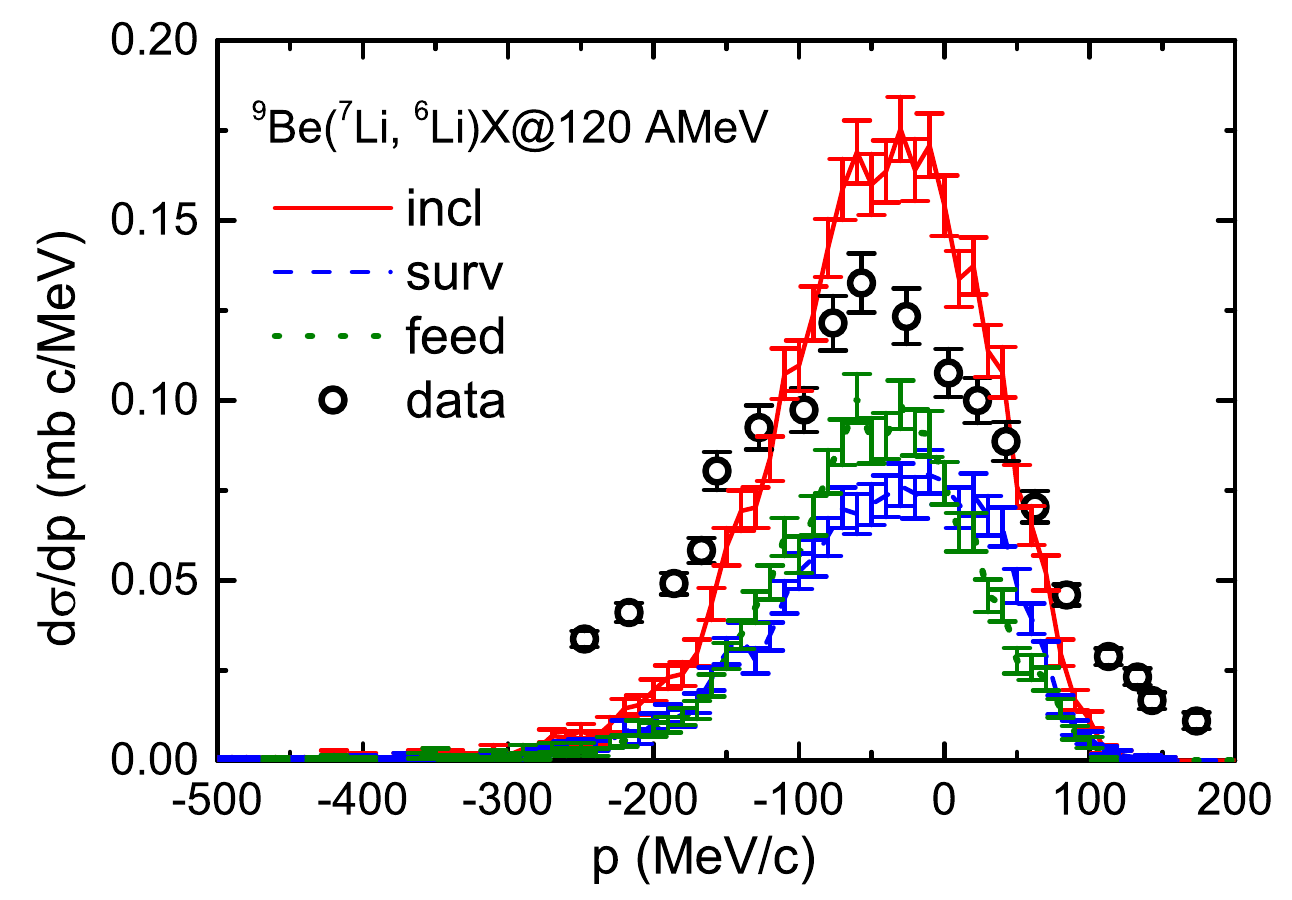}
\caption{Same as Fig. \ref{19C} but for $^{7}$Li + $^{9}$Be collisions at 120 AMeV \cite{grinyer2012systematic}.}
    \label{7Li}
\end{figure}

\begin{figure}[H]   
\centering
    \includegraphics[width=0.45\textwidth,angle=0]{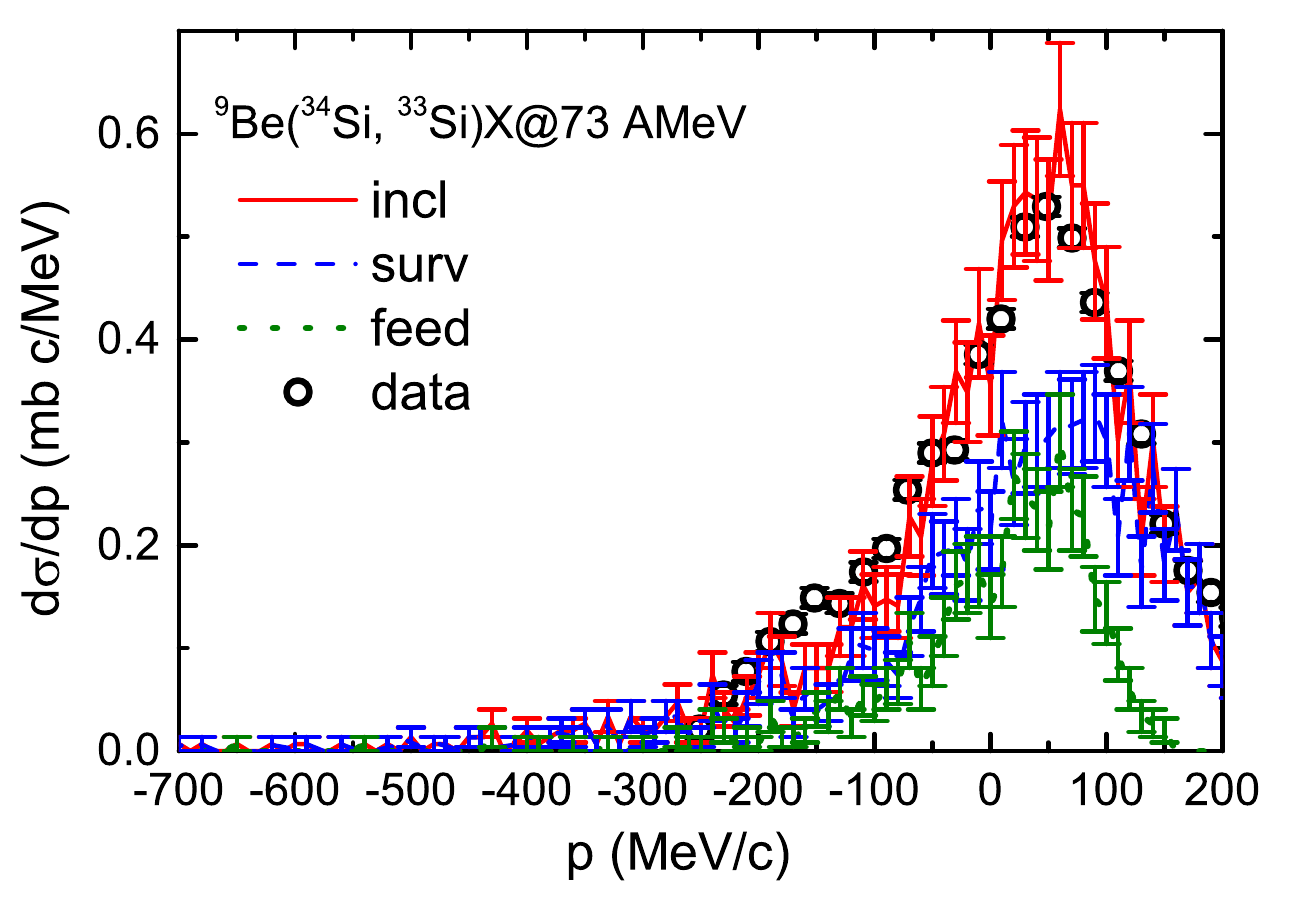}
\caption{Same as Fig. \ref{19C} but for $^{34}$Si + $^{9}$Be collisions at 73 AMeV \cite{enders2002single}.}
    \label{34Si}
\end{figure}

\begin{figure}[H]   
\centering
    \includegraphics[width=0.45\textwidth,angle=0]{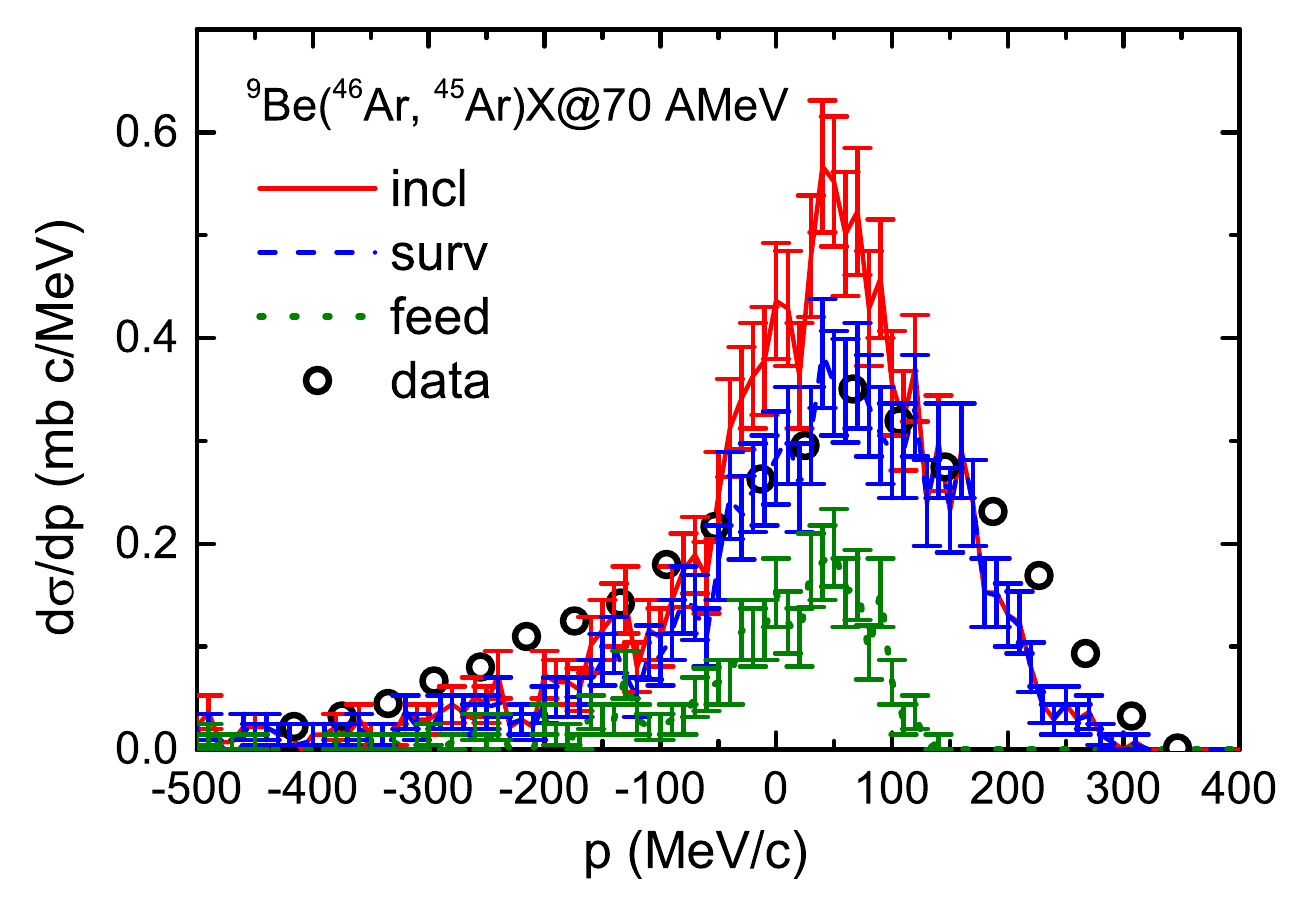}
\caption{Same as Fig. \ref{19C} but for $^{46}$Ar + $^{9}$Be collisions at 70 AMeV \cite{gade2005knockout}.}
    \label{46Ar}
\end{figure}

\begin{figure}[H]   
\centering
    \includegraphics[width=0.45\textwidth,angle=0]{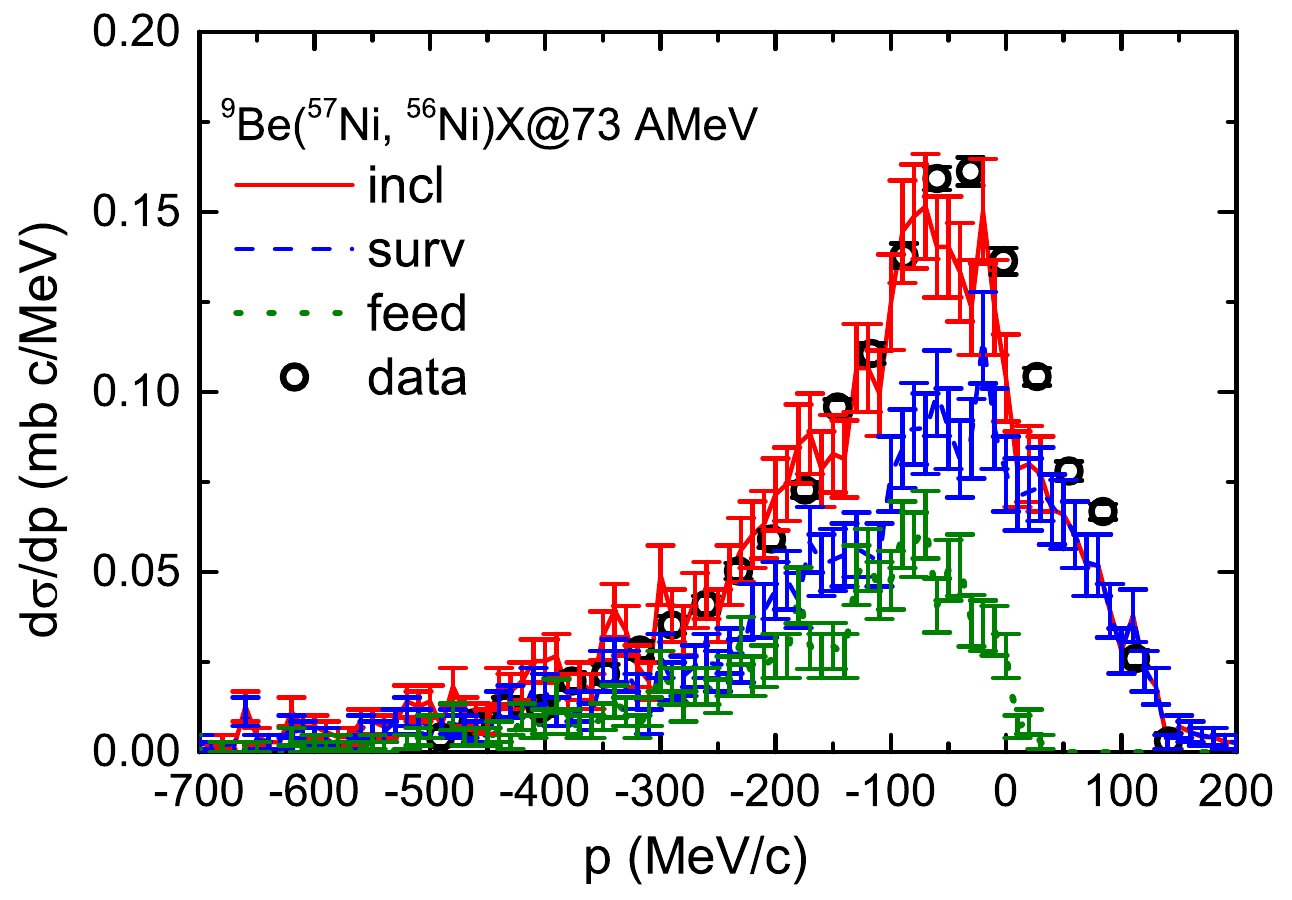}
\caption{Same as Fig. \ref{19C} but for $^{57}$Ni + $^{9}$Be collisions at 73 AMeV \cite{yurkewicz2006one}.}
    \label{57Ni}
\end{figure}

\begin{figure}[H]   
\centering
    \includegraphics[width=0.45\textwidth,angle=0]{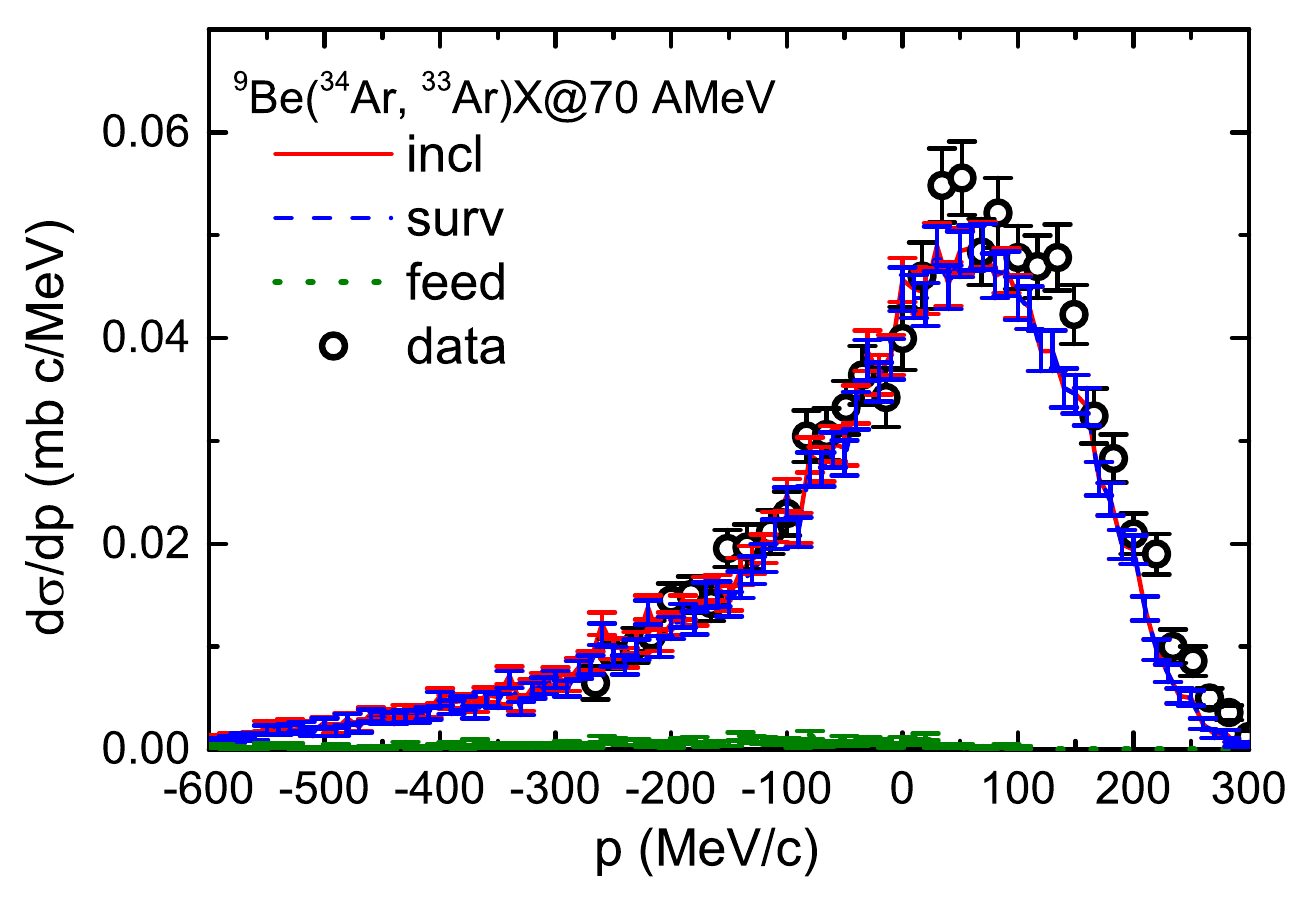}
\caption{Same as Fig. \ref{19C} but for $^{34}$Ar + $^{9}$Be collisions at 70 AMeV \cite{gade2004one}.}
    \label{34Ar}
\end{figure}

\begin{figure}[H]   
\centering
    \includegraphics[width=0.45\textwidth,angle=0]{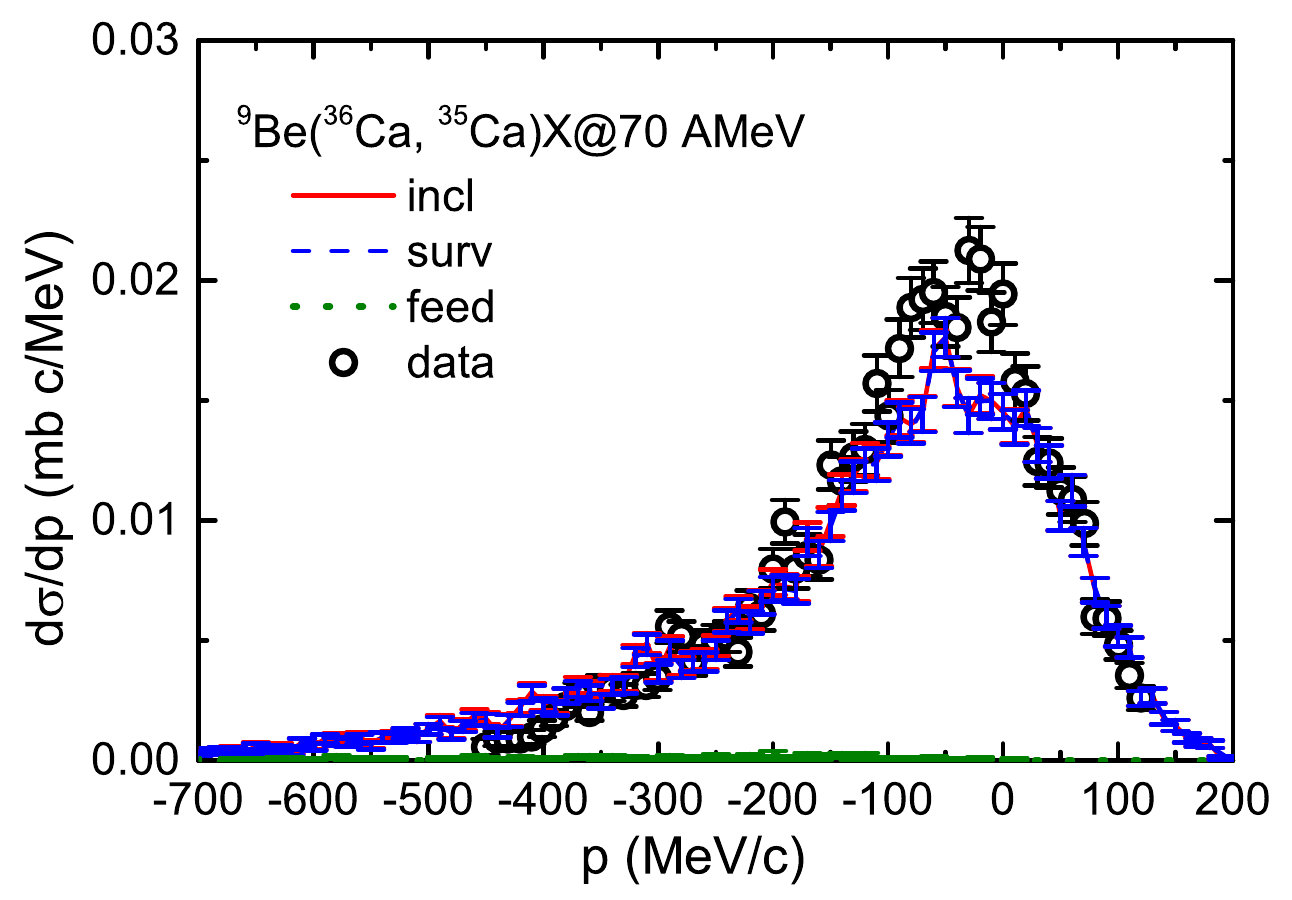}
\caption{Same as Fig. \ref{19C} but for $^{36}$Ca + $^{9}$Be collisions at 70 AMeV \cite{shane2012proton}.}
    \label{36Ca}
\end{figure}

\begin{figure}[H]   
\centering
    \includegraphics[width=0.45\textwidth,angle=0]{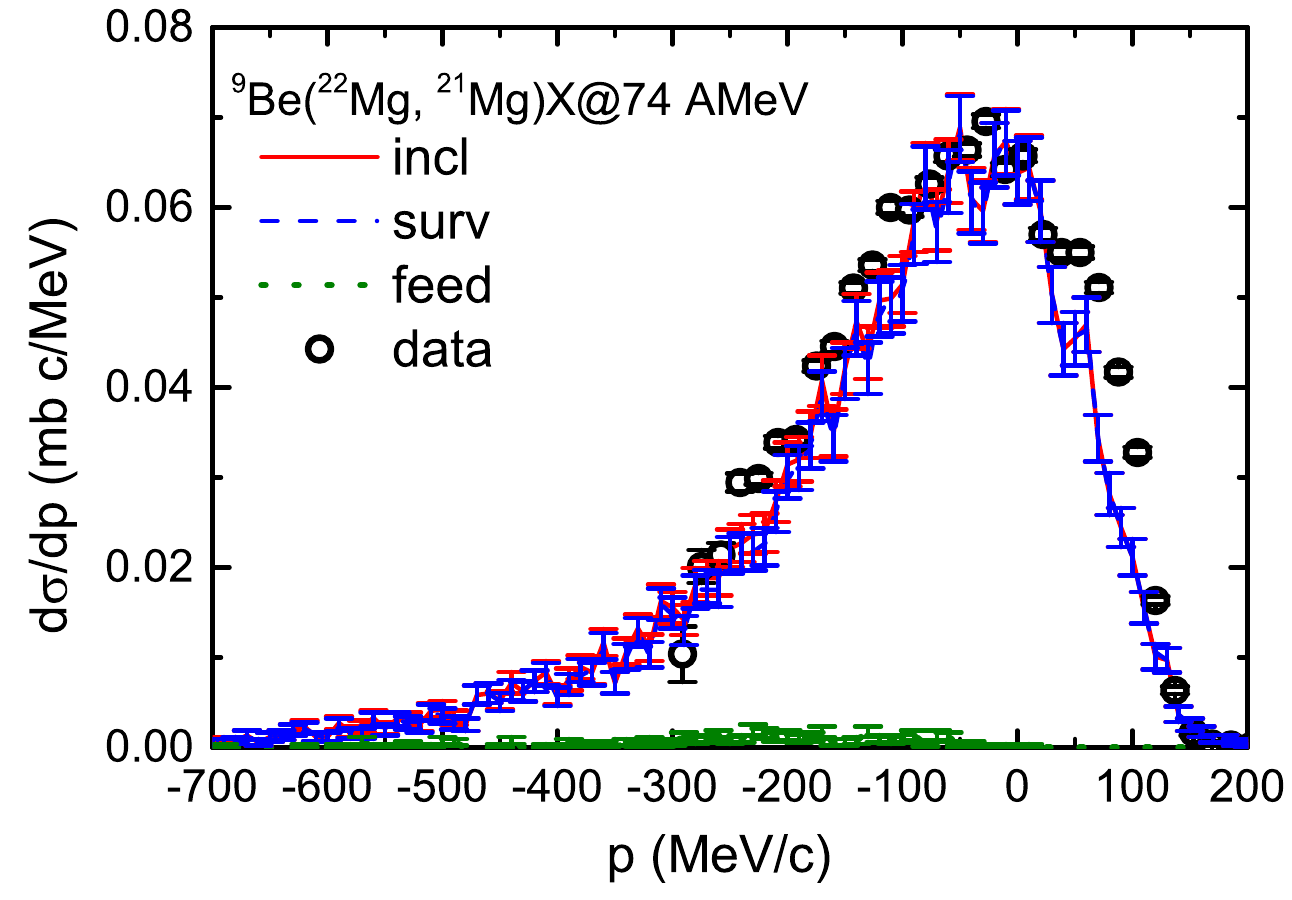}
\caption{Same as Fig. \ref{19C} but for $^{22}$Mg + $^{9}$Be collisions at 74 AMeV \cite{diget2008structure}.}
    \label{22Mg}
\end{figure}

\begin{figure}[H]   
\centering
    \includegraphics[width=0.45\textwidth,angle=0]{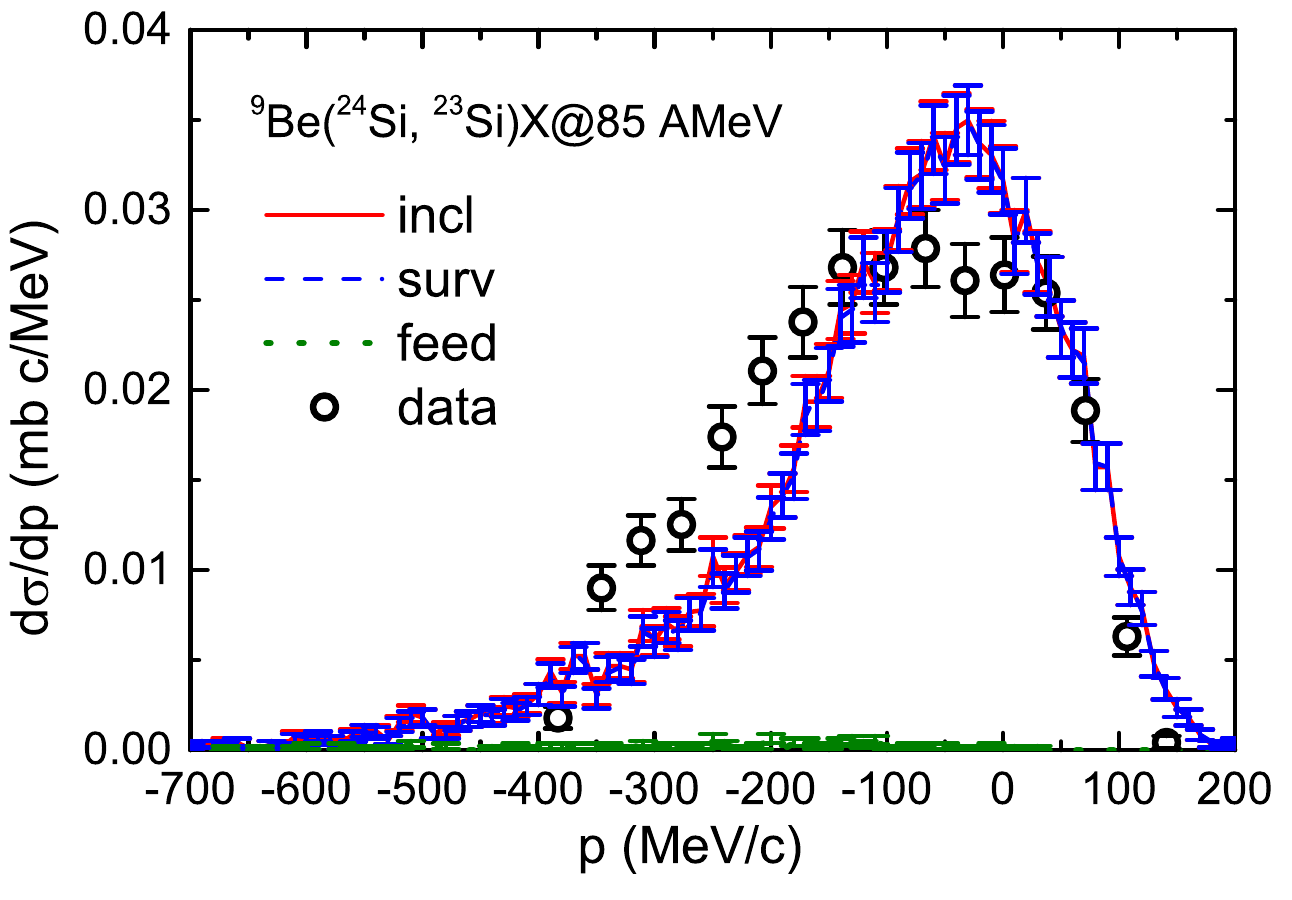}
\caption{Same as Fig. \ref{19C} but for $^{24}$Si + $^{9}$Be collisions at 85 AMeV \cite{gade2008reduction}.}
    \label{24Si}
\end{figure}

\begin{figure}[H]   
\centering
    \includegraphics[width=0.45\textwidth,angle=0]{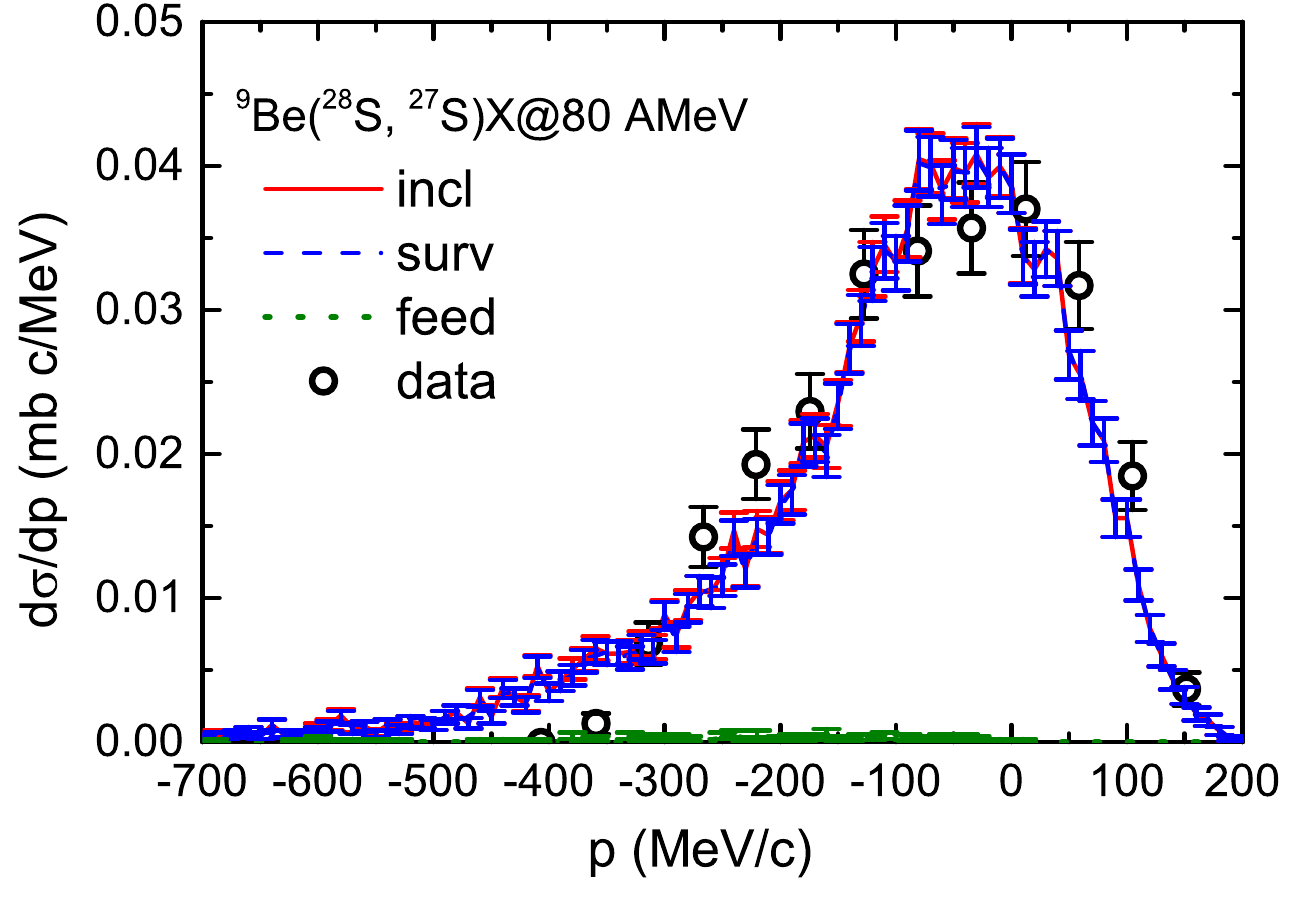}
\caption{Same as Fig. \ref{19C} but for $^{28}$S + $^{9}$Be collisions at 80 AMeV \cite{gade2008reduction}.}
    \label{28S}
\end{figure}

\begin{figure}[H]   
\centering
    \includegraphics[width=0.45\textwidth,angle=0]{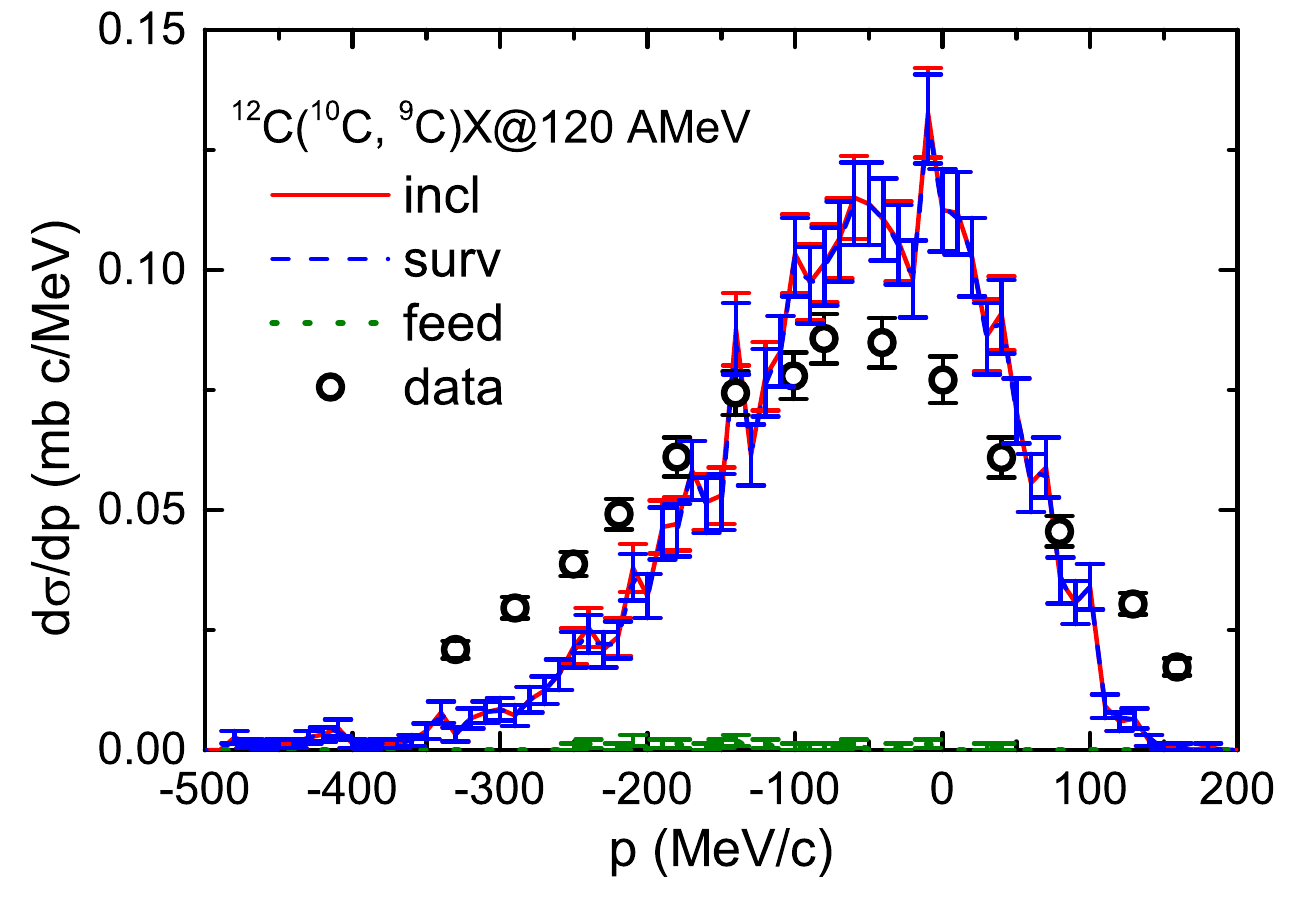}
\caption{Same as Fig. \ref{19C} but for $^{10}$C + $^{12}$C collisions at 120 AMeV \cite{grinyer2012systematic}.}
    \label{10C}
\end{figure}

\begin{figure}[H]   
\centering
    \includegraphics[width=0.45\textwidth,angle=0]{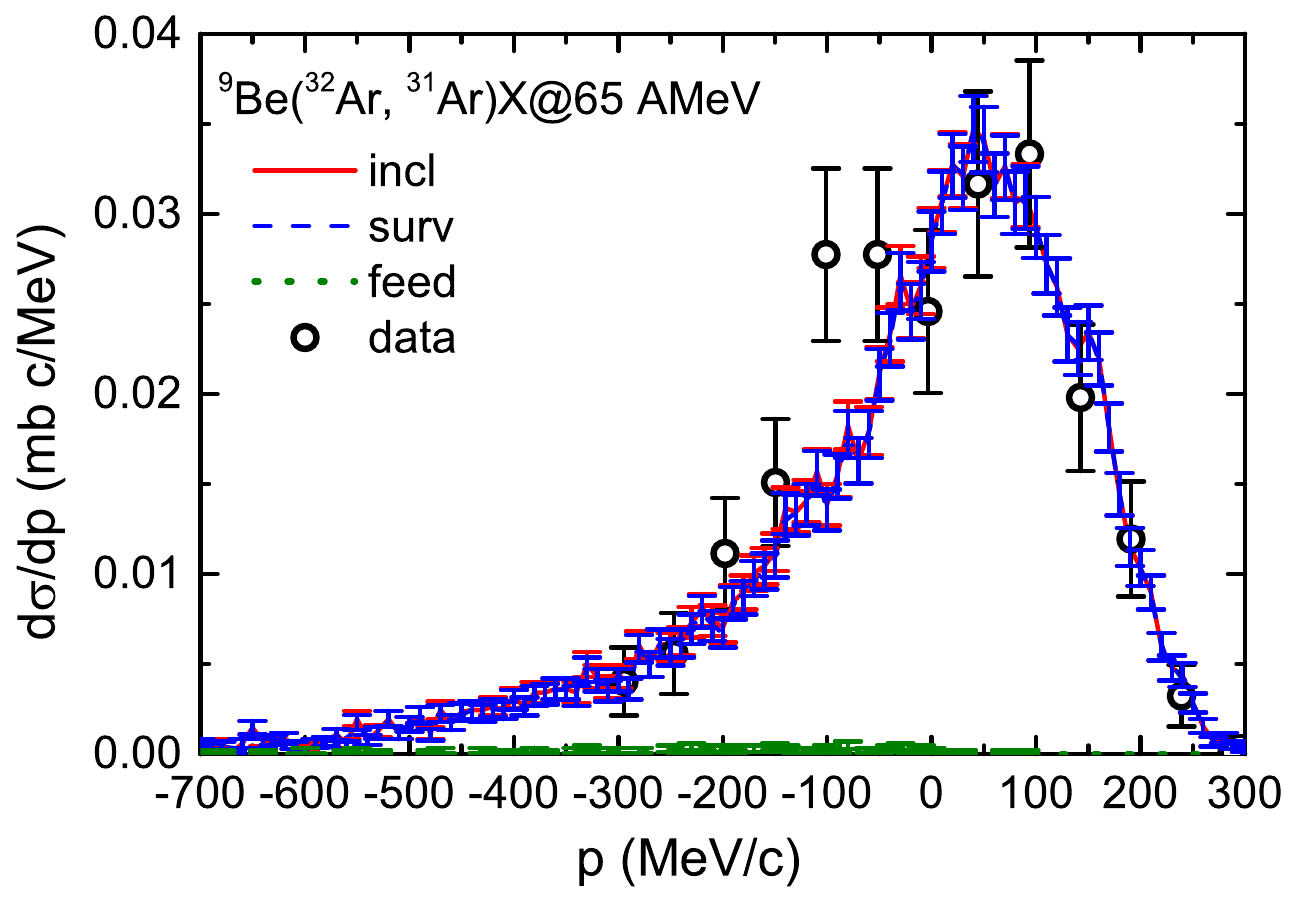}
\caption{Same as Fig. \ref{19C} but for $^{32}$Ar + $^{9}$Be collisions at 65 AMeV \cite{gade2004reduced}.}
    \label{32Ar}
\end{figure}

\begin{figure}[H]   
\centering
    \includegraphics[width=0.45\textwidth,angle=0]{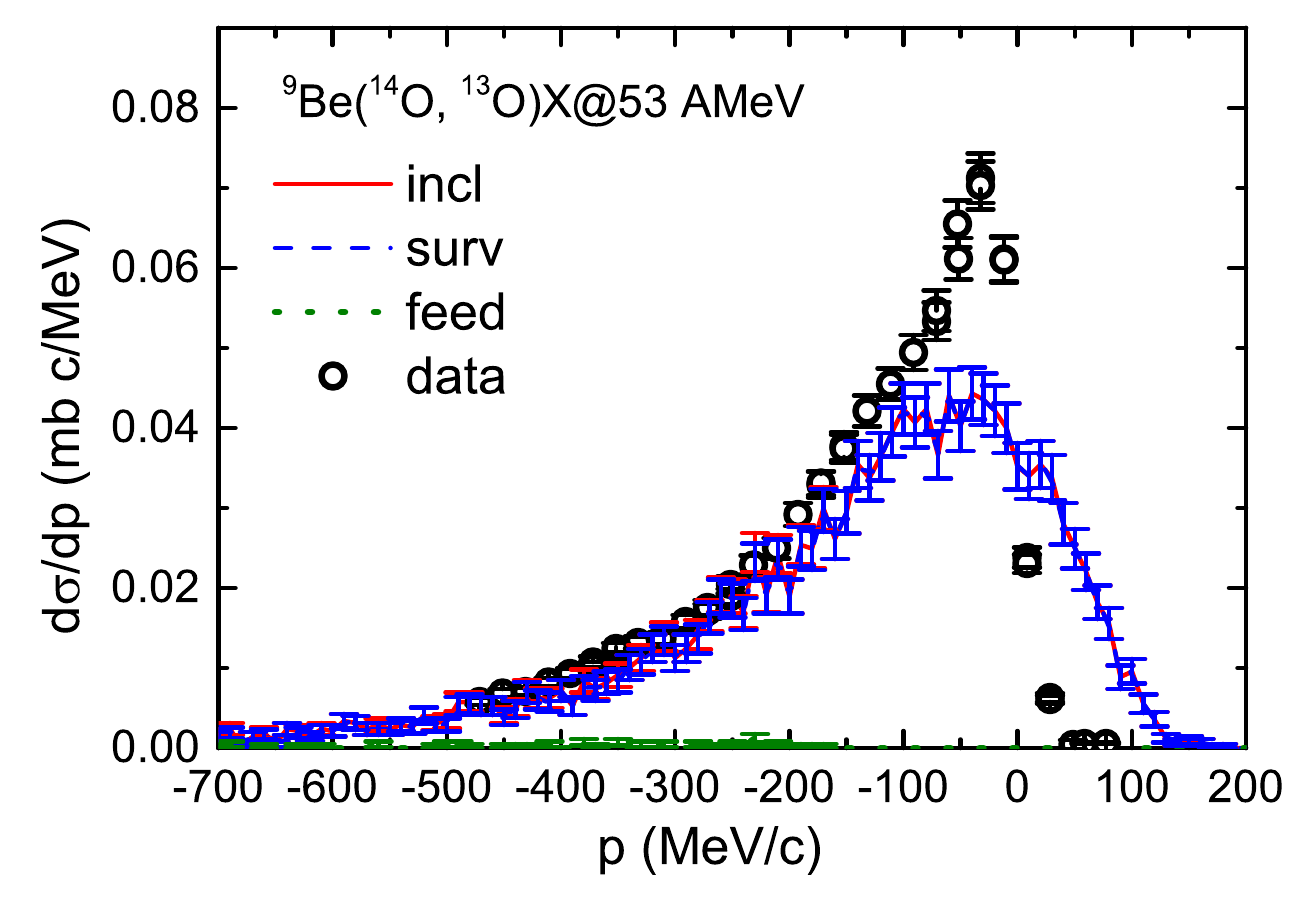}
\caption{Same as Fig. \ref{19C} but for $^{14}$O + $^{9}$Be collisions at 53 AMeV \cite{flavigny2012nonsudden}.}
    \label{14O}
\end{figure}

~\Cref{19C,19C57,19C64,29Ne,35Si,20C,30Ne,9Li,18C80,24O,24O92,16C,16C62,16C83,34Mg242.5,57Cr77,7Li,34Si,46Ar,57Ni,34Ar,36Ca,22Mg,24Si,28S,10C,32Ar,14O} compare the inclusive experimental residue parallel-momentum distributions with the corresponding IQMD+GEMINI calculations and their mechanism decomposition.
For a consistent comparison, both experimental and calculated spectra are presented in the projectile rest frame.
When the original publications report laboratory-frame momenta, we transform them to the projectile rest frame using the published beam kinematics.

For both experiment and theory, the overall normalization is chosen such that the spectrum integral reproduces the published inclusive one-neutron-removal cross section.
To enable a uniform extraction of longitudinal-momentum widths, we apply a light Gaussian smoothing to the spectra and, when different centering conventions are used in the experimental presentations, a small recentering is applied to align the peaks.
These preprocessing steps are used only to stabilize the width determination and do not affect the extracted full widths at half maximum (FWHM) within the quoted uncertainties; they also do not change the mechanism decomposition or the conclusions drawn from it.

\appsubsec{B.1. FWHM definition}

For the 28 systems with published residue parallel-momentum distributions, we extract FWHM and compare them with the corresponding IQMD+GEMINI calculations.
All spectra are analyzed on a common longitudinal-momentum grid $p_z$, and the same procedure is applied to the experimental spectra and to the calculated inclusive and mechanism-resolved spectra.

For a discrete spectrum $\{p_i,y_i\}$, we first apply a light Gaussian smoothing (in bin-index space) to suppress bin-to-bin fluctuations while preserving the overall line shape.
We use a seven-point Gaussian kernel with width $\sigma_{\rm idx}=1$, normalized to unity.
The smoothed spectrum $y_i^{\rm (sm)}$ is obtained by discrete convolution; any negative values (if present after smoothing) are set to zero.

We then locate the global maximum $y_{\max}$ at $p_{\max}$, define the half-maximum level $y_{\rm half}=y_{\max}/2$, and determine the left and right half-maximum crossing points by linear interpolation between the adjacent bins that bracket $y_{\rm half}$.
The FWHM is defined as
\begin{equation}
  W = p_R - p_L .
\end{equation}

\appsubsec{B.2. Uncertainties on the FWHM}

The statistical uncertainty on $W$ is obtained by Monte Carlo propagation of the bin-wise uncertainties on $d\sigma/dp_z$.
For each spectrum we generate an ensemble of pseudo-experiments,
\begin{equation}
  y_i^{(k)} = y_i + \mathcal{N}(0,\delta y_i),
\end{equation}
with $k=1,\dots,N_{\rm samp}$ and $N_{\rm samp}=400$.
For each pseudo-spectrum we set negative values to zero, apply the same smoothing kernel, and extract $W^{(k)}$ following Supplement~B.1.
The quoted central value is taken as the FWHM extracted from the smoothed mean spectrum, and the $1\sigma$ statistical uncertainty is taken as the sample standard deviation of the valid ensemble,
\begin{equation}
  \delta W \approx
  \sqrt{\frac{1}{N_{\rm eff}-1}
        \sum_{k=1}^{N_{\rm eff}}
        \bigl(W^{(k)}-\overline{W}\bigr)^2},
\end{equation}
where $N_{\rm eff}$ is the number of pseudo-experiments that yield a well-defined FWHM and $\overline{W}$ is the ensemble mean.
This procedure is applied identically to the experimental spectra and to the IQMD+GEMINI spectra.
Possible additional systematics associated with, e.g., the smoothing choice or finite bin width are not included; these effects are expected to be subleading for the width trends discussed in Fig.~3 of the main text.

\clearpage

\section*{Supplement C. Semi-experimental momentum spectra and two-template fits}
\addcontentsline{toc}{section}{Appendix C. Semi-experimental momentum spectra and two-template fits}

\setcounter{figure}{0}
\renewcommand{\thetable}{C\arabic{table}}
\renewcommand{\thefigure}{C\arabic{figure}}

~\Cref{19C243,S3_19C57,S3_19C64,S3_29Ne240,S3_35Si73,S3_30Ne228,S3_9Li80,S3_18C80,S3_24O920,S3_24O92-3,S3_16C75,S3_16C62,S3_16C83,S3_34Mg242.5,S3_57Cr77,S3_7Li120,S3_34Si73,S3_46Ar73,S3_57Ni73,S3_34Ar70,S3_36Ca70,S3_22Mg74,S3_24Si85,28S80,S3_10C120,S3_14O53} present the semi-experimental residue parallel-momentum distributions for the one-neutron-removal reactions analyzed in this work.
The two-template analysis in this appendix is an experimentally anchored spectral-shape consistency check, not a unique event-by-event separation of reaction mechanisms.
The IQMD+GEMINI feeding/survival decomposition used in the main text is obtained event-by-event from the dynamical $A-1$ prefragment branch and the subsequent GEMINI decay path, and does not rely on the template fits described here.
When the published observable is a yield (counts) spectrum as a function of residue parallel momentum, we convert it into a semi-experimental differential cross section by a single overall normalization.
Specifically, the spectrum is rescaled such that its integral over the reported momentum range equals the published inclusive one-nucleon removal cross section $\sigma_i$, i.e.,
\begin{equation}
  \left.\frac{{\rm d}\sigma}{{\rm d}p}\right|_{p_k}
  = \mathcal{N}\,Y(p_k),
  \qquad
  \mathcal{N}
  = \frac{\sigma_i}{\sum_k Y(p_k)\,\Delta p_k},
\end{equation}
where $Y(p_k)$ is the published yield in bin $k$ and $\Delta p_k$ is the corresponding bin width.
This procedure preserves the published spectral shape while enforcing the correct inclusive normalization.

For each system, the symbols show the measured $\mathrm{d}\sigma/\mathrm{d}p$ with statistical uncertainties.
The solid curve shows the best-fit two-template description,
\begin{equation}
\label{eq:dp}
\frac{{\rm d}\sigma}{{\rm d}p}=
\sigma_i\Biggl[
\frac{(1-f^{\rm exp}_{\rm feed})}{\lambda_{\rm surv}}\,T_{\rm surv}\!\left(\dfrac{p-p_1}{\lambda_{\rm surv}}\right)
+
\frac{f^{\rm exp}_{\rm feed}}{\lambda_{\rm feed}}\,T_{\rm feed}\!\left(\dfrac{p-p_2}{\lambda_{\rm feed}}\right)
\Biggr],
\end{equation}
where $\sigma_i=\int (\mathrm{d}\sigma/\mathrm{d}p)\,\mathrm{d}p$ is fixed by the spectrum integral and $0\le f^{\rm exp}_{\rm feed}\le 1$ defines the semi-experimental evaporation-feeding fraction (with $f_{\rm surv}=1-f^{\rm exp}_{\rm feed}$).
In the fit we parametrize $f_{\rm feed}=\sin^2\theta$ and $f_{\rm surv}=\cos^2\theta$ to enforce $f_{\rm feed}+f_{\rm surv}=1$.

The parameters $p_1$ and $p_2$ allow independent centroid shifts of the survival-like and feeding-like components, respectively, and absorb residual alignment differences between the measured spectrum and the fixed templates.
Equivalently, one may rewrite these shifts in terms of a common centroid $p_0$ and a relative displacement $\Delta p$ between the two components,
\begin{equation}
p_1=p_0+\Delta p, \qquad p_2=p_0,
\end{equation}
so that the survival-like component is shifted by an additional $\Delta p$ relative to the feeding-like component.
In our implementation we enforce $\Delta p>0$ by parametrizing $\Delta p=\exp(\phi)$.

The scale parameters $\lambda_{\rm surv}$ and $\lambda_{\rm feed}$ allow limited template-shape rescaling while preserving unit area through the explicit $1/\lambda$ factors.
In the fits we impose asymmetric bounds, $\lambda_{\rm surv}\in[0.7,1.0]$ and $\lambda_{\rm feed}\in[1.0,1.7]$, such that the survival-like component is always at least as narrow as the feeding-like component (i.e., it cannot be broader than the feeding component), consistent with the observed width hierarchy across the dataset.

The dashed and dotted curves indicate the corresponding survival-like and evaporation-like contributions,
$\sigma_i(1-f^{\rm exp}_{\rm feed})\,T_{\rm surv}((p-p_1)/\lambda_{\rm surv})/\lambda_{\rm surv}$ and
$\sigma_i f^{\rm exp}_{\rm feed}\,T_{\rm feed}((p-p_2)/\lambda_{\rm feed})/\lambda_{\rm feed}$, respectively.
For clarity, only the best-fit curves are displayed; the extracted $f^{\rm exp}_{\rm feed}$ values and their uncertainties are summarized in Fig.~3(a) of the main text.

The fixed templates $T_{\rm feed}$ and $T_{\rm surv}$ are constructed once from the benchmark spectra of the ${}^{20}$C and ${}^{32}$Ar systems, respectively, which correspond to the minimum and maximum $\Delta S$ values in our dataset.
Each benchmark spectrum is smoothed using an unweighted Gaussian-kernel procedure with a bandwidth set by the median spacing of the reported momentum points, then shifted such that its maximum lies at $p=0$ and normalized to unit area.
The resulting template shapes are held fixed for all systems.

For each reaction, the spectrum is described by the two-template form in Eq.~(\ref{eq:dp}) with fit parameters $(f_{\rm feed},p_0,\Delta p,\lambda_{\rm feed},\lambda_{\rm surv})$ (with $f_{\rm surv}=1-f_{\rm feed}$ and $\Delta p>0$).
The two benchmark reactions are included among the panels for completeness and are treated with the same fitting procedure as the other systems.

Uncertainties on the semi-experimental evaporation-feeding fractions $f^{\exp}_{\rm feed}$ shown in Fig.~3(a) of the main text are obtained by Monte Carlo resampling.
For each reaction, pseudo-data sets are generated by fluctuating the measured spectrum point-by-point within its statistical uncertainties, assuming independent Gaussian variations for each momentum bin.
Each pseudo-spectrum is then refitted with the fixed templates using the same two-template procedure and parameter bounds as for the central fit, with the inclusive normalization $\sigma_i$ recalculated for each pseudo-spectrum from its integral over the reported momentum range.
The quoted $1\sigma$ uncertainty on $f_{\rm feed}$ is taken from the central $68\%$ interval of the Monte Carlo distribution, $\delta f_{\rm feed}=\tfrac{1}{2}(f_{\rm feed}^{84\%}-f_{\rm feed}^{16\%})$.
These panels therefore illustrate the quality of the two-template description and the evolution of the evaporation-like and survival-like components across the dataset.

\begin{figure}[H]
\centering
    \includegraphics[width=0.85\textwidth,angle=0]{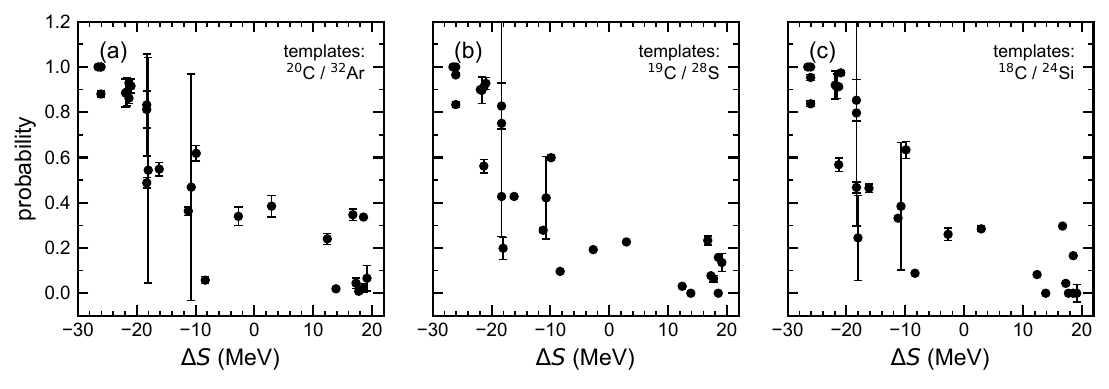}
\caption{\label{fig:template_sets_feed}
Evaporation-feeding fraction $f^{\rm exp}_{\rm feed}$ extracted from residue parallel-momentum spectra as a function of the asymmetry $\Delta S$ using different template pairs: (a) ${}^{20}$C/${}^{32}$Ar, (b) ${}^{19}$C/${}^{28}$S, and (c) ${}^{18}$C/${}^{24}$Si.
The comparison shows that the extracted feeding trend is stable against the tested template choices within the uncertainties of the two-template ansatz.}
\end{figure}

\begin{figure}[H]   
\centering
    \includegraphics[width=0.45\textwidth,angle=0]{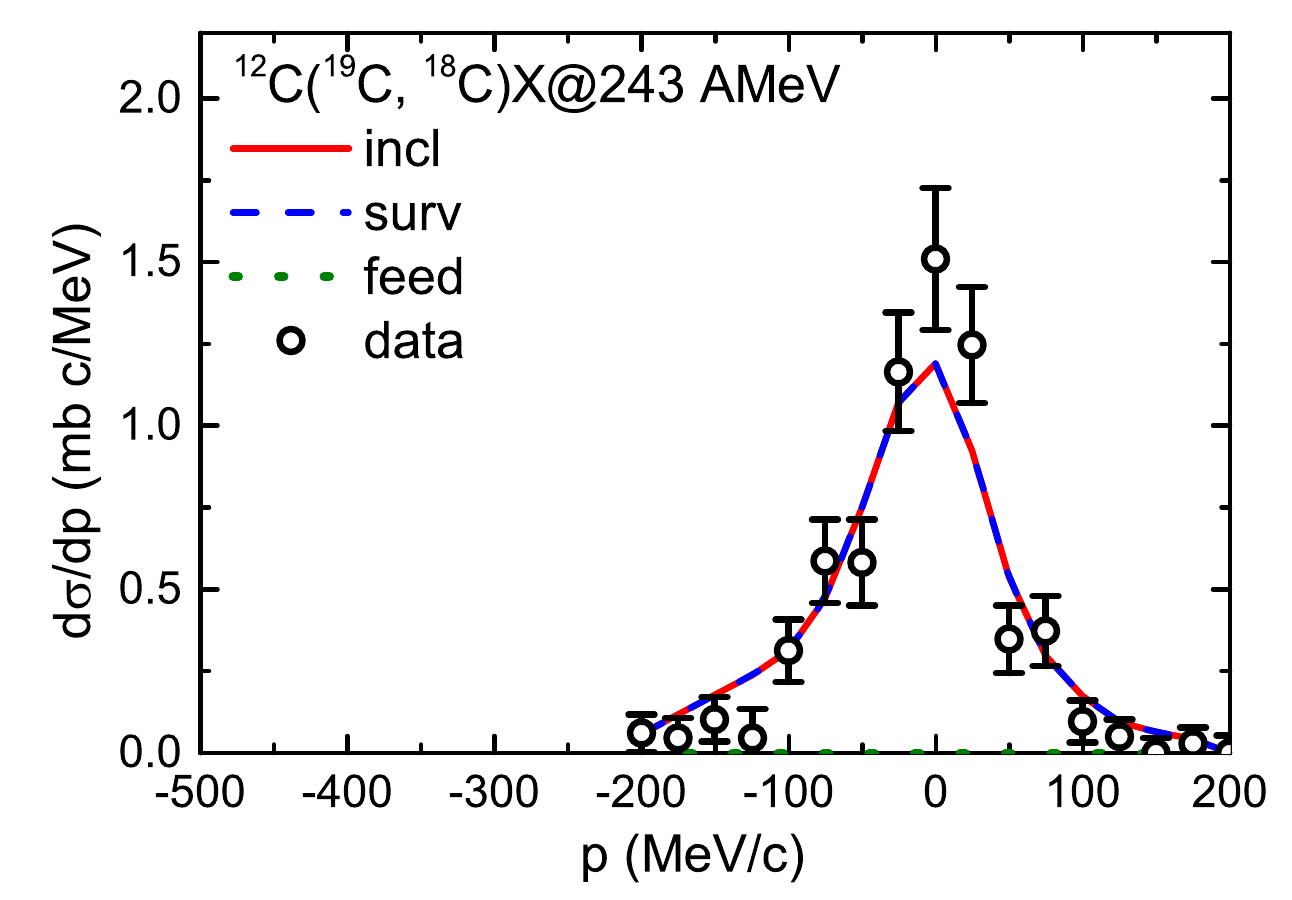}
\caption{Residue parallel-momentum distribution for one-neutron removal from $^{19}$C+$^{12}$C at 243~AMeV~\cite{kobayashi2012one}.
Circles show the published spectrum, renormalized to a semi-experimental differential cross section whose integral matches the inclusive removal cross section.
The solid curve is the best-fit two-template description.
The dashed and dotted curves show the decomposed survival-like and evaporation-feeding components, respectively.}

    \label{19C243}
\end{figure}

\begin{figure}[H]   
\centering
    \includegraphics[width=0.45\textwidth,angle=0]{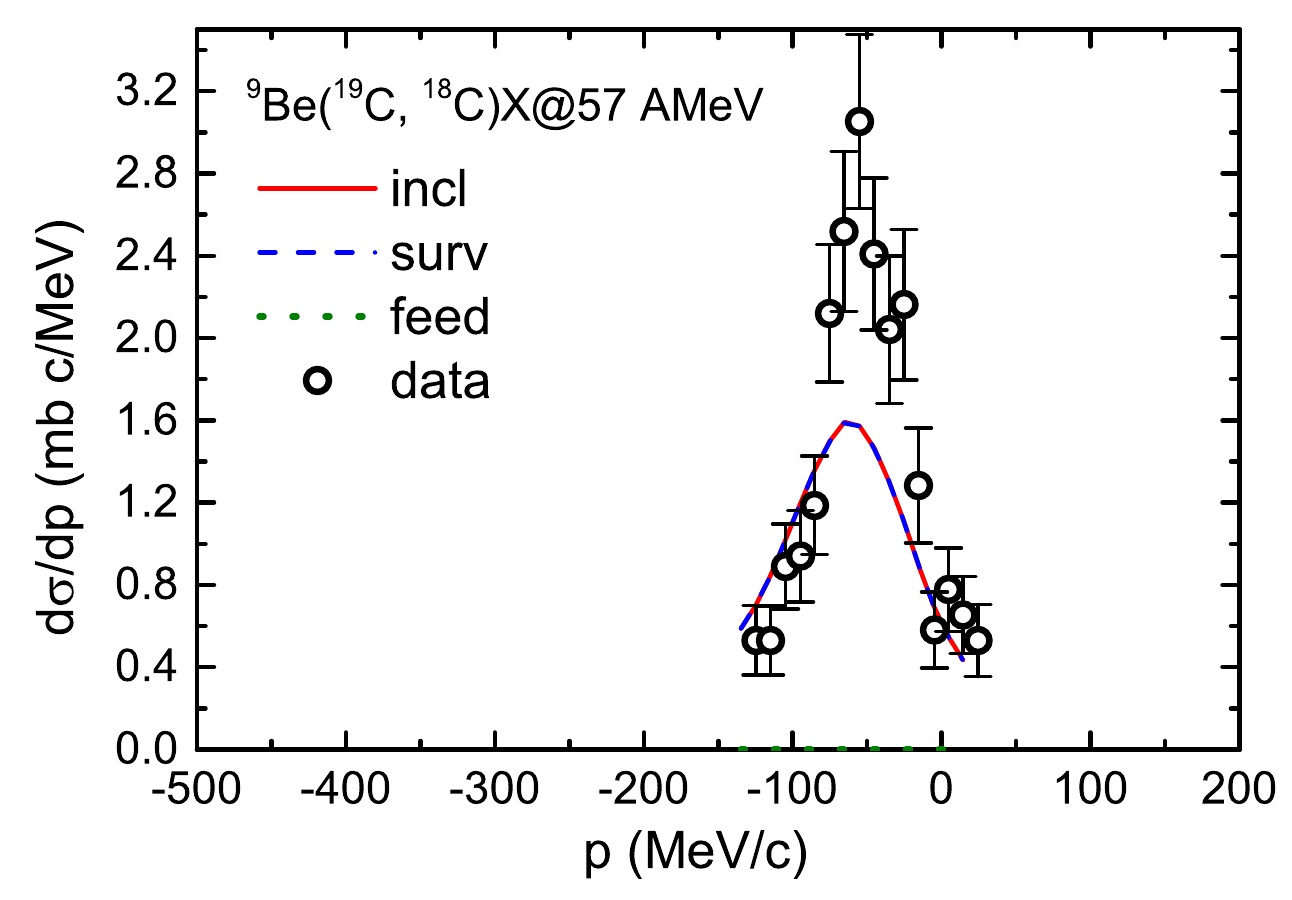}
\caption{Same as Fig. \ref{19C243} but for $^{19}$C + $^{9}$Be collisions at 57 AMeV \cite{maddalena2001single}. }
    \label{S3_19C57}
\end{figure}

\begin{figure}[H]   
\centering
    \includegraphics[width=0.45\textwidth,angle=0]{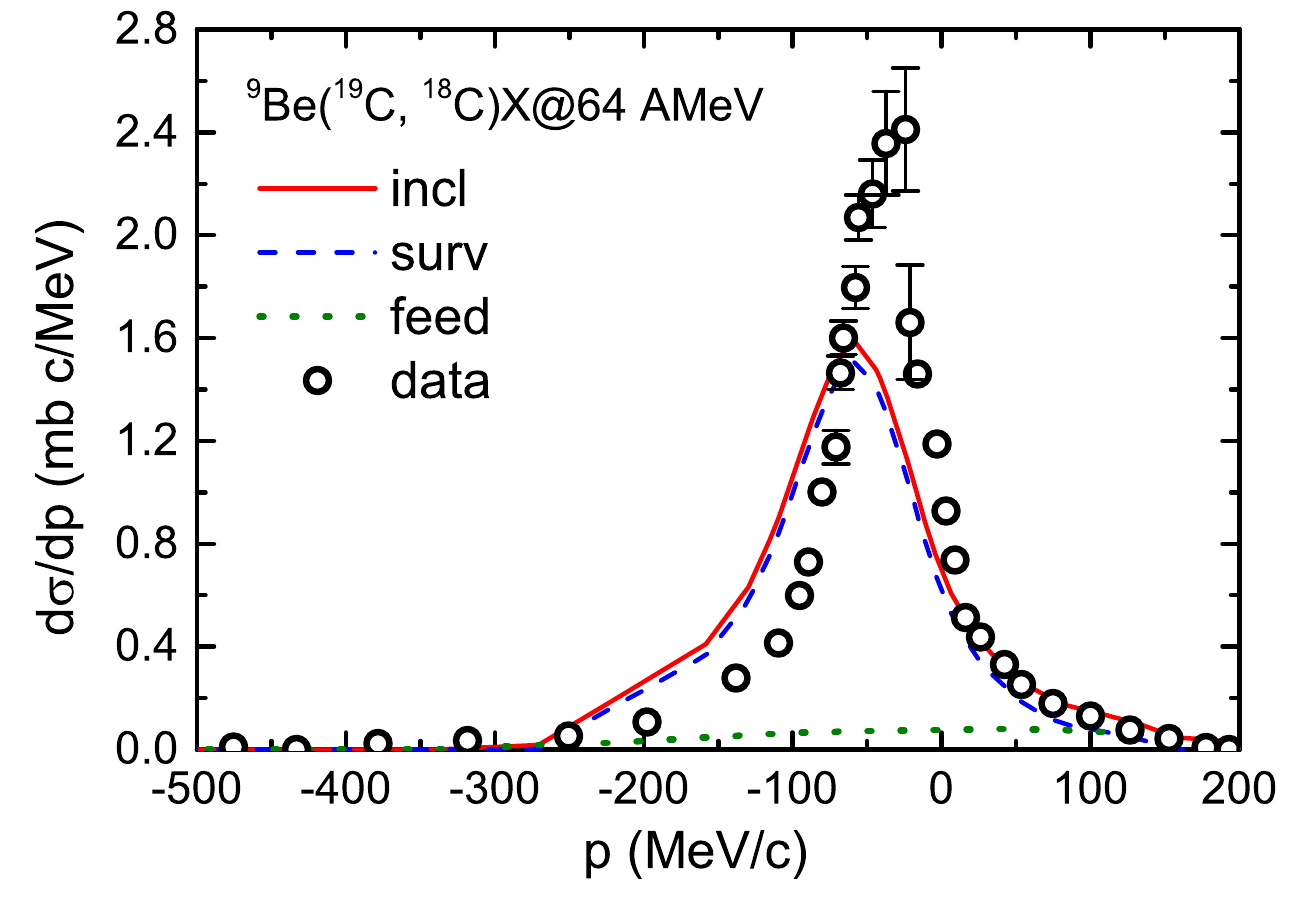}
\caption{Same as Fig. \ref{19C243} but for $^{19}$C + $^{9}$Be collisions at 64 AMeV \cite{simpson2009one}. }
    \label{S3_19C64}
\end{figure}

\begin{figure}[H]   
\centering
    \includegraphics[width=0.45\textwidth,angle=0]{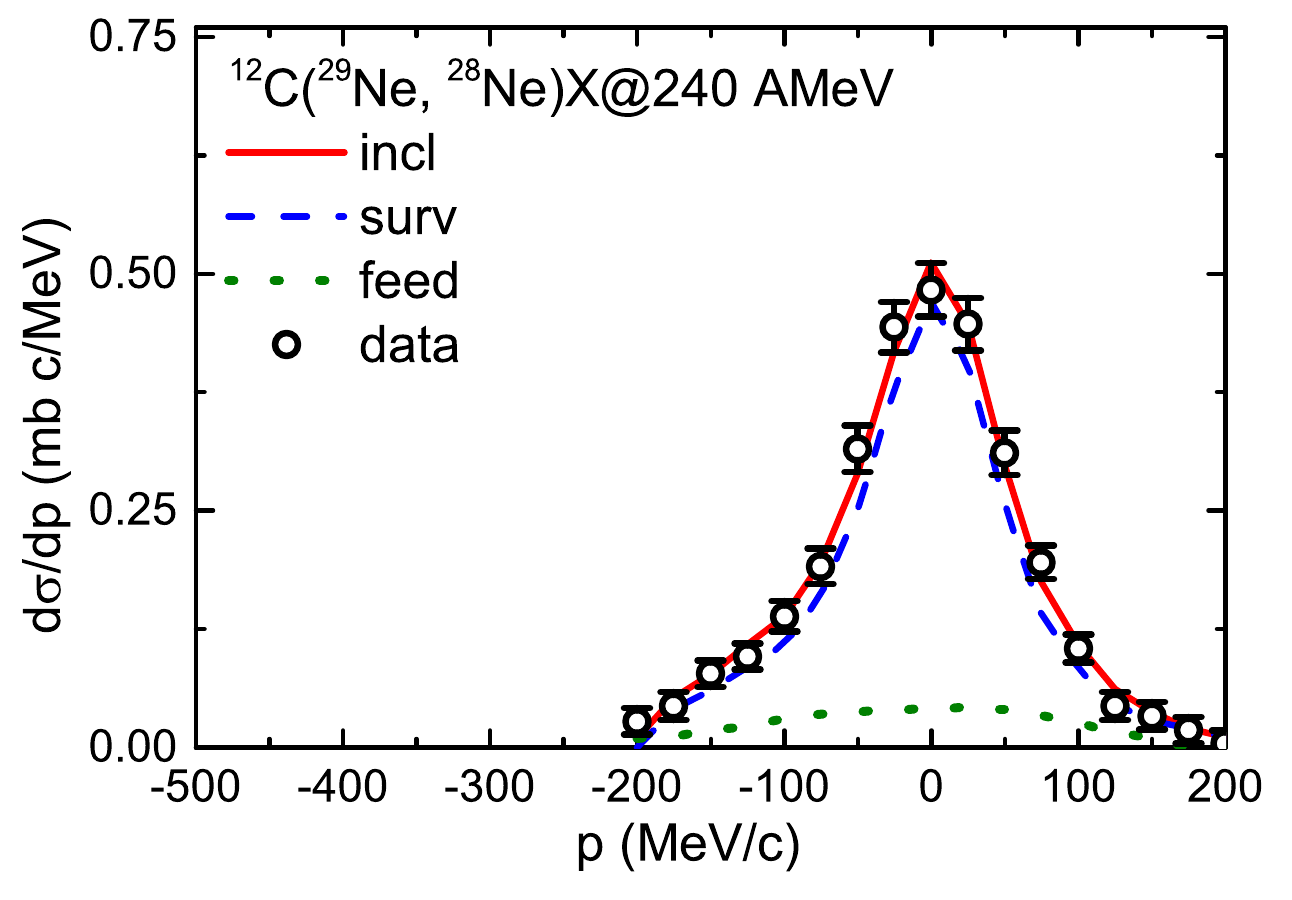}
\caption{Same as Fig. \ref{19C243} but for $^{29}$Ne + $^{12}$C collisions at 240 AMeV \cite{kobayashi2016one}.}
    \label{S3_29Ne240}
\end{figure}

\begin{figure}[H]   
\centering
    \includegraphics[width=0.45\textwidth,angle=0]{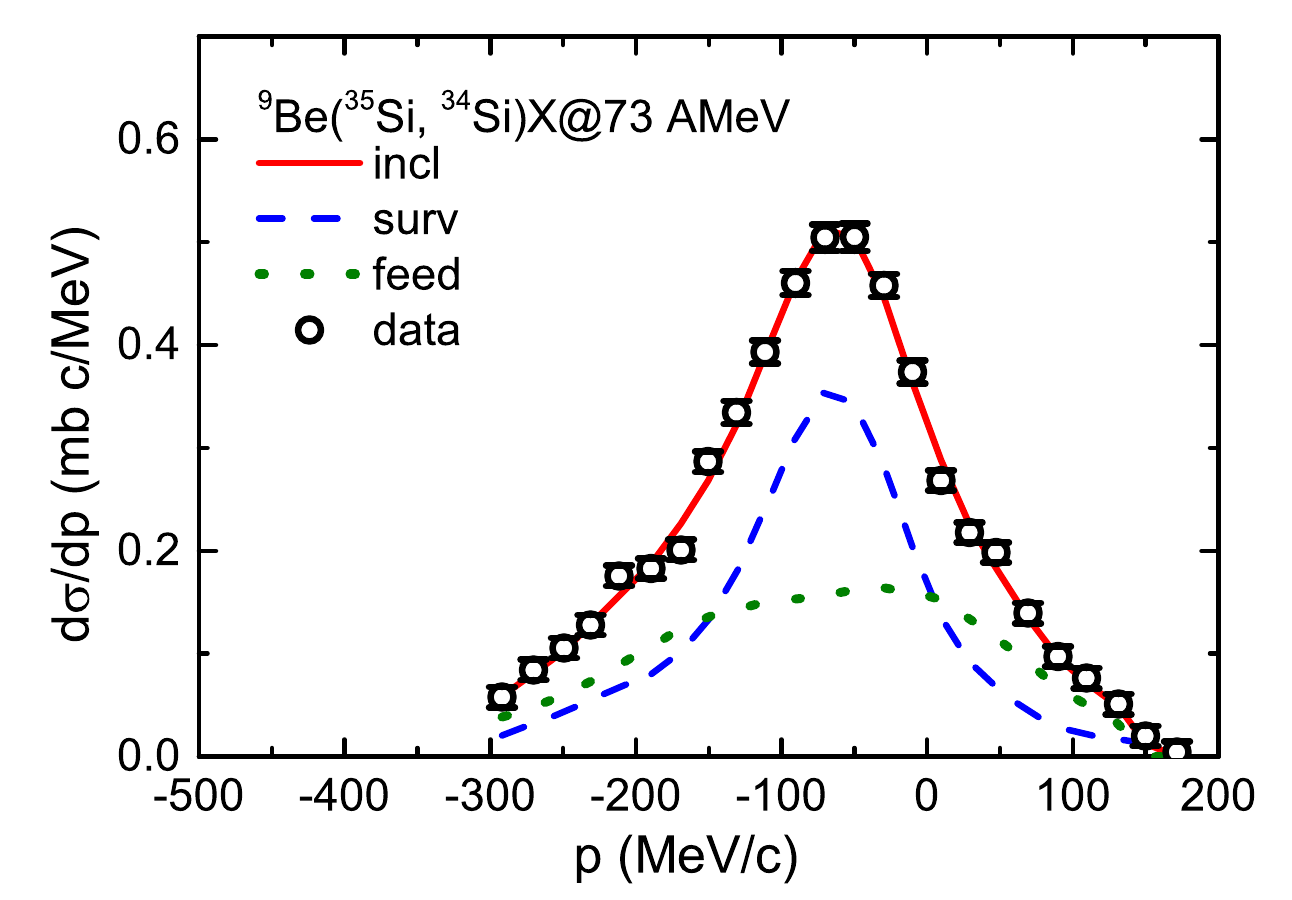}
\caption{Same as Fig. \ref{19C243} but for $^{35}$Si + $^{9}$Be collisions at 73 AMeV \cite{enders2002single}.}
    \label{S3_35Si73}
\end{figure}

\begin{figure}[H]   
\centering
    \includegraphics[width=0.45\textwidth,angle=0]{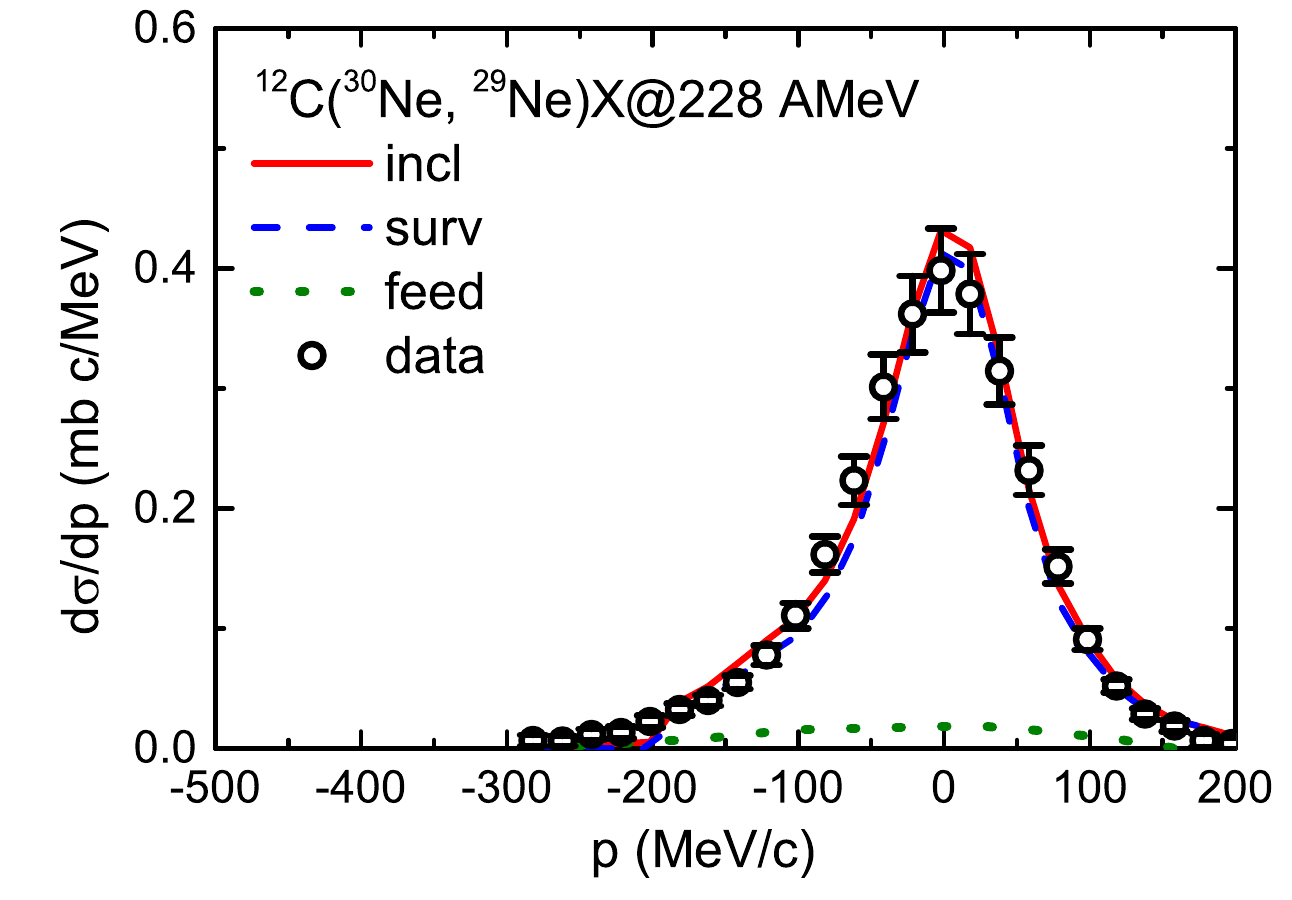}
\caption{Same as Fig. \ref{19C243} but for $^{30}$Ne + $^{12}$C collisions at 228 AMeV \cite{liu2017intruder}.}
    \label{S3_30Ne228}
\end{figure}

\begin{figure}[H]   
\centering
    \includegraphics[width=0.45\textwidth,angle=0]{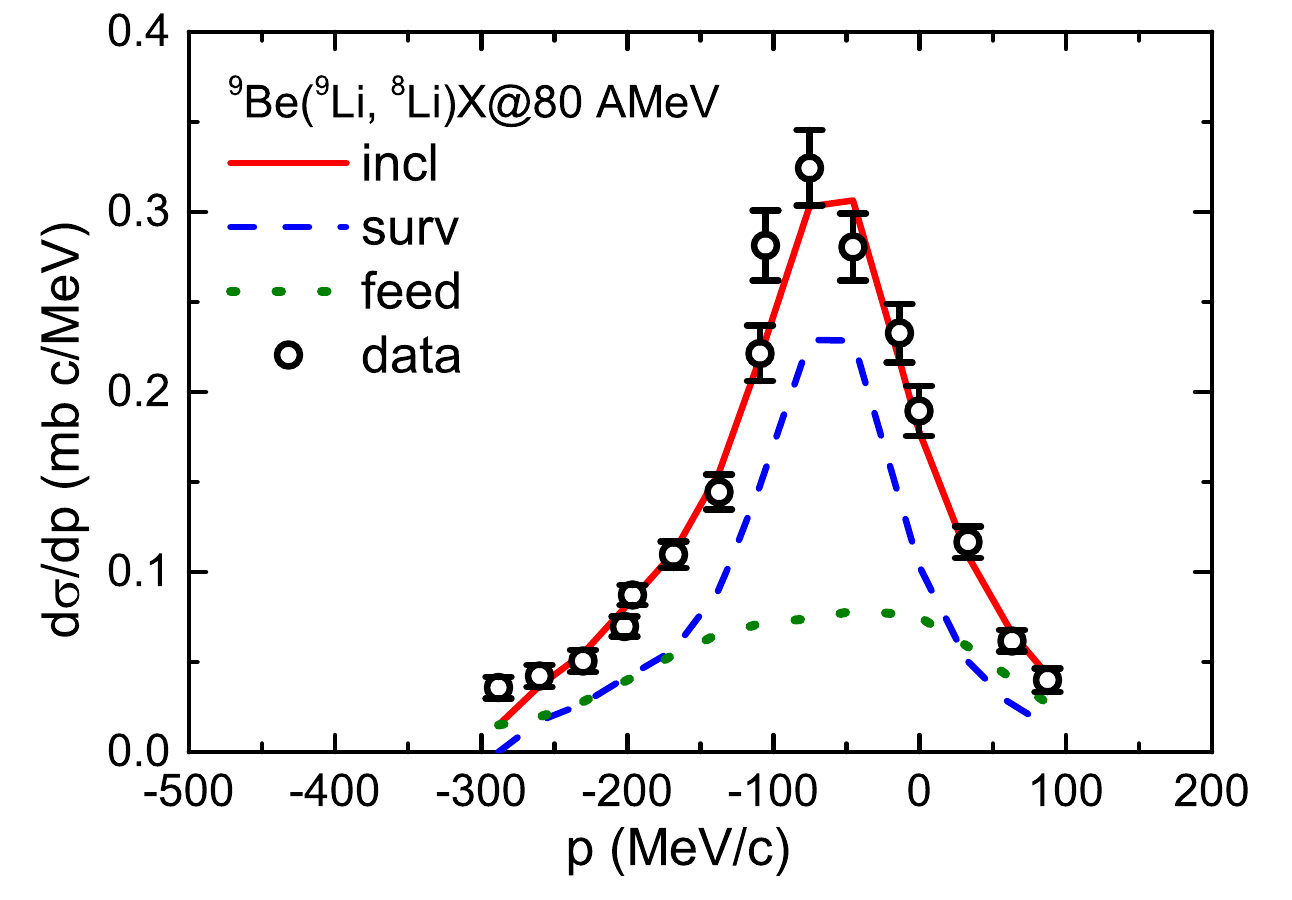}
\caption{Same as Fig. \ref{19C243} but for $^{9}$Li + $^{9}$Be collisions at 80 AMeV \cite{grinyer2012systematic}.}
    \label{S3_9Li80}
\end{figure}

\begin{figure}[H]   
\centering
    \includegraphics[width=0.45\textwidth,angle=0]{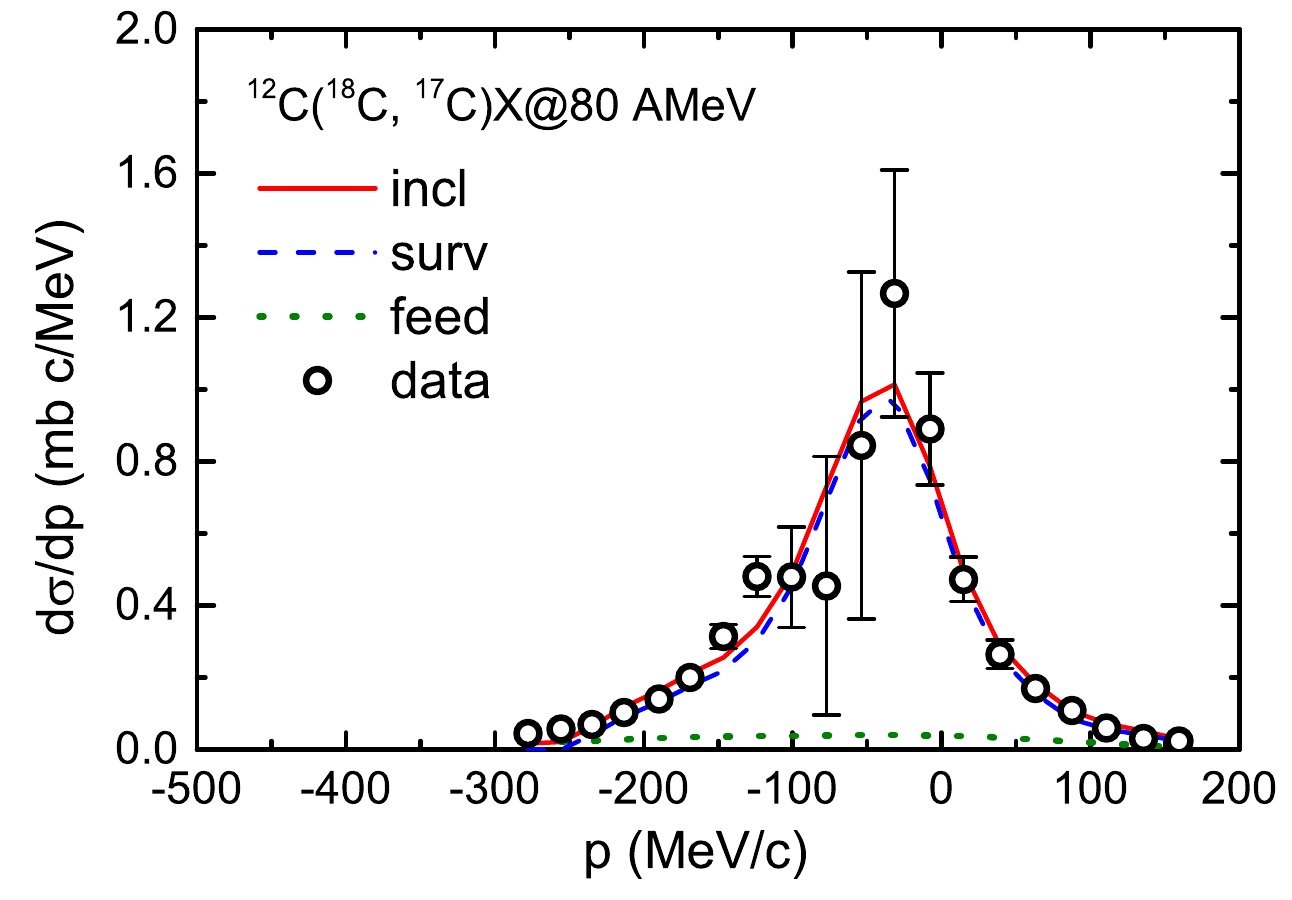}
\caption{Same as Fig. \ref{19C243} but for $^{18}$C + $^{12}$C collisions at 80 AMeV \cite{ozawa2008measurement}.}
    \label{S3_18C80}
\end{figure}

\begin{figure}[H]   
\centering
    \includegraphics[width=0.45\textwidth,angle=0]{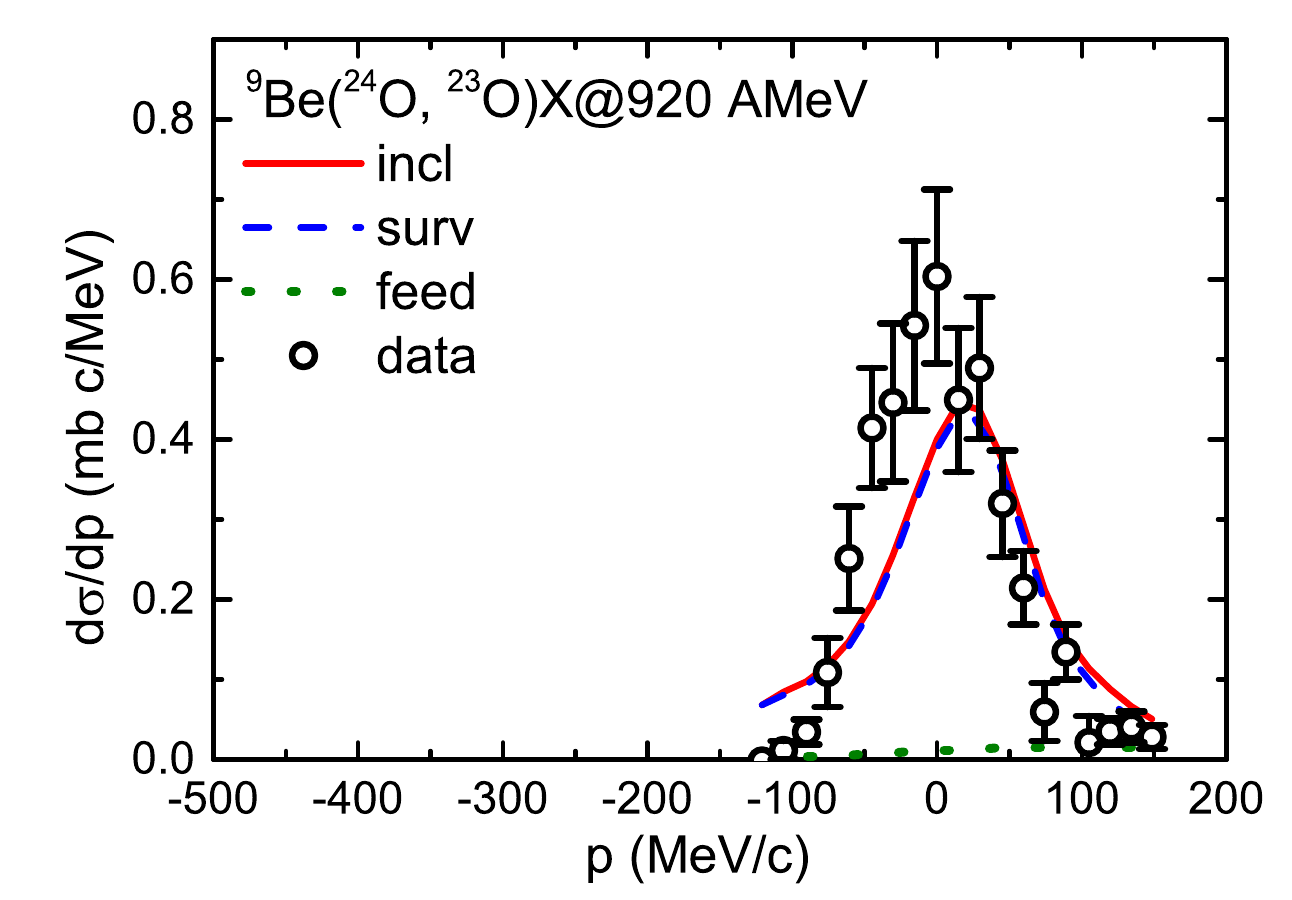}
\caption{Same as Fig. \ref{19C243} but for $^{24}$O + $^{12}$C collisions at 920 AMeV \cite{kanungo2009one}.}
    \label{S3_24O920}
\end{figure}

\begin{figure}[H]   
\centering
    \includegraphics[width=0.45\textwidth,angle=0]{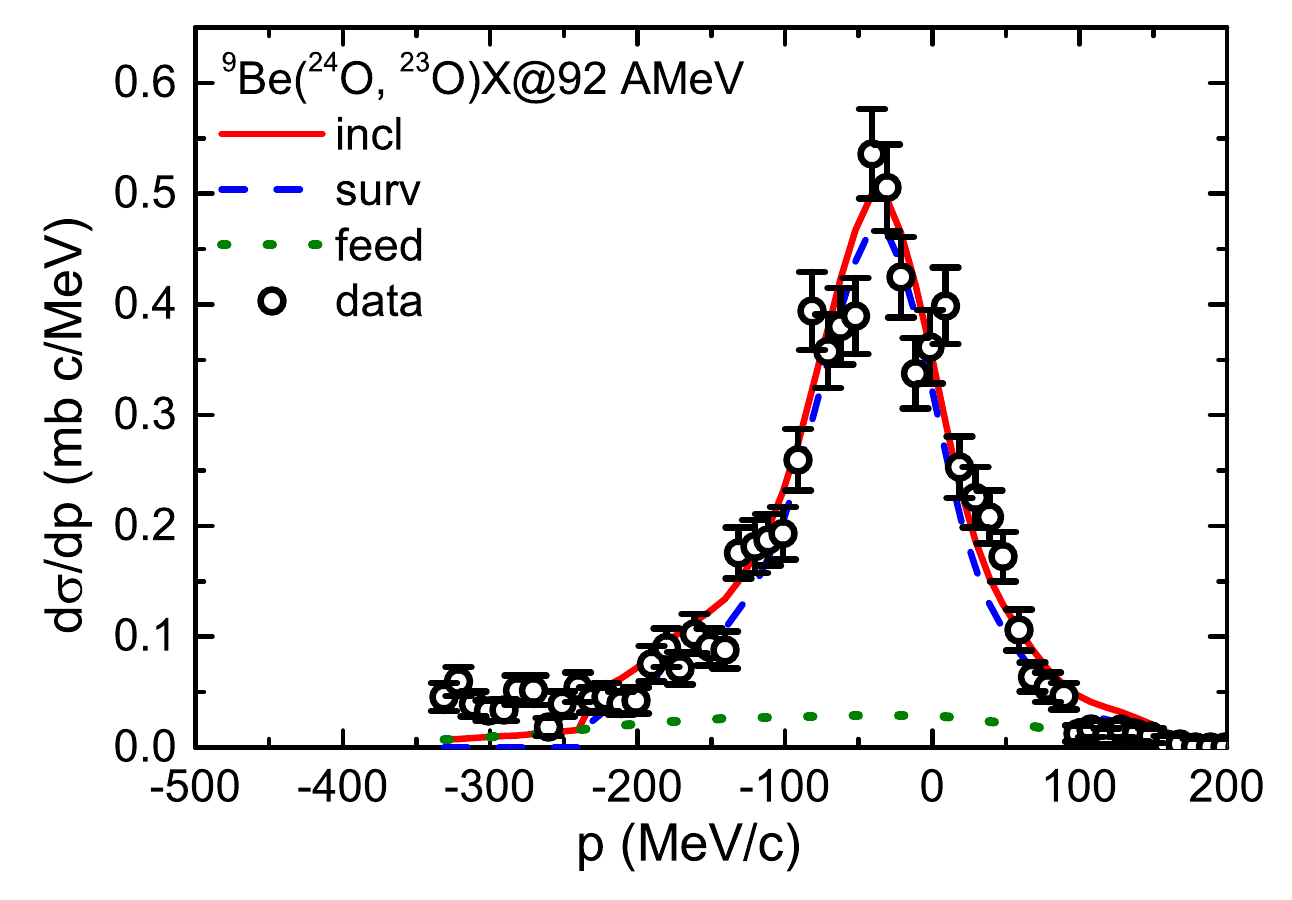}
\caption{Same as Fig. \ref{19C243} but for $^{24}$O + $^{9}$Be collisions at 92 AMeV \cite{divaratne2018one}.}
    \label{S3_24O92-3}
\end{figure}

\begin{figure}[H]   
\centering
    \includegraphics[width=0.45\textwidth,angle=0]{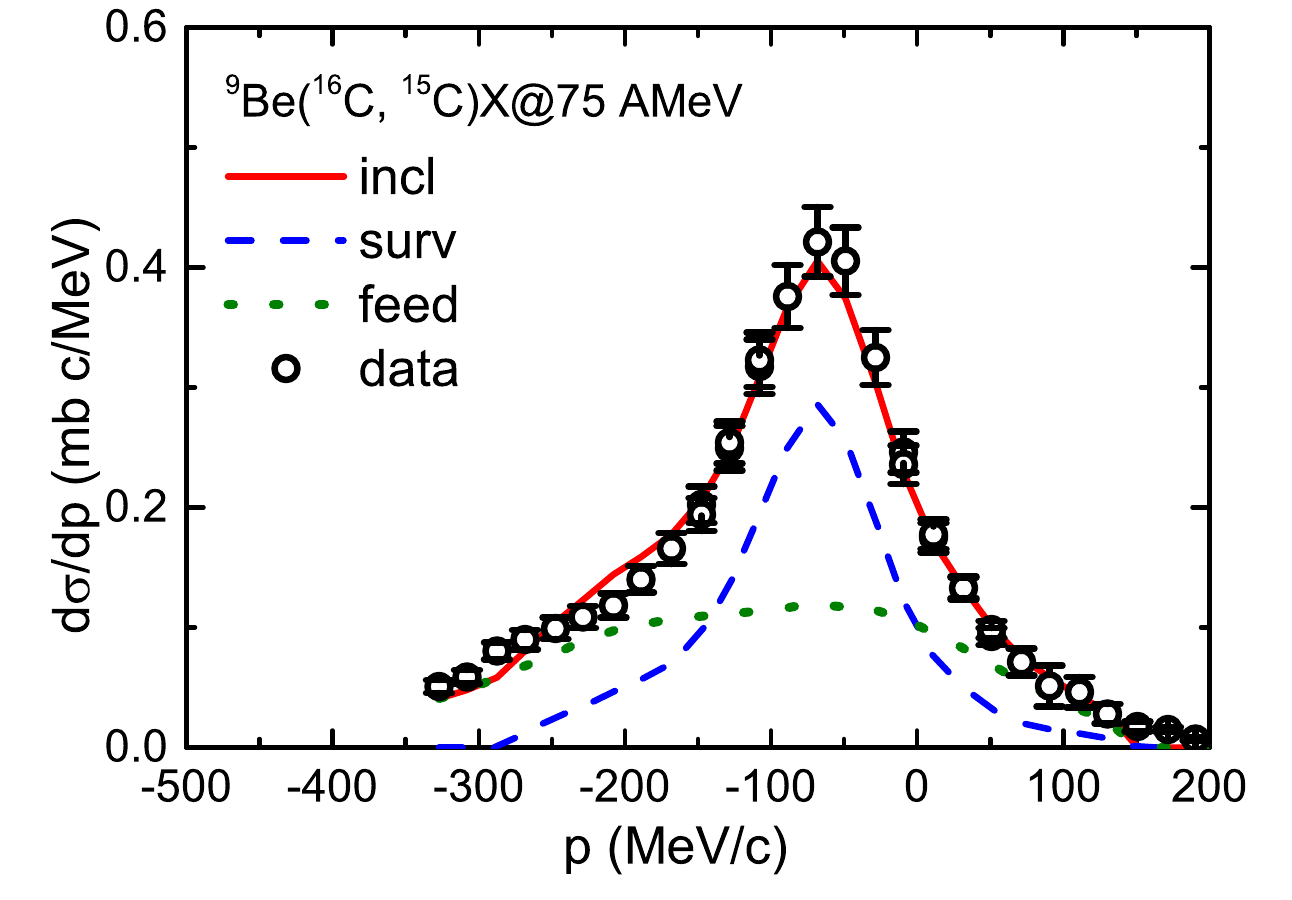}
\caption{Same as Fig. \ref{19C243} but for $^{16}$C + $^{9}$Be collisions at 75 AMeV \cite{flavigny2012nonsudden}.}
    \label{S3_16C75}
\end{figure}

\begin{figure}[H]   
\centering
    \includegraphics[width=0.45\textwidth,angle=0]{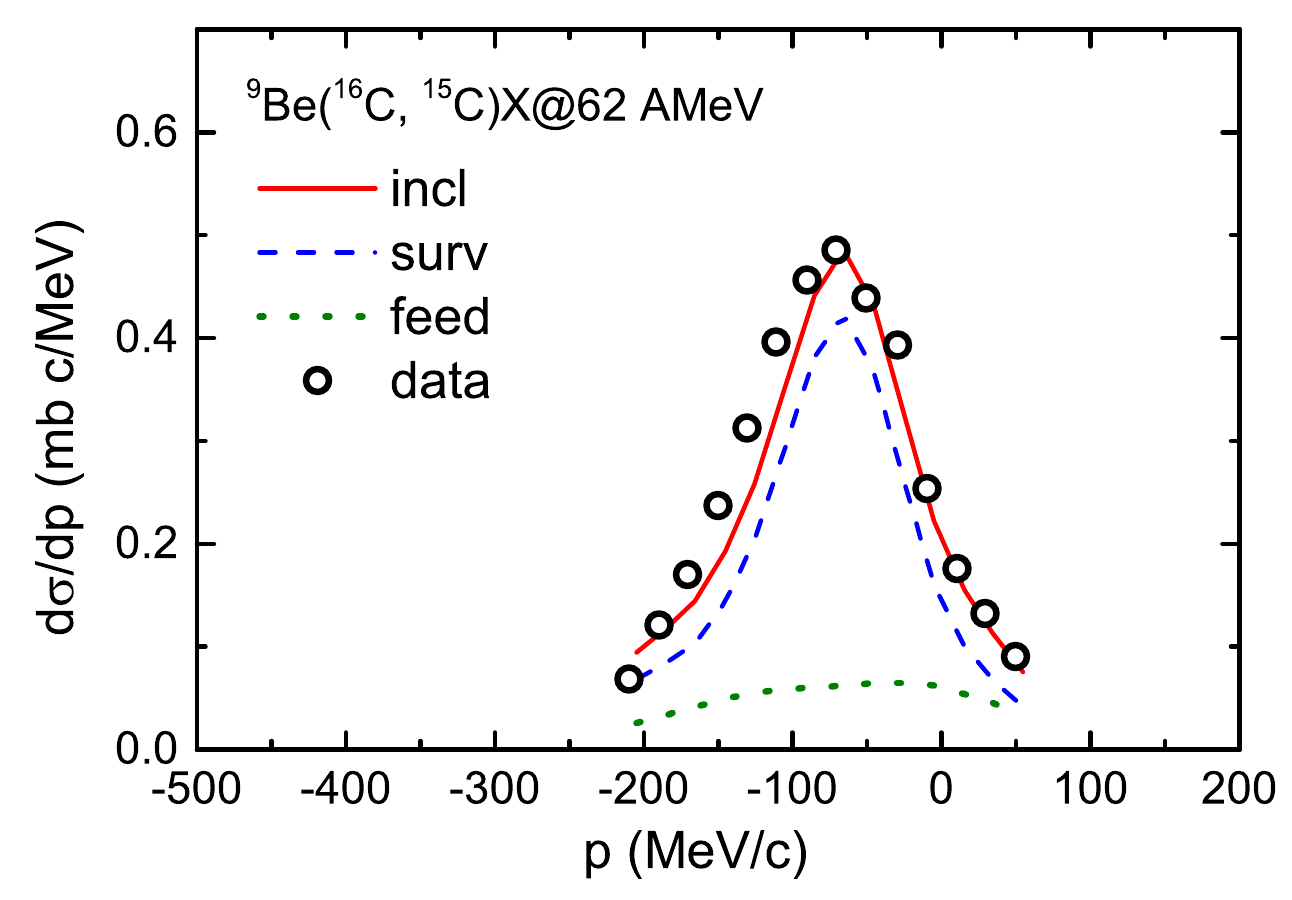}
\caption{Same as Fig. \ref{19C243} but for  $^{16}$C + $^{9}$Be collisions at 62 AMeV \cite{maddalena2001single}.}
    \label{S3_16C62}
\end{figure}

\begin{figure}[H]   
\centering
    \includegraphics[width=0.45\textwidth,angle=0]{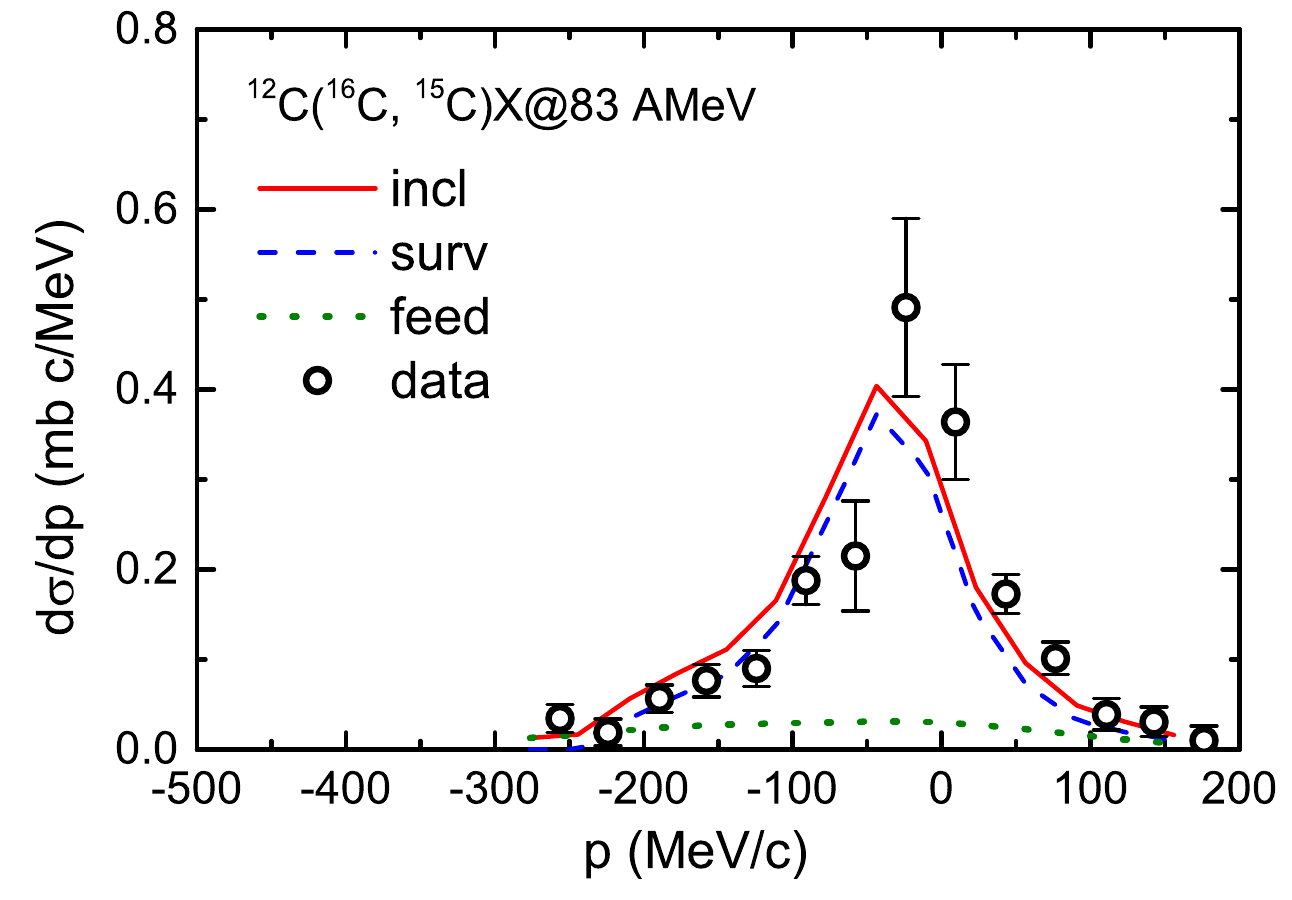}
\caption{Same as Fig. \ref{19C243} but for  $^{16}$C + $^{12}$C collisions at 83 AMeV \cite{yamaguchi2004neutron}.}
    \label{S3_16C83}
\end{figure}

\begin{figure}[H]   
\centering
    \includegraphics[width=0.45\textwidth,angle=0]{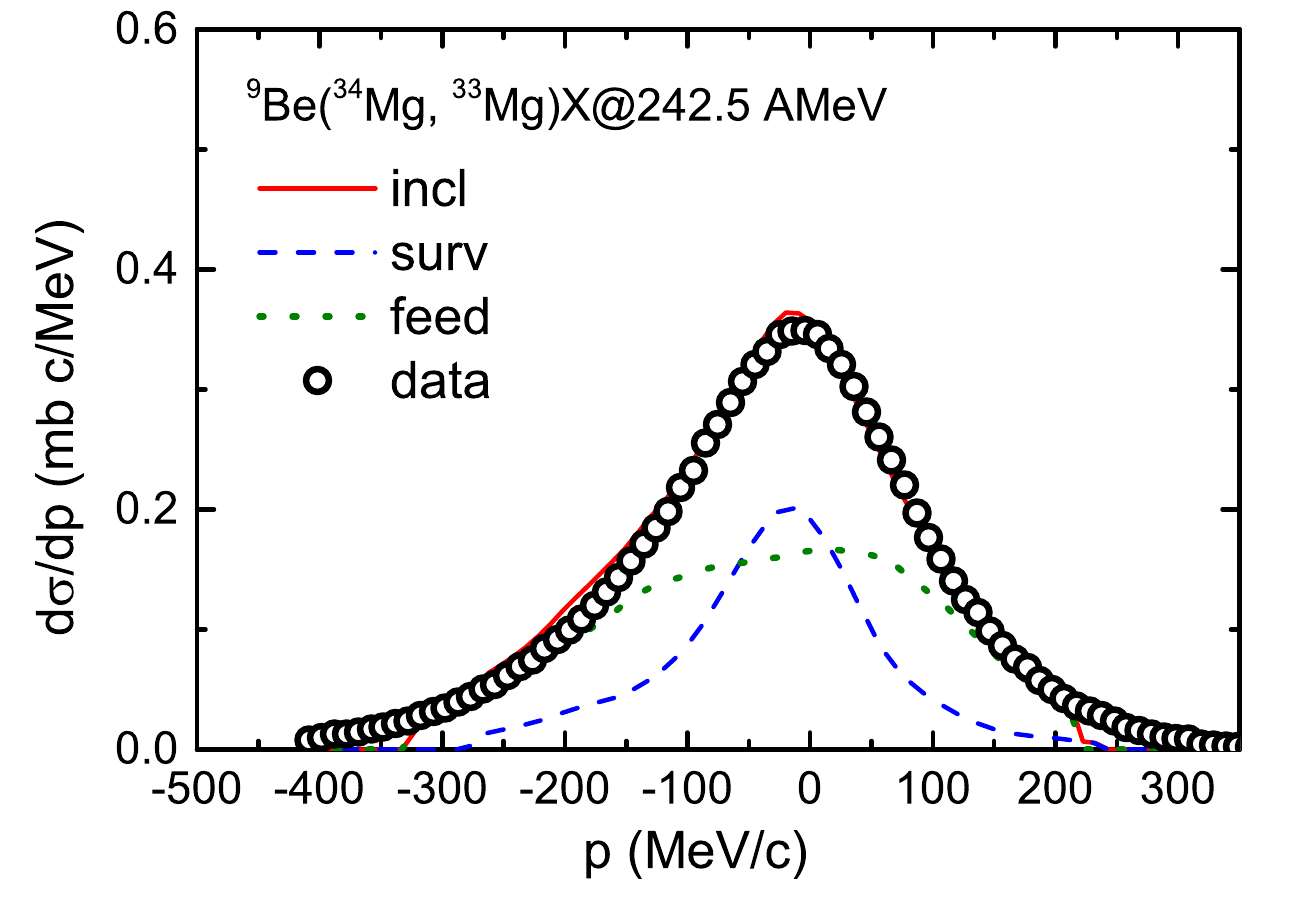}
\caption{Same as Fig. \ref{19C243} but for  $^{34}$Mg + $^{9}$Be collisions at 242.5 AMeV \cite{bazin2021spectroscopy}.}
    \label{S3_34Mg242.5}
\end{figure}

\begin{figure}[H]   
\centering
    \includegraphics[width=0.45\textwidth,angle=0]{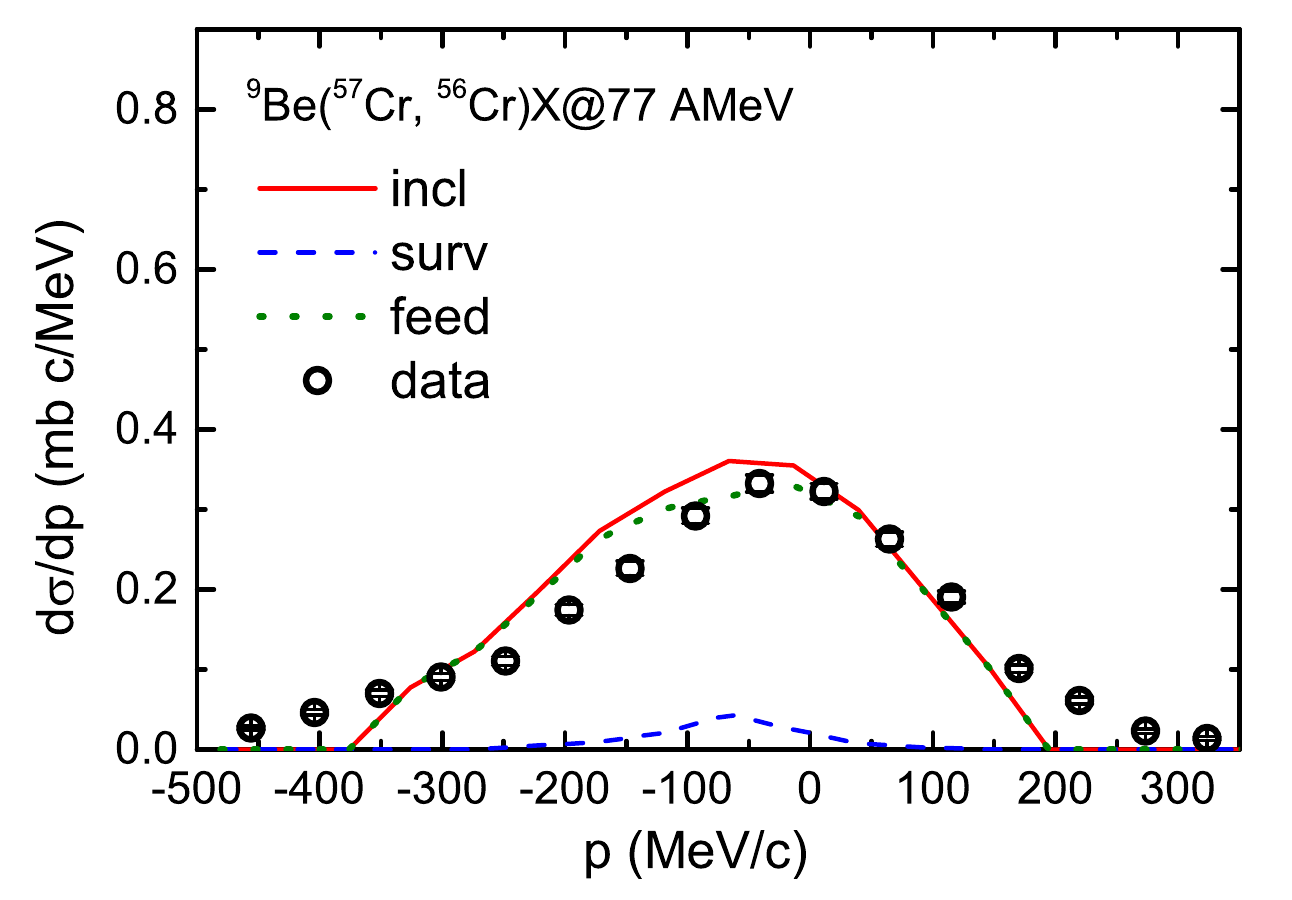}
\caption{Same as Fig. \ref{19C243} but for  $^{57}$Cr + $^{9}$Be collisions at 77 AMeV \cite{gade2006one}.}
    \label{S3_57Cr77}
\end{figure}

\begin{figure}[H]   
\centering
    \includegraphics[width=0.45\textwidth,angle=0]{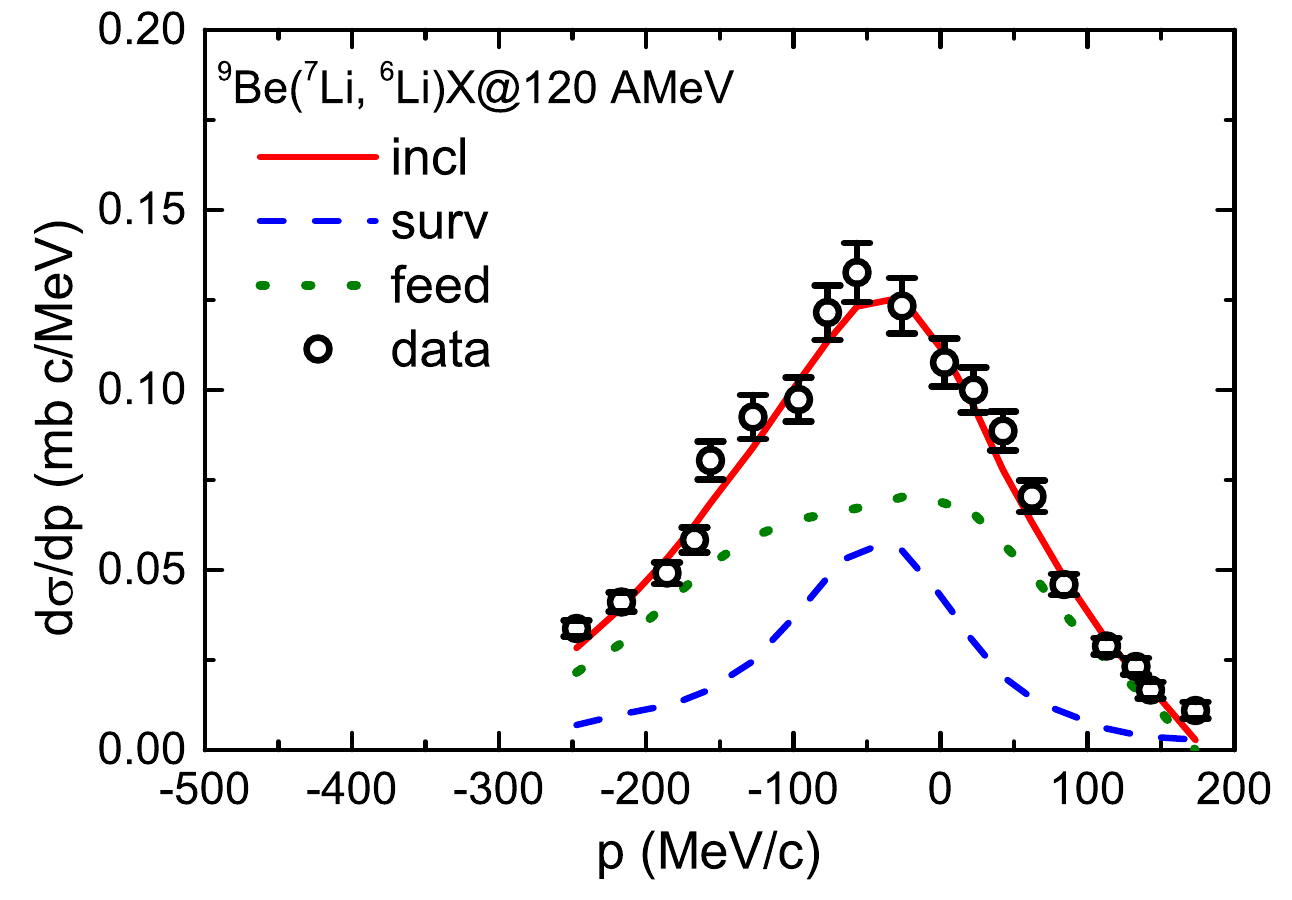}
\caption{Same as Fig. \ref{19C243} but for $^{7}$Li + $^{9}$Be collisions at 120 AMeV \cite{grinyer2012systematic}.}
    \label{S3_7Li120}
\end{figure}

\begin{figure}[H]   
\centering
    \includegraphics[width=0.45\textwidth,angle=0]{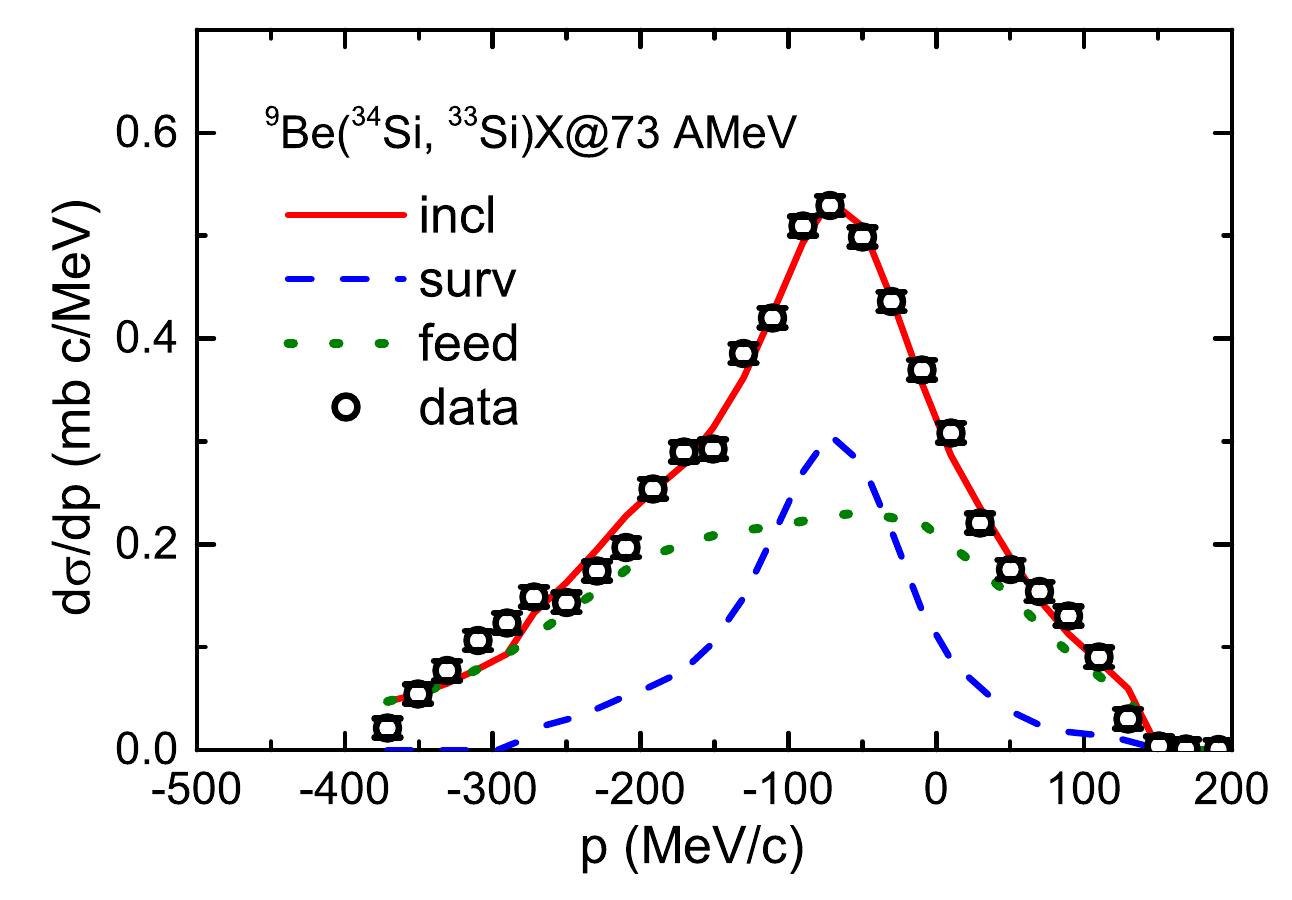}
\caption{Same as Fig. \ref{19C243} but for $^{34}$Si + $^{9}$Be collisions at 73 AMeV \cite{enders2002single}.}
    \label{S3_34Si73}
\end{figure}

\begin{figure}[H]   
\centering
    \includegraphics[width=0.45\textwidth,angle=0]{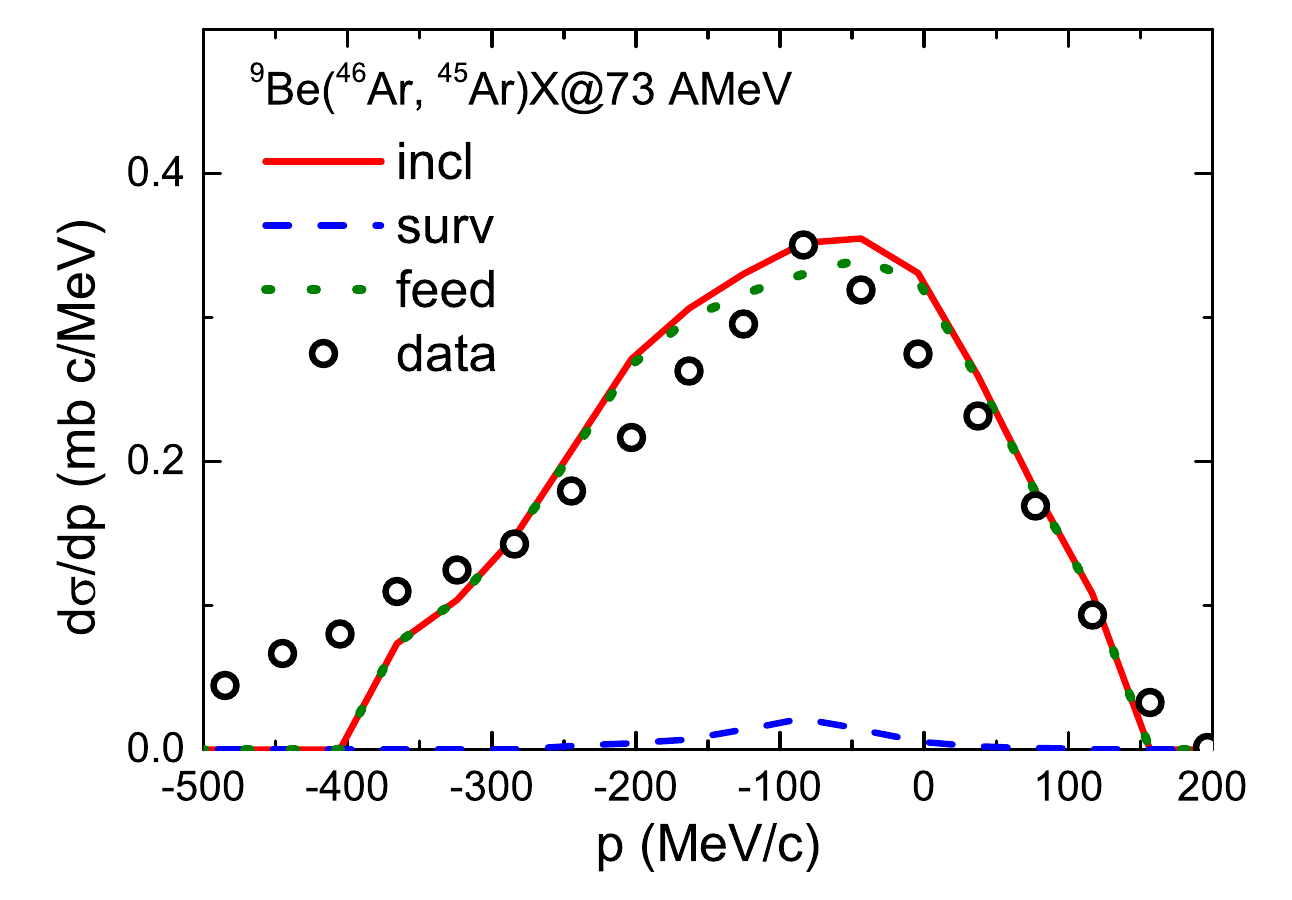}
\caption{Same as Fig. \ref{19C243} but for $^{46}$Ar + $^{9}$Be collisions at 73 AMeV \cite{gade2005knockout}.}
    \label{S3_46Ar73}
\end{figure}

\begin{figure}[H]   
\centering
    \includegraphics[width=0.45\textwidth,angle=0]{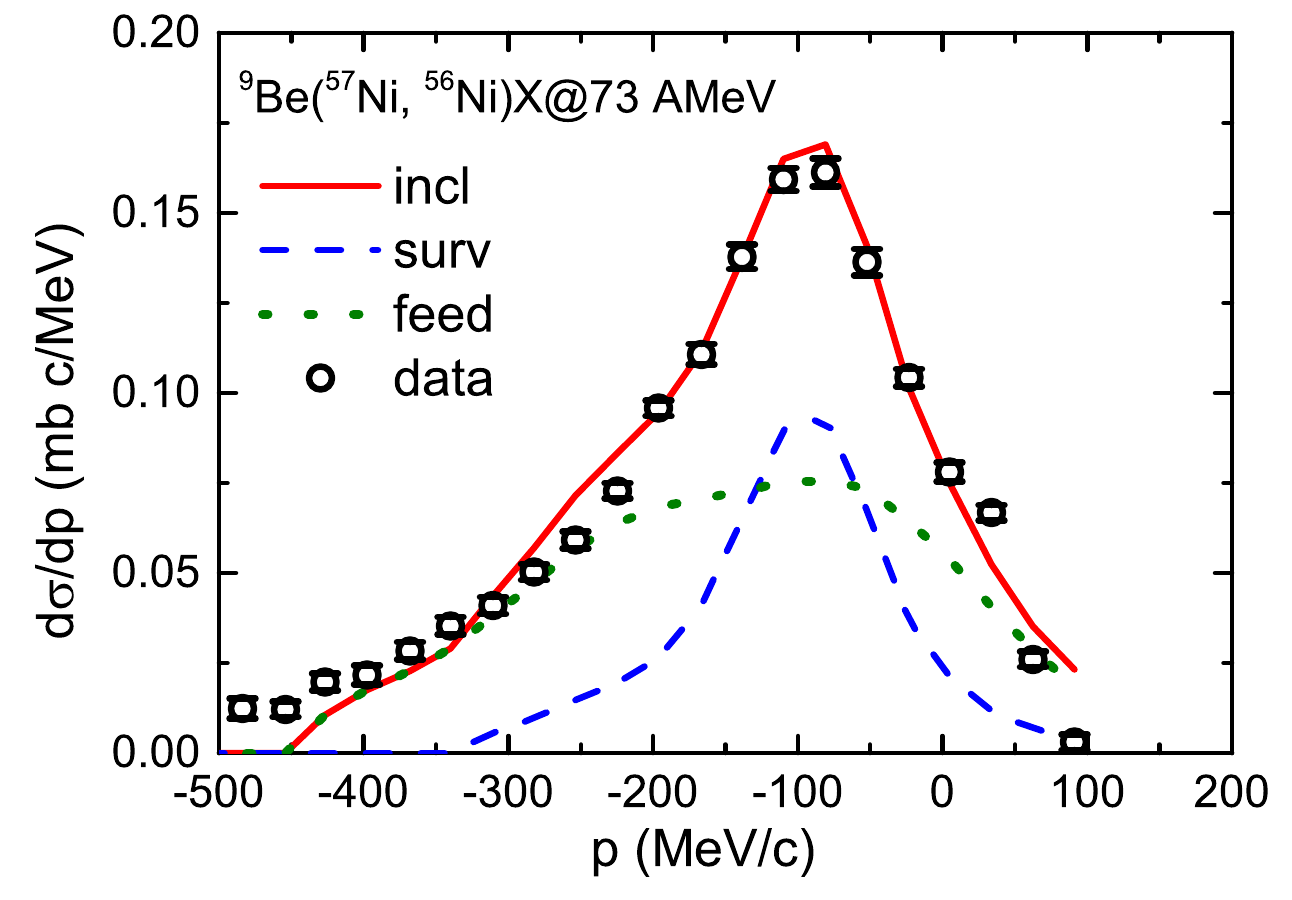}
\caption{Same as Fig. \ref{19C243} but for $^{57}$Ni + $^{9}$Be collisions at 73 AMeV \cite{yurkewicz2006one}.}
    \label{S3_57Ni73}
\end{figure}

\begin{figure}[H]   
\centering
    \includegraphics[width=0.45\textwidth,angle=0]{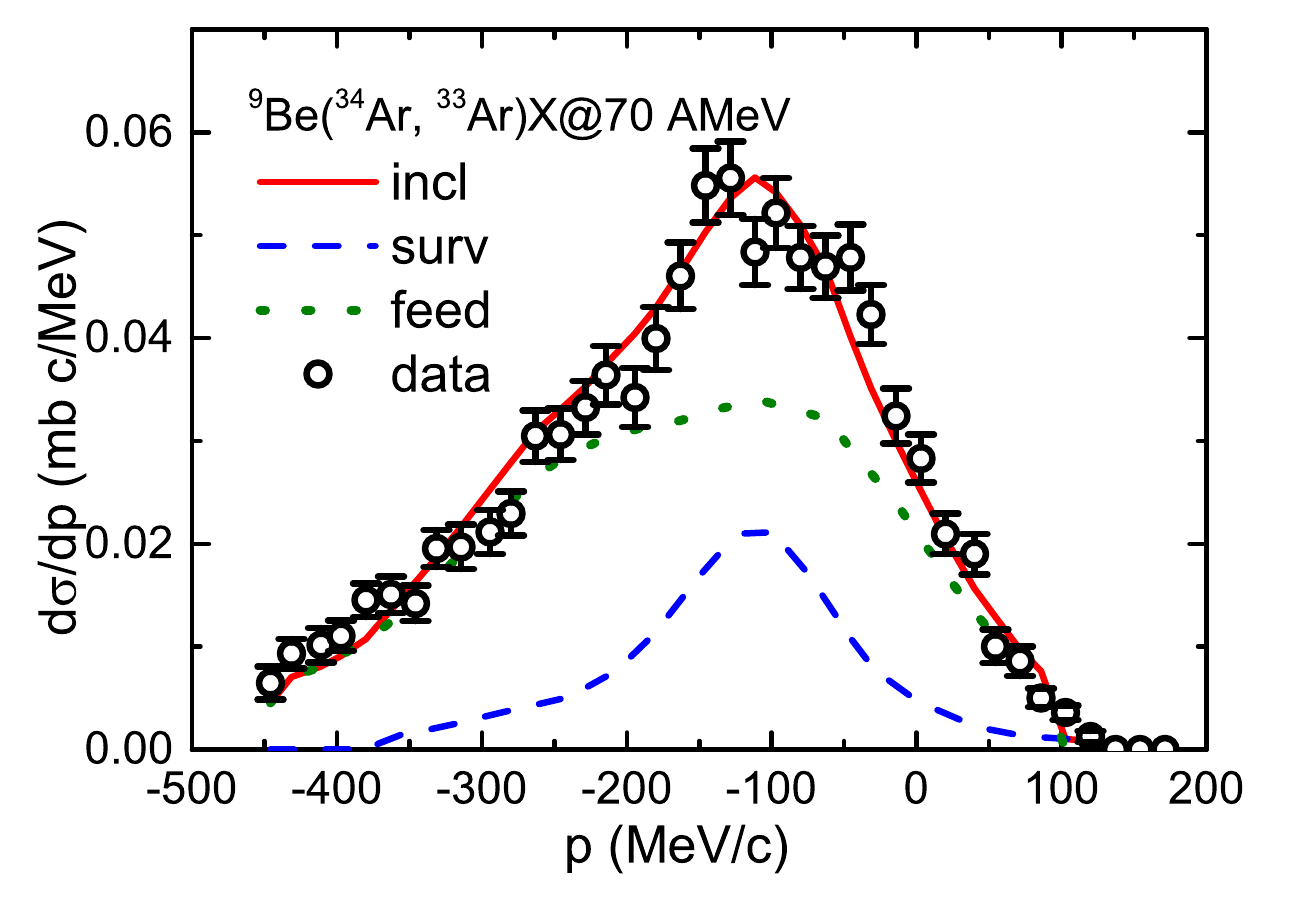}
\caption{Same as Fig. \ref{19C243} but for $^{34}$Ar + $^{9}$Be collisions at 70 AMeV \cite{gade2004one}.}
    \label{S3_34Ar70}
\end{figure}

\begin{figure}[H]   
\centering
    \includegraphics[width=0.45\textwidth,angle=0]{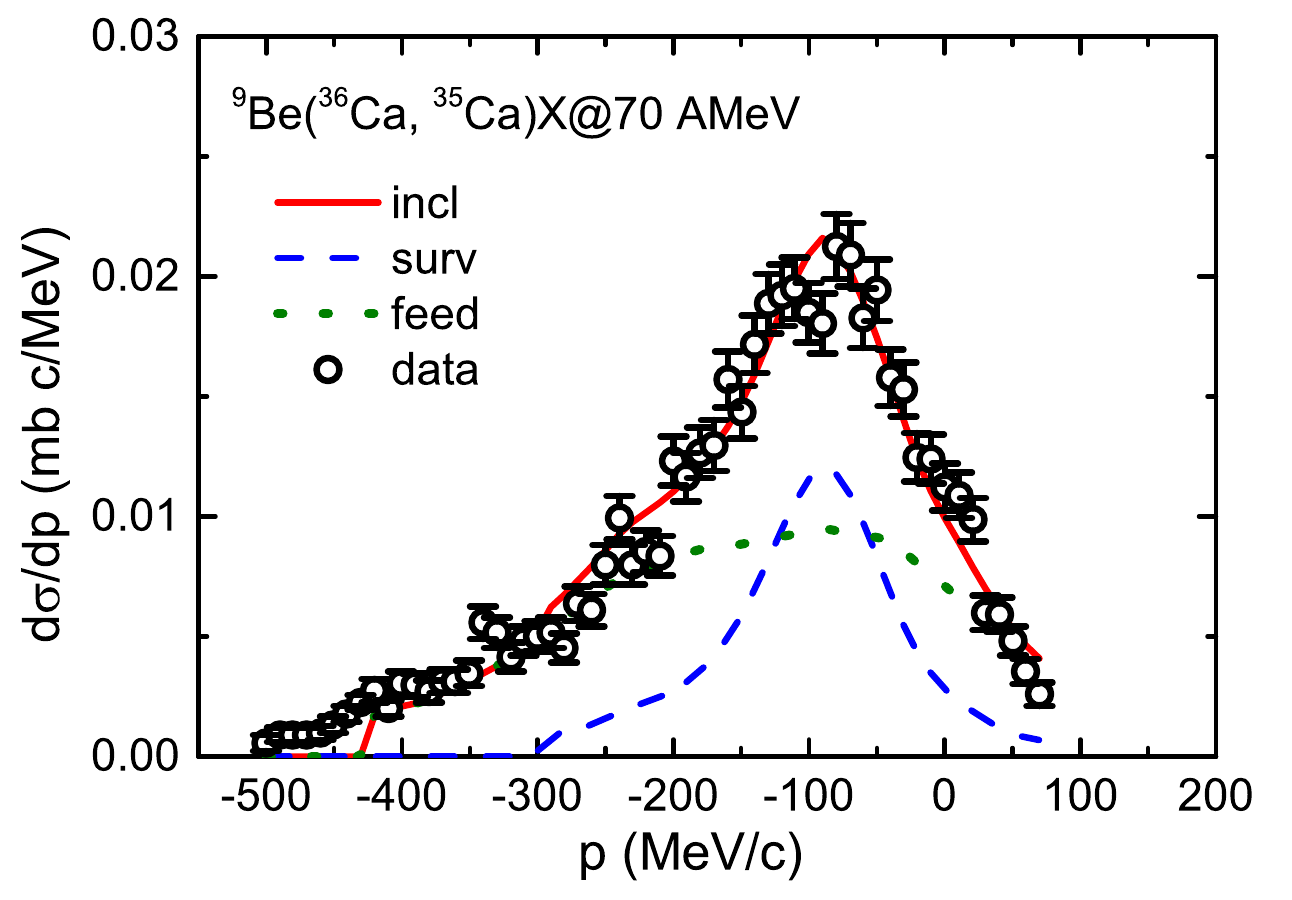}
\caption{Same as Fig. \ref{19C243} but for $^{36}$Ca + $^{9}$Be collisions at 70 AMeV \cite{shane2012proton}.}
    \label{S3_36Ca70}
\end{figure}

\begin{figure}[H]   
\centering
    \includegraphics[width=0.45\textwidth,angle=0]{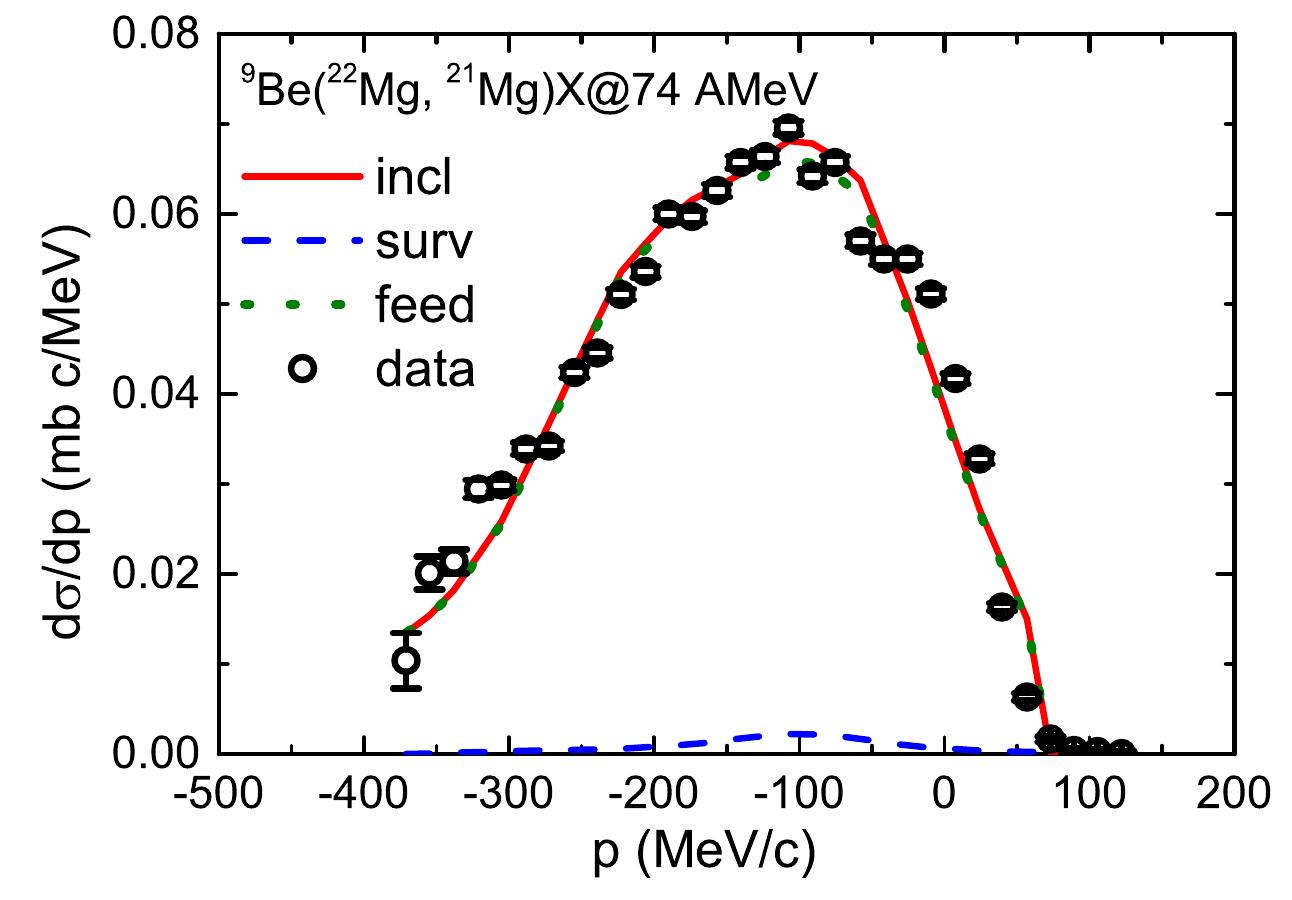}
\caption{Same as Fig. \ref{19C243} but for $^{22}$Mg + $^{9}$Be collisions at 74 AMeV \cite{diget2008structure}.}
    \label{S3_22Mg74}
\end{figure}

\begin{figure}[H]   
\centering
    \includegraphics[width=0.45\textwidth,angle=0]{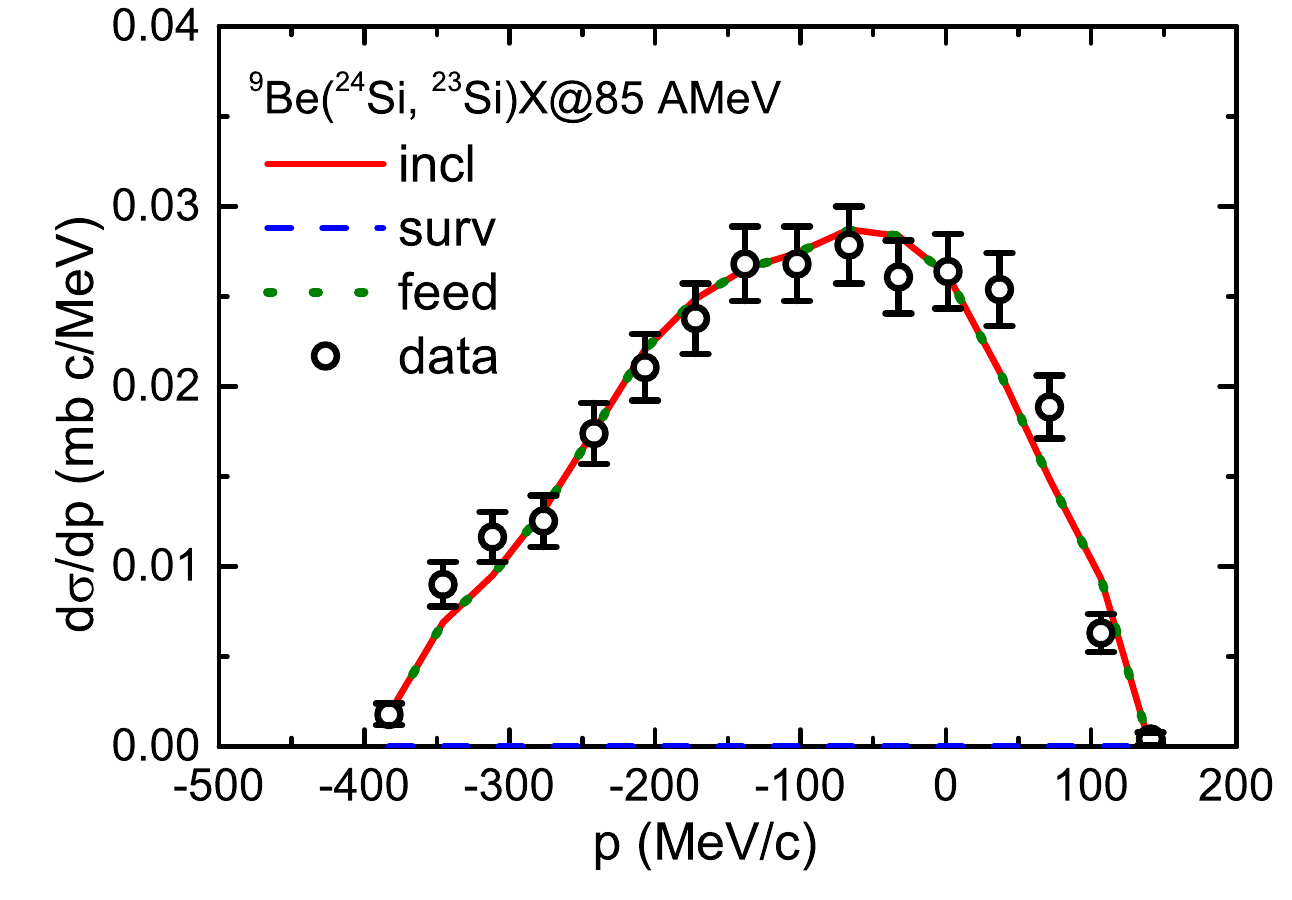}
\caption{Same as Fig. \ref{19C243} but for $^{24}$Si + $^{9}$Be collisions at 85 AMeV \cite{gade2008reduction}.}
    \label{S3_24Si85}
\end{figure}

\begin{figure}[H]   
\centering
    \includegraphics[width=0.45\textwidth,angle=0]{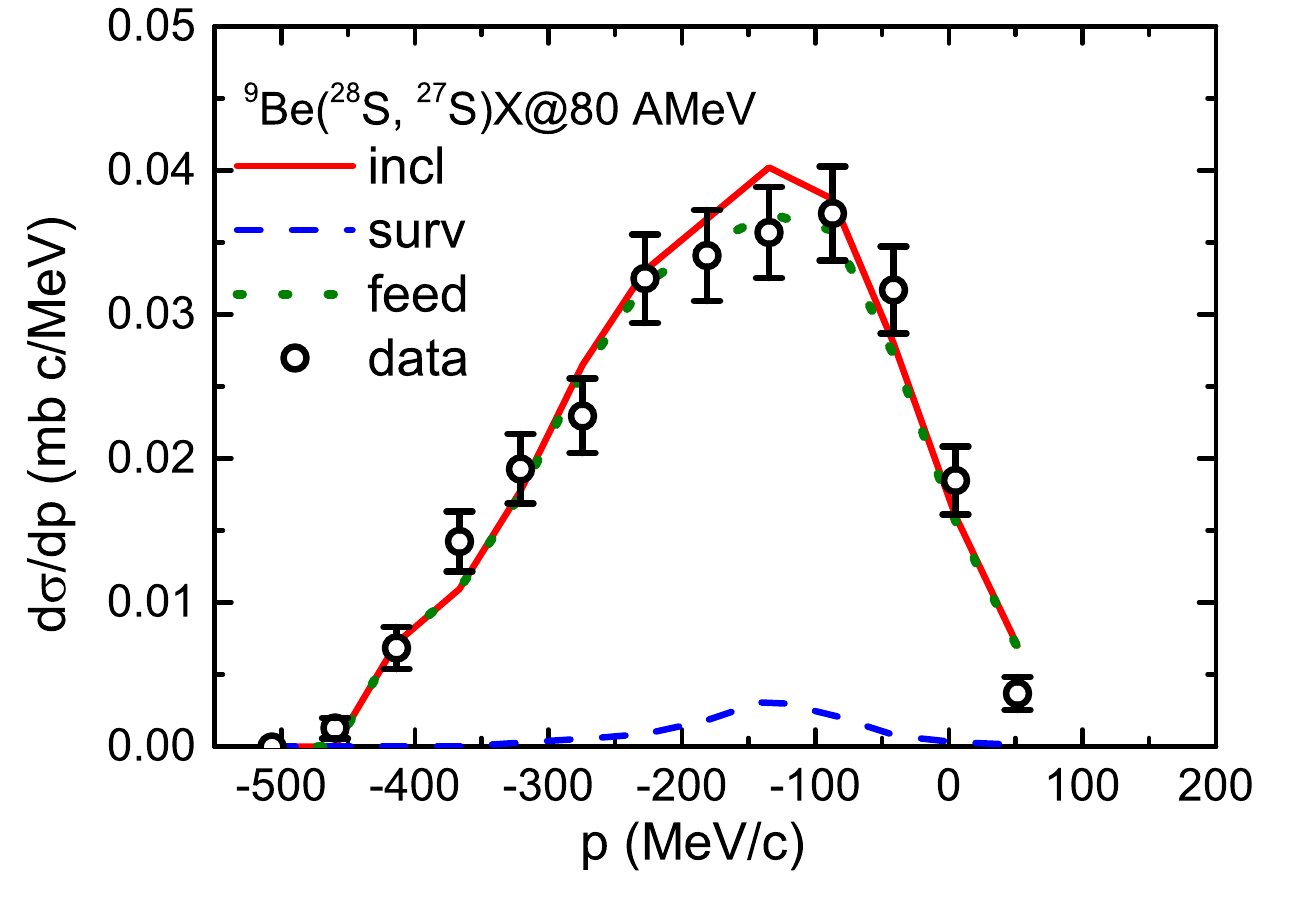}
\caption{Same as Fig. \ref{19C243} but for $^{28}$S + $^{9}$Be collisions at 80 AMeV \cite{gade2008reduction}.}
    \label{28S80}
\end{figure}

\begin{figure}[H]   
\centering
    \includegraphics[width=0.45\textwidth,angle=0]{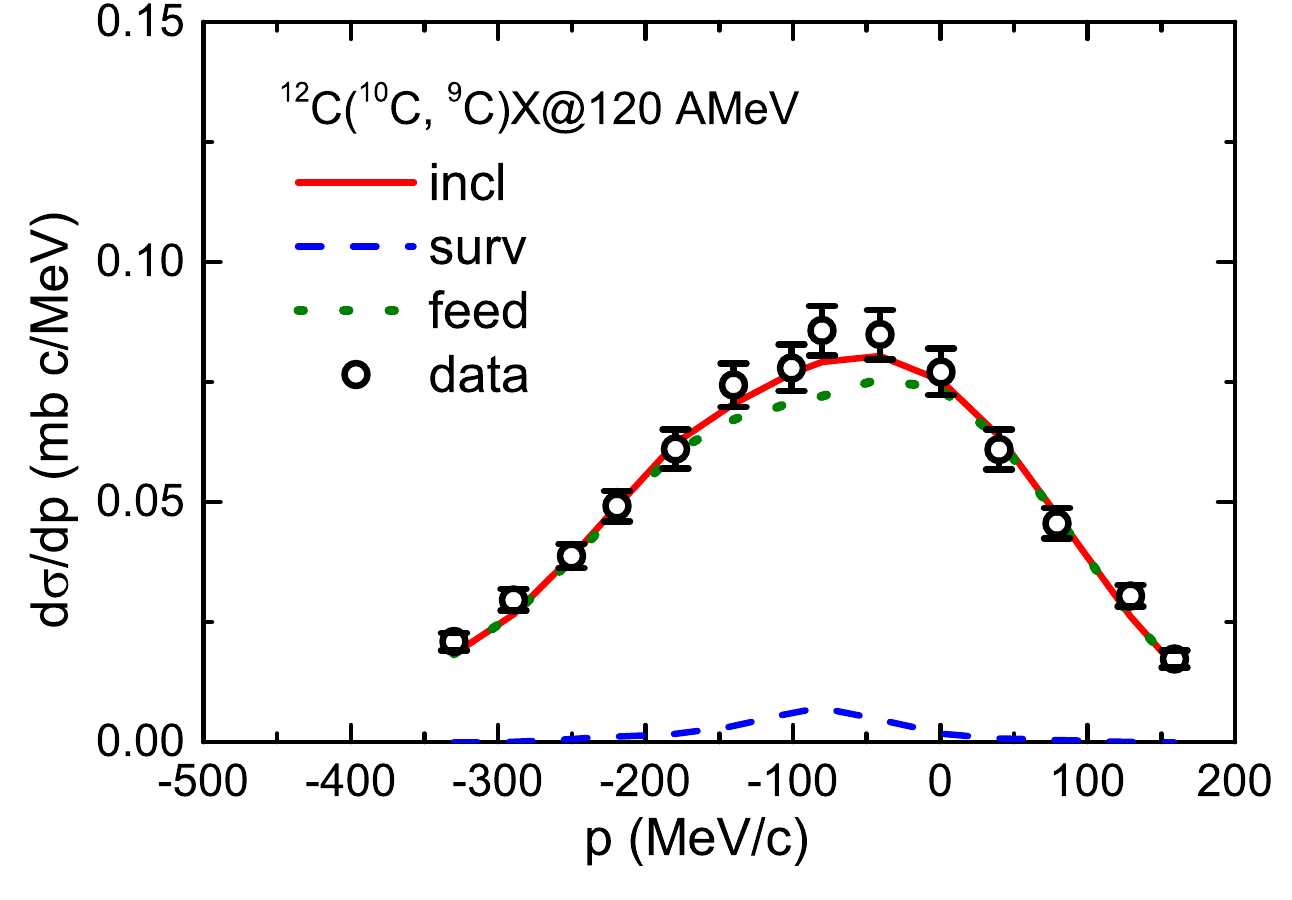}
\caption{Same as Fig. \ref{19C243} but for $^{10}$C + $^{12}$C collisions at 120 AMeV \cite{grinyer2012systematic}.}
    \label{S3_10C120}
\end{figure}

\begin{figure}[H]   
\centering
    \includegraphics[width=0.45\textwidth,angle=0]{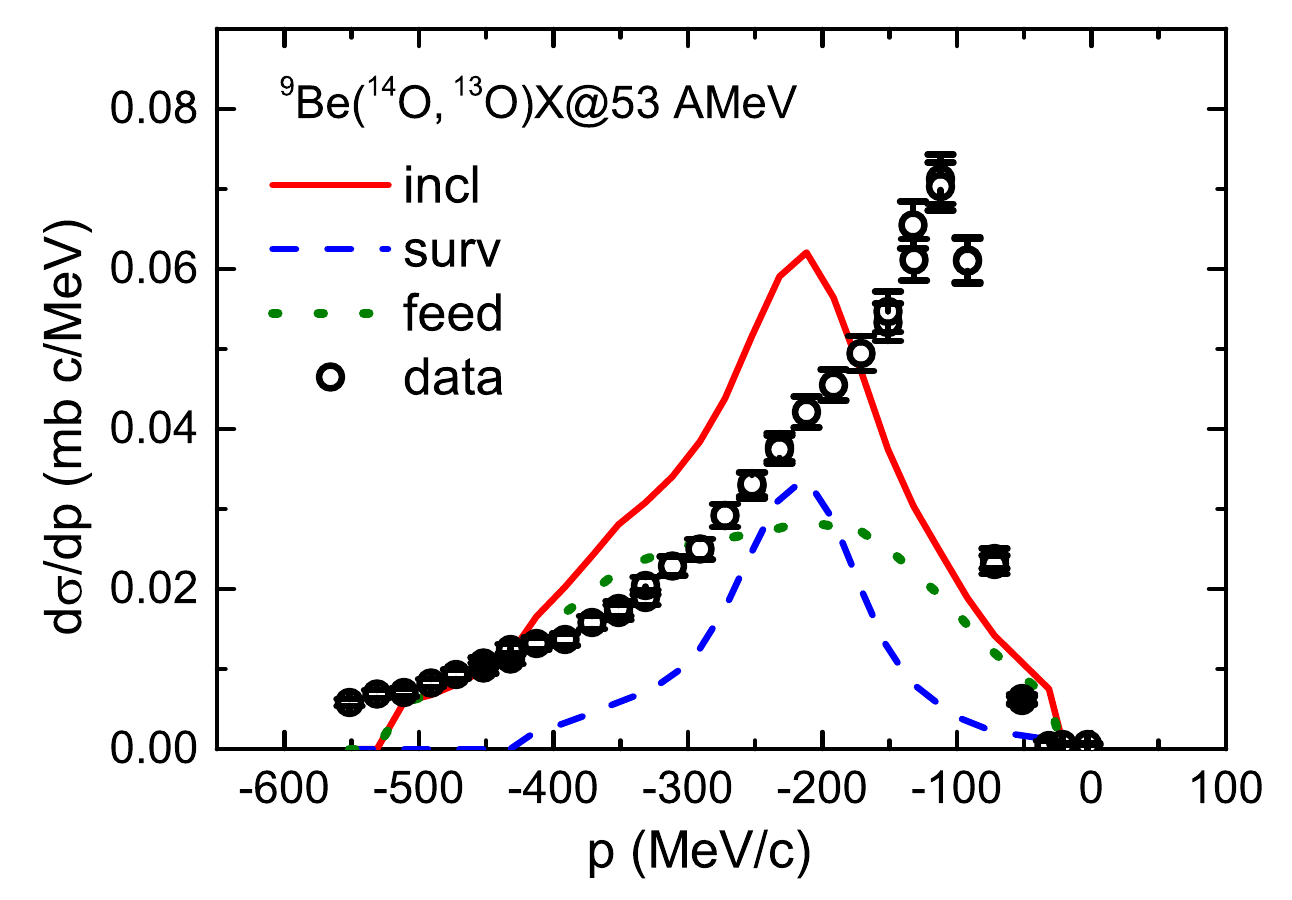}
\caption{Same as Fig. \ref{19C243} but for $^{14}$O + $^{9}$Be collisions at 53 AMeV \cite{flavigny2012nonsudden}.}
    \label{S3_14O53}
\end{figure}

To test the sensitivity of the extracted feeding fraction to the choice of anchor spectra, we repeated the same two-template extraction using two additional template pairs, ${}^{19}$C/${}^{28}$S and ${}^{18}$C/${}^{24}$Si, in addition to the default ${}^{20}$C/${}^{32}$Ar pair.
The resulting $f^{\rm exp}_{\rm feed}$ values are compared in Fig.~\ref{fig:template_sets_feed}.
Although individual points show variations associated with the adopted ansatz, the three anchor choices give consistent $\Delta S$ dependence within uncertainties: a large feeding contribution on the neutron-rich side that decreases toward the neutron-deficient side.

\clearpage

\section*{Appendix D. A threshold--tail model for the evaporation-loss probability $P_{\rm loss}$}
\addcontentsline{toc}{section}{Appendix D. A threshold--tail model for the evaporation-loss probability $P_{\rm loss}$}

The purpose of this appendix is to provide a compact effective estimate for the evaporation-loss
probability within the class of events that populate the excited prefragment
$^{A-1}_{~~~Z}X^\ast$ immediately after the dynamical step.
This threshold--tail model is not intended to be a microscopic description of all possible dynamical $A-1$
production mechanisms (e.g., transfer-like, inelastic, or quasi-free pathways) and their detailed
branching; rather, it supplies a minimal, separation-energy--controlled estimate of the fraction of
$^{A-1}_{~~~Z}X^\ast$ that fails to reach a particle-stable residue without particle evaporation.

We estimate the evaporation-loss probability
\begin{equation}
P_{\rm loss}\equiv \frac{\sigma_{\rm loss}}{\sigma_{\rm dyn}},
\label{eq:Ploss_def_appD}
\end{equation}
where $\sigma_{\rm dyn}$ is the dynamical $A-1$ prefragment-production cross section, and $\sigma_{\rm loss}$ denotes the subset of $\sigma_{\rm dyn}$
that undergoes statistical evaporation of at least one particle before reaching a particle-stable residue.
The survival fraction of the dynamical $A-1$ residue (zero-particle evaporation) is $P_{\rm surv}=1-P_{\rm loss}$.

Let the excitation energy $E^\ast$ of $^{A-1}_{~~~Z}X^\ast$ follow a normalized distribution $w(E^\ast)$,
\begin{equation}
\int_{0}^{\infty} w(E^\ast)\,dE^\ast = 1.
\label{eq:w_norm_appD}
\end{equation}
We approximate the condition ``evaporate at least one particle'' by a single effective threshold $E_{\rm th}$:
for $E^\ast<E_{\rm th}$ particle evaporation is suppressed (dominantly $\gamma$ de-excitation),
whereas for $E^\ast\ge E_{\rm th}$ at least one nucleon emission is likely.
This yields the tail-area estimate
\begin{equation}
P_{\rm loss}\approx \int_{E_{\rm th}}^{\infty} w(E^\ast)\,dE^\ast,
\qquad
P_{\rm surv}=1-P_{\rm loss}\approx \int_{0}^{E_{\rm th}} w(E^\ast)\,dE^\ast.
\label{eq:Ploss_tail_appD}
\end{equation}

We stress that Eq.~(\ref{eq:Ploss_tail_appD}) is deliberately formulated as a minimal surrogate:
while a full statistical-decay calculation would express the survival probability through ratios of
energy-dependent partial widths (as implemented in GEMINI), here we adopt a one-parameter
threshold criterion to keep the estimate transparent and independent of detailed GEMINI decay parameters.
This avoids introducing a second detailed evaporation model and preserves the role of
IQMD+GEMINI as an external baseline for cross-checking.

For the light nuclei considered here, we neglect $\alpha$ emission and take
\begin{equation}
E_{\rm th}=\min\!\Big(E_n,\;E_p\Big),
\qquad
E_n = S_n\!\big(^{A-1}_{~~~Z}X\big),
\qquad
E_p = S_p\!\big(^{A-1}_{~~~Z}X\big)+V_C^{(p)}.
\label{eq:Eth_appD}
\end{equation}
All separation energies are evaluated for the same mother nucleus $^{A-1}_{~~~Z}X$.
Here $E_{\rm th}$ should be viewed as an effective single-particle emission threshold;
for proton emission we approximate the barrier effect by an additive shift $V_C^{(p)}$,
sufficient for capturing the dominant $Z$-dependence in a compact closed form.
For proton emission, we use a contact-barrier estimate
\begin{equation}
V_C^{(p)} \approx
\frac{1.44\,(Z-1)}{r_0\big((A-2)^{1/3}+1\big)}.
\label{eq:Vc_p_appD}
\end{equation}
with a standard radius parametrization $R=r_0A^{1/3}$ and $r_0\simeq 1.2$~fm,
so that $1.44$ denotes $e^2$ in MeV$\cdot$fm.

Adopting the minimal exponential tail used in the main text,
\begin{equation}
w(E^\ast)=\frac{1}{E_{0}}\exp\!\left(-\frac{E^\ast}{E_{0}}\right),\qquad E^\ast\ge 0,
\label{eq:w_exp_appD}
\end{equation}
the tail integral in Eq.~(\ref{eq:Ploss_tail_appD}) gives
\begin{equation}
P_{\rm loss}\approx \exp\!\left(-\frac{E_{\rm th}}{E_{0}}\right),
\qquad
P_{\rm surv}\approx 1-\exp\!\left(-\frac{E_{\rm th}}{E_{0}}\right).
\label{eq:Ploss_closed_appD}
\end{equation}
Combining the above,
\begin{equation}
P_{\rm loss}\approx \exp\!\left[
-\frac{1}{E_{0}}\min\!\Big(S_n,\;S_p+V_C^{(p)}\Big)
\right],
\label{eq:Ploss_final_appD}
\end{equation}
where $S_n$ and $S_p$ are separation energies of the prefragment nucleus $^{A-1}_{~~~Z}X$.
Thus $P_{\rm loss}$ depends only on the effective threshold $E_{\rm th}$ and a single tail-scale parameter $E_{0}$.
The scale $E_0$ is the only parameter of this effective model.
In this work $E_0$ is not adjusted to reproduce the IQMD+GEMINI results; instead,
we use representative values to illustrate the magnitude of $P_{\rm loss}$ and to demonstrate that,
over a broad range of $E_0$, the qualitative $\Delta S$-dependence is set chiefly by the effective
threshold $E_{\rm th}$, while varying $E_0$ mainly rescales the overall magnitude (and thus the
sensitivity) of the leakage probability.
A dedicated scan over $E_0=1$--$9$~MeV is presented in Fig.~4 of main text.

In the global mechanism-decomposition adopted in the main text, pathways occurring during the
dynamical stage (up to $\sim 100$~fm/$c$) are not further subdivided by their microscopic labels.
Their impact on the inclusive one-neutron-removal yield is accounted for through the state reached
after the fast stage: events that populate $^{A-1}X^\ast$ contribute to the dynamical $A-1$ branch and may leak
via evaporation (quantified here by $P_{\rm loss}$), whereas events that first populate $^{A}X^\ast$
and subsequently evaporate into the $^{A-1}X$ residues are collected in the feeding component
discussed in the main text.
Accordingly, this appendix focuses only on estimating the evaporation-loss fraction within the
dynamical $A-1$ branch.

\FloatBarrier    

\bibliographystyle{elsarticle-num} 
\bibliography{onr_ref2}